\documentclass[tightenlines,floats,aps,nofootinbib,prd,onecolumn,preprintnumbers]{revtex4}

\usepackage[dvipdfmx]{graphicx}
\usepackage{amsmath,amssymb,algorithm,algorithmic,bm,color,mathrsfs}
\allowdisplaybreaks[1]

\newcommand{\dd}{\partial}
\newcommand{\xx}{\boldsymbol{x}}

\newcommand{\HH}{\mathcal{H}}

\newcommand{\Mpl}{M_{\rm pl}}

\begin{document}

\title{Testing gravity theories with cosmic microwave background in the degenerate higher-order scalar-tensor theory}

\author{Takashi Hiramatsu$^1$\footnote{Email: hiramatz@rikkyo.ac.jp} and Daisuke Yamauchi$^2$\footnote{Email: yamauchi@jindai.jp}}
\affiliation{
$^1$ Department of Physics, Rikkyo University, Toshima, Tokyo 171-8501, Japan\\
$^2$ Faculty of Engineering, Kanagawa University, Kanagawa, 221-8686, Japan
}

\begin{abstract}
We study the cosmic microwave background (CMB) in the framework of the degenerate higher-order
scalar-tensor (DHOST) theory to test gravity theories. This theoretical framework includes
the wide class of dark energy models such as the Horndeski theory and its extensions as certain 
limits, and the general relativity can be also recovered. In this study, to test gravity theories 
with CMB, we formulate the linear perturbations of gravity and matter in the theory and their 
effective description parameterised by time-dependent effective field theory (EFT) parameters, 
$\alpha_i$ $(i=B,K,T,M,H,L)$ and $\beta_i$ $(i=1,2,3)$. Based on the resultant DHOST framework, 
we develop a numerical code to solve Boltzmann equations consistently. We then show that the 
angular power spectra of the CMB temperature anisotropies, E-mode and lensing potential 
as a demonstration and find that the parameter characterising the DHOST theory, 
$\beta_1$, provides the larger modifications of the spectra, compared with other EFT 
parameters. We also show the results in the case of a specific model in which the cosmic 
expansion as well as the EFT parameters are consistently determined.
\end{abstract}

\preprint{RUP-20-14}

\maketitle

%%%%%%%%%%%%%%%%%%%%%%%%%%%%%%%%%%%%%%%%%%%%%%%%%%%%%%%%%%%%%%%%%%%%%%%%%%%%%%%
%%%%%%%%%%%%%%%%%%%%%%%%%%%%%%%%%%%%%%%%%%%%%%%%%%%%%%%%%%%%%%%%%%%%%%%%%%%%%%%
\section{Introduction}
\label{sec:intro}
%%%%%%%%%%%%%%%%%%%%%%%%%%%%%%%%%%%%%%%%%%%%%%%%%%%%%%%%%%%%%%%%%%%%%%%%%%%%%%%
%%%%%%%%%%%%%%%%%%%%%%%%%%%%%%%%%%%%%%%%%%%%%%%%%%%%%%%%%%%%%%%%%%%%%%%%%%%%%%%

The nearly simultaneous detection of the gravitational event GW170817 and its optical counterpart GRB170817A 
has put a strong constraint on the speed of gravitational waves propagating from a neutron star binary such that
it should not deviate from that of light, $|c_{\rm gw}-c|\leq 10^{-15}$\cite{TheLIGOScientific:2017qsa,Monitor:2017mdv,GBM:2017lvd},
which forbids any extensions of General Relativity (GR) predicting a large deviation of $c_{\rm gw}$.
This measurement can therefore put constraints on scalar-tensor theory as alternative to dark 
energy~\cite{Creminelli:2017sry,Ezquiaga:2017ekz,Baker:2017hug,Sakstein:2017xjx,Langlois:2017dyl}.
To explore the theories of gravity beyond GR, 
the degenerate higher-order scalar-tensor (DHOST) theory
\cite{Langlois:2015cwa,Langlois:2015skt} (see \cite{Langlois:2018dxi} for review and references therein)
is useful since most of known theories of gravity so far, such as the Horndeski 
theory~\cite{Horndeski:1974wa,Deffayet:2011gz,Kobayashi:2011nu} 
and the beyond-Horndeski theory~\cite{Gleyzes:2014dya,Gleyzes:2014qga}, are included.
The DHOST theory has eight arbitrary functions of 
the scalar field $\phi$ and $X=\partial_\mu\phi\partial^\mu\phi$, dubbed as
$P(\phi,X), Q(\phi,X), f_2(\phi,X)$ and $a_i(\phi,X)$ with $i=1,\cdots,5$.
There are three degeneracy conditions eliminating the unwanted higher-order time-derivative terms.
The measurement of GW170817 strongly implies that the propagation speed of gravitational waves and 
the speed of light strictly coincide, that is $c_{\rm gw}=c$.
Even when imposing this condition, a certain subclass of type-I quadratic DHOST theory
survived~\cite{Creminelli:2017sry,Langlois:2017dyl}.
This theory is still phenomenologically interesting because the Vainshtein screening mechanism 
is successfully implemented outside matter, whereas its partial breaking occurs 
inside \cite{Langlois:2017dyl,Kobayashi:2014ida,Crisostomi:2017lbg,Dima:2017pwp,Hirano:2019scf}. 
This phenomenon can be used to put the additional constraints on the DHOST theory.
Moreover, several theoretical constraints on the DHOST theory have been discussed in the 
literatures \cite{Creminelli:2018xsv,Creminelli:2019kjy}. 

However, the propagation of gravitational waves from GW170817 as well as the local
measurement of gravity can put any constraints on gravity theories 
at the relatively low redshift, $z\lesssim 0.01$. 
In this sense, there is still a large viability of extended gravity
theories whose deviation from GR emerges at high redshift, say,
$z\gtrsim 1$. One of well-established experiments at such a high
redshift is the cosmic microwave background (CMB). 
As of the Planck experiments, we know that the $\Lambda$CDM model can
well describe our Universe~\cite{Aghanim:2018eyx}.
Hence the extended gravity theories are required to satisfy that the background
evolution should be almost the same as $\Lambda$CDM. %In this sense, 
We then need to explore the dynamics of linear perturbations of gravity and matter
contents in the extended theories of gravity with keeping the background the fiducial one.

In this paper, we investigate the time-evolutions of the metric perturbations, density/velocity perturbations 
of fluid components and the perturbation of the scalar field after reheating.
To do so, we employ the effective description of the DHOST theory, following
the approach called the Effective Field Theory (EFT) of Dark Energy  
\cite{Bloomfield:2012ff,Gubitosi:2012hu,Gleyzes:2013ooa,Bloomfield:2013efa,Piazza:2013coa,Gleyzes:2014rba,Gleyzes:2015pma,DAmico:2016ntq,Langlois:2017mxy}.
The EFT describing the DHOST theory has nine time-dependent parameters, and 
the degeneracy conditions mentioned later reduces them to six parameters,
$\alpha_i(t)$ with $i=K,B,T,M,H$ and $\beta_1(t)$, which are defined as the coefficients of 
terms such as $\delta K^{ij}\delta K_{ij}$ and $\delta K\delta R$ in the ADM Lagrangian and vanish in GR.
The EFT parameters are frequently assumed to scale as (e.g., \cite{DAmico:2016ntq})
%%%%%%%%%%%%%%%%%%%%%%%%%%%%%%%%%%%%%%%%%%%%%%%%%%%%%%%%%%%%%%%%%%%%%%%%%%%%%%%
%
\begin{align}
 \alpha_i(t) = \alpha_{i,0}\frac{\Omega_{\rm DE}(t)}{\Omega_{\rm DE,0}},\quad
 \beta_1(t) = \beta_{1,0}\frac{\Omega_{\rm DE}(t)}{\Omega_{\rm DE,0}},\label{eq:scaling1}
\end{align}
%
%%%%%%%%%%%%%%%%%%%%%%%%%%%%%%%%%%%%%%%%%%%%%%%%%%%%%%%%%%%%%%%%%%%%%%%%%%%%%%%
where $\Omega_{\rm DE}(t)$ and $\Omega_{\rm DE,0}$ denote the fractional energy density of the dark 
energy and its present value. 
Following this parametrisation, the EFT parameters are negligible in the early Universe where 
$\Omega_{\rm DE}(t)$ is quite small, recovering GR. Therefore we do not need to consider 
the modification of the initial perturbations in solving the Boltzmann equation 
from very high redshift. Then we compute the angular power spectra of 
the CMB temperature anisotropies ($C^{TT}_\ell$), 
E-mode polarisation ($C^{EE}_\ell$), and the lensing potential ($C^{\phi\phi}_\ell$).

The EFT approach is useful to for us to know the impacts of modification of
each term in the Lagrangian on the time-evolution of the perturbations in the fixed background.
Strictly speaking, however, the background geometry is not consistently
treated in the EFT approach. In the DHOST theory, the time-evolution of
background scalar field $\dot{\phi}_0(t)$ modifies the Friedmann equation and
determines how the EFT parameters evolve in time.
To demonstrate a consistent way to describe both the background and the perturbations,
we also solve the set of equations with the EFT parameters and the cosmic expansion history
computed from the DHOST theory with the parametrisation of the arbitrary functions therein proposed by
Crisostomi and Koyama \cite{Crisostomi:2017pjs}.

This paper is organised as follows. In Sec.~\ref{sec:theories}, we derive the evolution equations of 
the background and linear perturbations in the DHOST theory. In Sec.~\ref{sec:EFT}, we derive them 
in the effective description of the type-I DHOST theory, and explicitly show the relations 
between the EFT parameters, $\alpha_i$ and $\beta_1$, and the scalar field.
In Sec.~\ref{sec:setup}, we briefly explain the setup for the numerical calculations. 
In Sec.~\ref{sec:results_eft}, we show the angular power spectra and how precisely we can estimate 
the EFT parameters according to the Fisher analysis.
In Sec.~\ref{sec:results_ck}, we demonstrate a consistent treatment 
of the background geometry and the perturbations with a specific model.
Finally, we conclude in Sec.~\ref{sec:conclusion}. 
Since the derived equations are too long to show in the 
main text, Appendix supplements the main text. Throughout the paper, we use the unit with
$c=\hbar=1$, and $\Mpl^{-2} := 8\pi G$ where $G$ is the usual Newton constant.

%%%%%%%%%%%%%%%%%%%%%%%%%%%%%%%%%%%%%%%%%%%%%%%%%%%%%%%%%%%%%%%%%%%%%%%%%%%%%%%
%%%%%%%%%%%%%%%%%%%%%%%%%%%%%%%%%%%%%%%%%%%%%%%%%%%%%%%%%%%%%%%%%%%%%%%%%%%%%%%
\section{DHOST theory}
\label{sec:theories}
%%%%%%%%%%%%%%%%%%%%%%%%%%%%%%%%%%%%%%%%%%%%%%%%%%%%%%%%%%%%%%%%%%%%%%%%%%%%%%%
%%%%%%%%%%%%%%%%%%%%%%%%%%%%%%%%%%%%%%%%%%%%%%%%%%%%%%%%%%%%%%%%%%%%%%%%%%%%%%%

%%==============================================================
%%==============================================================
\subsection{Basics}
%%==============================================================
%%==============================================================

We consider the quadratic DHOST theory, whose action is given as \cite{Langlois:2015cwa}
%%%%%%%%%%%%%%%%%%%%%%%%%%%%%%%%%%%%%%%%%%%%%%%%%%%%%%%%%%%%%%%%%%%%%%%%%%%%%%%
%
\begin{align}
 S &= \int\!d^4x\,\sqrt{-g}\,\mathcal{L}_{\rm DHOST} + \int\!d^4x\,\sqrt{-g}\,\mathcal{L}_{m},
\end{align}
%
%%%%%%%%%%%%%%%%%%%%%%%%%%%%%%%%%%%%%%%%%%%%%%%%%%%%%%%%%%%%%%%%%%%%%%%%%%%%%%%
where we assume that the Lagrangian for matter, $\mathcal{L}_m$, minimally couples to gravity, and
%%%%%%%%%%%%%%%%%%%%%%%%%%%%%%%%%%%%%%%%%%%%%%%%%%%%%%%%%%%%%%%%%%%%%%%%%%%%%%%
%
\begin{align}
 \mathcal{L}_{\rm DHOST} &:= P(\phi,X) + Q(\phi,X)\Box\phi + f_2(\phi,X){}^{(4)}R
    + \sum_{i=1}^5a_{i}(\phi,X)\mathcal{L}_{i}, \label{eq:DHOST_action}
\end{align}
%
%%%%%%%%%%%%%%%%%%%%%%%%%%%%%%%%%%%%%%%%%%%%%%%%%%%%%%%%%%%%%%%%%%%%%%%%%%%%%%%
with $P,Q,f_2$ and $a_{i}$ being arbitrary functions of $\phi$ and 
$X:=\partial_\mu\phi\partial^\mu\phi$.
The Lagrangians for derivative couplings of the scalar field are described as 
%%%%%%%%%%%%%%%%%%%%%%%%%%%%%%%%%%%%%%%%%%%%%%%%%%%%%%%%%%%%%%%%%%%%%%%%%%%%%%%
%
\begin{align}
 \mathcal{L}_1 &:= \phi_{\mu\nu}\phi^{\mu\nu}, \quad
 \mathcal{L}_2 := (\Box\phi)^2, \quad
 \mathcal{L}_3 := (\Box\phi)\phi^\mu\phi_{\mu\nu}\phi^\nu,\quad
 \mathcal{L}_4 := \phi^\mu\phi_{\mu\rho}\phi^{\rho\nu}\phi_\nu, \quad
 \mathcal{L}_5 := (\phi^\mu\phi_{\mu\nu}\phi^\nu)^2,
\end{align}
%
%%%%%%%%%%%%%%%%%%%%%%%%%%%%%%%%%%%%%%%%%%%%%%%%%%%%%%%%%%%%%%%%%%%%%%%%%%%%%%%
with $\phi_\mu:=\nabla_\mu\phi$ and $\phi_{\mu\nu}:=\nabla_\mu\nabla_\nu\phi$.
In this paper, we consider the type-I degeneracy condition to avoid the ghost instability, which is 
given by the following three conditions~\cite{Langlois:2015cwa}:
%%%%%%%%%%%%%%%%%%%%%%%%%%%%%%%%%%%%%%%%%%%%%%%%%%%%%%%%%%%%%%%%%%%%%%%%%%%%%%%
%
\begin{align}
&a_2=-a_1\,,\notag\\
&a_4=\frac{1}{8\left(f_2+a_2X\right)^2}\Bigl[
16Xa_2^2+4\left( 3f+16Xf_X\right) a_2^2+\left( 16X^2f_X-12Xf\right) a_3a_2-X^2fa_3^2
\notag\\
&\quad\quad\quad\quad\quad\quad\quad\quad\quad
+16f_X\left( 3f+4Xf_X\right) a_2+8f\left( Xf_X-f\right)a_3+48ff_X^2
\Bigr]\,,\label{eq:degeneracy_DHOST}\\
&a_5=\frac{\left( 4f_X+2a_2+Xa_3\right)\left( -2a_2^2+3Xa_2a_3-4f_Xa_2+4fa_3\right)}{8\left(f+Xa_2\right)^2}
\,,\notag
\end{align}
%
%%%%%%%%%%%%%%%%%%%%%%%%%%%%%%%%%%%%%%%%%%%%%%%%%%%%%%%%%%%%%%%%%%%%%%%%%%%%%%%
where the subscripts $\phi$ and $X$ denote the derivatives with respect to them.
Since the DHOST theory contains the higher-order derivatives, it is useful to introduce the following quantity 
as the variation of the Lagrangian in the gravity sector with a variable A:
%%%%%%%%%%%%%%%%%%%%%%%%%%%%%%%%%%%%%%%%%%%%%%%%%%%%%%%%%%%%%%%%%%%%%%%%%%%%%%%
%
\begin{align}
  \mathcal{E}_A &:= \frac{1}{\sqrt{-g}}\sum_{j=0}(-1)^j\partial_{\mu_1}\cdots\partial_{\mu_j}\frac{\delta (\sqrt{-g}\mathcal{L}_{\rm DHOST})}{\delta\,\partial_{\mu_1}\cdots\partial_{\mu_j} A}\,.
\label{eq:def_EA}
\end{align}
%
%%%%%%%%%%%%%%%%%%%%%%%%%%%%%%%%%%%%%%%%%%%%%%%%%%%%%%%%%%%%%%%%%%%%%%%%%%%%%%%

%%==============================================================
%%==============================================================
\subsection{Background equations}
%%==============================================================
%%==============================================================

We assume that the background metric is a flat Friedmann-Lema\^{i}tre-Robertson-Walker metric (FLRW), 
%%%%%%%%%%%%%%%%%%%%%%%%%%%%%%%%%%%%%%%%%%%%%%%%%%%%%%%%%%%%%%%%%%%%
%
\begin{align}
  ds^2 = -N^2dt^2 + a^2\delta_{ij}dx^idx^j.
\label{eq:background metric}
\end{align}
%
%%%%%%%%%%%%%%%%%%%%%%%%%%%%%%%%%%%%%%%%%%%%%%%%%%%%%%%%%%%%%%%%%%%%
By the use of the quantity defined in Eq.~\eqref{eq:def_EA}, one can easily write the background equation-of-motion 
for the lapse, the scale factor and the scalar field. The explicit expressions for 
$\mathcal{E}_N, \mathcal{E}_a, \mathcal{E}_\phi$ are shown in Appendix \ref{appsec:background}.
With this, the governing equation of the scalar field is simply written as
%%%%%%%%%%%%%%%%%%%%%%%%%%%%%%%%%%%%%%%%%%%%%%%%%%%%%%%%%%%%%%%%%%%%%%%%%%%%%%%
%
\begin{align}
 \mathcal{E}_\phi = 0.
\label{eq:background_phi0}
\end{align}
%
%%%%%%%%%%%%%%%%%%%%%%%%%%%%%%%%%%%%%%%%%%%%%%%%%%%%%%%%%%%%%%%%%%%%%%%%%%%%%%%
To get the evolution equation in the gravity sector, we have to take into account the matter content.
We assume that the matter content is described as fluids. Hence the energy-momentum tensor is given as
%%%%%%%%%%%%%%%%%%%%%%%%%%%%%%%%%%%%%%%%%%%%%%%%%%%%%%%%%%%%%%%%%%%%%%%%%%%%%%%
%
\begin{align}
 T^{\mu\nu} = \frac{2}{\sqrt{-g}}\frac{\delta\sqrt{-g}\mathcal{L}_m}{\delta g_{\mu\nu}}
=\sum_{I=B,C,\gamma,\nu}(\rho_I+p_I)u_I^\mu u_I^\nu + p_Ig^{\mu\nu}, \label{eq:def_T}
\end{align}
%
%%%%%%%%%%%%%%%%%%%%%%%%%%%%%%%%%%%%%%%%%%%%%%%%%%%%%%%%%%%%%%%%%%%%%%%%%%%%%%%
where $\rho_I$ and $p_I$ are the energy density and pressure of $I=b$ (baryon), $c$ (CDM), $\gamma$ (photon), $\nu$ (massless neutrinos), and 
$u^\mu_I$ is the 4-velocity of $I$, $u_I^\mu:=(u_I^0,v_I^i/a)$ with the velocity perturbation $v^i_I$.
The zero-th order of the energy-momentum tensor are calculated as
%%%%%%%%%%%%%%%%%%%%%%%%%%%%%%%%%%%%%%%%%%%%%%%%%%%%%%%%%%%%%%%%%%%%%%%%%%%%%%%
%
\begin{alignat}{2}
 T^{00} &= \rho_{\rm s} := \sum_I\rho_I, 
&\quad T^{ij} &= \frac{\delta_{ij}}{a^2}p_{\rm s}, \quad p_{\rm s}:=\sum_I p_I,
\end{alignat}
%
%%%%%%%%%%%%%%%%%%%%%%%%%%%%%%%%%%%%%%%%%%%%%%%%%%%%%%%%%%%%%%%%%%%%%%%%%%%%%%%
which satisfy the conservation law, $\dot{\rho}_s + 3H(\rho_{s}+p_{s})=0$, where a dot denote
the derivative with respect to $t$. Because
the variations with respect to $N$ and $a$ can be rewritten in terms of those with respect to
metric, $\delta/\delta N = -2\delta/\delta g_{00}$ and $\delta/\delta a = 2a\delta_{ij}(\delta/\delta g_{ij})$, 
we obtain the extended Friedmann equation and acceleration equation in the DHOST theory,
%%%%%%%%%%%%%%%%%%%%%%%%%%%%%%%%%%%%%%%%%%%%%%%%%%%%%%%%%%%%%%%%%%%%%%%%%%%%%%%
%
\begin{align}
\mathcal{E}_N &= \rho_{\rm s}, \quad -\frac{a}{3}\mathcal{E}_a = p_{\rm s}, \label{eq:background_Na}
\end{align}
%
%%%%%%%%%%%%%%%%%%%%%%%%%%%%%%%%%%%%%%%%%%%%%%%%%%%%%%%%%%%%%%%%%%%%%%%%%%%%%%%
where the left-hand sides are defined in Eqs.~(\ref{eq:var_N})(\ref{eq:var_a}).

%%==============================================================
%%==============================================================
\subsection{Euler-Lagrange equations for perturbations}
%%==============================================================
%%==============================================================

In this subsection, we briefly discuss the Euler-Lagrange equation derived from the full DHOST Lagrangian Eq.~\eqref{eq:DHOST_action}.
Focusing on the scalar perturbations, we consider
the metric perturbations in the Newton-gauge form, which is defined as
%%%%%%%%%%%%%%%%%%%%%%%%%%%%%%%%%%%%%%%%%%%%%%%%%%%%%%%%%%%%%%%%%%%%%%%%%%%%%%%
%
\begin{align}
  ds^2 = -(1+2\Psi)dt^2 + 2a^2\partial_i\xi dtdx^i + a^2\left[(1+2\Phi)\delta_{ij} +
 \left(\partial_i\partial_i-\frac{1}{3}\delta_{ij}\triangle\right)\eta\right]dx^idx^j,
\label{eq:Newton gauge}
\end{align}
%
%%%%%%%%%%%%%%%%%%%%%%%%%%%%%%%%%%%%%%%%%%%%%%%%%%%%%%%%%%%%%%%%%%%%%%%%%%%%%%%
and the perturbation of the scalar field as
%%%%%%%%%%%%%%%%%%%%%%%%%%%%%%%%%%%%%%%%%%%%%%%%%%%%%%%%%%%%%%%%%%%%%%%%%%%%%%%
%
\begin{align}
 \phi(t,\xx) = \phi_0(t) + \delta\phi(t,\xx).
\end{align}
%
%%%%%%%%%%%%%%%%%%%%%%%%%%%%%%%%%%%%%%%%%%%%%%%%%%%%%%%%%%%%%%%%%%%%%%%%%%%%%%%
The Euler-Lagrange equation Eq.~\eqref{eq:def_EA} for the perturbed variables, $\{\Psi,\Phi,\xi,\eta,\delta\phi\}$, 
can be derived by expanding the full action Eq.~\eqref{eq:DHOST_action} up to the second-order and varying the second-order action
with respect to each variable. 
Although we do not show the explicit expression for each $\mathcal{E}_A$, 
we use the resultant Euler-Lagrange equations
to determine the relation between the DHOST functions, 
$\{P(\phi,X), Q(\phi,X), f_2(\phi,X), a_i(\phi,X)\}$, and the EFT parameters, $\{\alpha_i(t),\beta_i(t)\}$
to be introduced in the later section.

%%%%%%%%%%%%%%%%%%%%%%%%%%%%%%%%%%%%%%%%%%%%%%%%%%%%%%%%%%%%%%%%%%%%%%%%%%%%%%%%
%%%%%%%%%%%%%%%%%%%%%%%%%%%%%%%%%%%%%%%%%%%%%%%%%%%%%%%%%%%%%%%%%%%%%%%%%%%%%%%%
\section{Linear perturbations in effective description of DHOST}
\label{sec:EFT}
%%%%%%%%%%%%%%%%%%%%%%%%%%%%%%%%%%%%%%%%%%%%%%%%%%%%%%%%%%%%%%%%%%%%%%%%%%%%%%%%
%%%%%%%%%%%%%%%%%%%%%%%%%%%%%%%%%%%%%%%%%%%%%%%%%%%%%%%%%%%%%%%%%%%%%%%%%%%%%%%%

In this section, we reformulate 
the linear perturbations of gravity and matter in the DHOST theory,
following the approach called the Effective Field Theory of Dark 
Energy~\cite{Bloomfield:2012ff,Gubitosi:2012hu,Gleyzes:2013ooa,Bloomfield:2013efa,Piazza:2013coa,Gleyzes:2014rba,Gleyzes:2015pma,DAmico:2016ntq,Langlois:2017mxy}.

%%==============================================================
%%==============================================================
\subsection{Effective quadratic action and EFT parameters}
%%==============================================================
%%==============================================================

In the context of the EFT, the metric is usually written in the ADM form:
%%%%%%%%%%%%%%%%%%%%%%%%%%%%%%%%%%%%%%%%%%%%%%%%%%%%%%%%%%%%%%%%%%%%%%%%%%%%%%%
%
\begin{align}
 ds^2 = -N^2dt^2+h_{ij}(dx^i+N^idt)(dx^j+N^jdt),
\end{align}
%
%%%%%%%%%%%%%%%%%%%%%%%%%%%%%%%%%%%%%%%%%%%%%%%%%%%%%%%%%%%%%%%%%%%%%%%%%%%%%%%
To study the linear perturbations for gravity and matter, we need to expand the action up to the second order
around the flat FLRW background given in Eq.~\eqref{eq:background metric} with the gauge $N=1$.
In the unitary gauge, the perturbed variables are the lapse $\delta N\equiv N-1$, the extrinsic curvature 
$\delta K_{ij}=K_{ij}-Hh_{ij}$ and the three-dimensional Ricci curvature ${}^{(3)}R_{ij}$.
To describe the effective action for the DHOST in the EFT language, we need to introduce the time and space derivatives 
of $\delta N$ in the effective Lagrangian.
The effective quadratic action in gravity sector is given as \cite{Langlois:2017mxy}
%%%%%%%%%%%%%%%%%%%%%%%%%%%%%%%%%%%%%%%%%%%%%%%%%%%%%%%%%%%%%%%%%%%%%%%%%%%%%%%
%
\begin{align}
 S^{(2)} &= \int\!d^4x\,\sqrt{-g}\mathcal{L}^{(2)},
\end{align}
%
%%%%%%%%%%%%%%%%%%%%%%%%%%%%%%%%%%%%%%%%%%%%%%%%%%%%%%%%%%%%%%%%%%%%%%%%%%%%%%%
with
%%%%%%%%%%%%%%%%%%%%%%%%%%%%%%%%%%%%%%%%%%%%%%%%%%%%%%%%%%%%%%%%%%%%%%%%%%%%%%%
%
\begin{align}
\mathcal{L}^{(2)} &:= \frac{M^2}{2}
 \left\{
  \delta K_{ij}\delta K^{ij}
 - \left(1+\frac{2}{3}\alpha_L\right)\delta K^2
 +(1+\alpha_T)\left({}^{(3)}R\frac{\delta\sqrt{h}}{a^3}+\delta_2{}^{(3)}R\right)
  \right. \notag \\ &\quad \left.
 + H^2\alpha_K\delta N^2
 + 4H\alpha_B\delta K\delta N
 + (1+\alpha_H)R\delta N
 + 4\beta_1\delta K\dot{\delta N}
 + \beta_2\dot{\delta N}^2
 + \frac{\beta_3}{a^2}(\partial\delta N)^2
 \right\}, \label{eq:quad_action}
\end{align}
%
%%%%%%%%%%%%%%%%%%%%%%%%%%%%%%%%%%%%%%%%%%%%%%%%%%%%%%%%%%%%%%%%%%%%%%%%%%%%%%%
where $H$ is the Hubble parameter, $M$ is the effective Planck mass, and 
$\delta_2$ extracts the second-order terms of the metric perturbations.
We have introduced the eight time-varying parameters characterising the effective quadratic Lagrangian, 
labelled as $\{\alpha_L,\alpha_T,\alpha_K,\alpha_B,\alpha_H,\beta_1,\beta_2,\beta_3\}$.
In addition to them, we introduce a parameter characterising
the time-variation of the effective Planck mass,
%%%%%%%%%%%%%%%%%%%%%%%%%%%%%%%%%%%%%%%%%%%%%%%%%%%%%%%%%%%%%%%%%%%%%%%%%%%%%%%
%
\begin{align}
	\alpha_M=\frac{1}{HM^2}\frac{dM^2}{\dd t}
	\,.
\end{align}
%
%%%%%%%%%%%%%%%%%%%%%%%%%%%%%%%%%%%%%%%%%%%%%%%%%%%%%%%%%%%%%%%%%%%%%%%%%%%%%%%
With these nine EFT parameters, we can fully specify the linear perturbations in the DHOST class of gravity theories.
In the unitary gauge, the scalar perturbations can be defined as
%%%%%%%%%%%%%%%%%%%%%%%%%%%%%%%%%%%%%%%%%%%%%%%%%%%%%%%%%%%%%%%%%%%%%%%%%%%%%%%
%
\begin{align}
 \delta N = N-1, \quad 
N^i=\delta^{ij}\partial_i\psi, \quad 
h_{ij}=a^2e^{2\zeta}\delta_{ij}+a^2\left(\partial_i\partial_i-\frac{1}{3}\delta_{ij}\triangle\right)\eta.
\label{eq:unitary gauge}
\end{align}
%
%%%%%%%%%%%%%%%%%%%%%%%%%%%%%%%%%%%%%%%%%%%%%%%%%%%%%%%%%%%%%%%%%%%%%%%%%%%%%%%
It would be convenient to change the gauge to compare the results derived in the previous section
where the scalar perturbation, $\delta\phi(t,\xx)$, is exposed.
To recover the scalar degree of freedom, we perform the time-coordinate
transformation $t\to t+\pi(t,\xx)$.
In general, 
the infinitesimal coordinate transformation, $x^\mu \to \overline{x}^\mu = x^\mu+\epsilon^\mu$,
for the metric perturbation $\delta g_{\mu\nu}$ is given as
%%%%%%%%%%%%%%%%%%%%%%%%%%%%%%%%%%%%%%%%%%%%%%%%%%%%%%%%%%%%%%%%%%%%%%%%%%%%%%%
%
\begin{align}
 \overline{\delta g}_{\mu\nu}(\overline{x})
 = \delta g_{\mu\nu}(x) - \nabla_{\mu}\epsilon_{\nu} - \nabla_{\nu}\epsilon_{\mu}.
\end{align}
%
%%%%%%%%%%%%%%%%%%%%%%%%%%%%%%%%%%%%%%%%%%%%%%%%%%%%%%%%%%%%%%%%%%%%%%%%%%%%%%%
For the infinitesimal time translation, $t \to \overline{t}=t+\pi(t,\xx)$, 
the displacement vector is given as $\epsilon^{\mu}=(\pi,0)$, and its dual vector is
$\epsilon_\mu=g_{\mu\nu}\epsilon^\nu=(-\pi,0)$, where we truncate the expansion
at the first order of the perturbations. 
Hence the gauge transformation implies
the relation between Eqs.~\eqref{eq:Newton gauge} and \eqref{eq:unitary gauge} as
%%%%%%%%%%%%%%%%%%%%%%%%%%%%%%%%%%%%%%%%%%%%%%%%%%%%%%%%%%%%%%%%%%%%%%%%%%%%%%%
%
\begin{align}
\delta N&=\Psi+\dot{\pi}, \quad \zeta=\Phi+H\pi, \quad \psi =\xi -\frac{1}{a^2}\pi,
\end{align}
%
%%%%%%%%%%%%%%%%%%%%%%%%%%%%%%%%%%%%%%%%%%%%%%%%%%%%%%%%%%%%%%%%%%%%%%%%%%%%%%%
where $\eta$ leaves unchanged under the gauge transformation. 
Rewriting the quadratic Lagrangian (\ref{eq:quad_action})
in terms of the new perturbative quantities, we have
%%%%%%%%%%%%%%%%%%%%%%%%%%%%%%%%%%%%%%%%%%%%%%%%%%%%%%%%%%%%%%%%%%%%%%%%%%%%%%%
%
\begin{align}
\mathcal{L}^{(2)}=\mathcal{L}^{(2)}_0+\frac{1}{a^2}\mathcal{L}^{(2)}_2+\frac{1}{a^4}\mathcal{L}^{(2)}_4,
\label{eq:qlag}
\end{align}
%
%%%%%%%%%%%%%%%%%%%%%%%%%%%%%%%%%%%%%%%%%%%%%%%%%%%%%%%%%%%%%%%%%%%%%%%%%%%%%%%
where $\mathcal{L}^{(2)}_i$ are shown in Appendix \ref{appsec:qlag_term}.

As a result of the time-coordinate transformation, the homogeneous scalar field 
in the unitary gauge acquires the spatial dependence, $\phi(t) \to \phi_0(t) + \delta\phi(t,\xx)$.
Thus
we can identify the spatial fluctuation as \cite{Bellini:2014fua},
%%%%%%%%%%%%%%%%%%%%%%%%%%%%%%%%%%%%%%%%%%%%%%%%%%%%%%%%%%%%%%%%%%%%%%%%%%%%%%%
%
\begin{align}
  \pi := -\frac{\delta\phi}{\dot{\phi}_0}. \label{eq:def_pi}
\end{align}
%
%%%%%%%%%%%%%%%%%%%%%%%%%%%%%%%%%%%%%%%%%%%%%%%%%%%%%%%%%%%%%%%%%%%%%%%%%%%%%%%

%%==============================================================
%%==============================================================
\subsection{Euler-Lagrange equations for perturbations with EFT parameters}
%%==============================================================
%%==============================================================
 
Varying Eq.~\eqref{eq:qlag} 
with respect to $\Psi, \Phi, \xi, \eta$ and $\pi$, we obtain the Euler-Lagrange equations for 
them in the Newton gauge. 
Respecting the relation Eq.~(\ref{eq:def_pi}), these Euler-Lagrange equations in the EFT can 
describe those derived from the original DHOST action 
{\it without} the degeneracy conditions in Eq.~(\ref{eq:degeneracy_DHOST}).
Comparing the coefficients of these equations in the two difference approaches, 
one can easily find the correspondence between the EFT parameters, $\{\alpha_i(t),\beta_i(t)\}$,
and the functions in the DHOST theory, $\{P(\phi ,X),Q(\phi ,X),f_2(\phi ,X),a_i(\phi ,X)\}$.
We found the following relations:
%%%%%%%%%%%%%%%%%%%%%%%%%%%%%%%%%%%%%%%%%%%%%%%%%%%%%%%%%%%%%%%%%%%%%%%%%%%%%%%
%
\begin{align}
M^2 &= 2\left(f_{2}+a_{1}\dot{\phi}_{0}^2\right),\label{eq:param_M}\\
M^2\beta_1 &= \frac{1}{2}\dot{\phi}_{0}^2\left(-2a_{2}-4f_{2X}+a_{3}\dot{\phi}_{0}^2\right),\\
M^2\beta_2 &= 2\dot{\phi}_{0}^2\left(a_{1}+a_{2}-(a_{3}+a_{4})\dot{\phi}_{0}^2+a_{5}\dot{\phi}_{0}^4\right),\\
M^2\beta_3 &= 2\dot{\phi}_{0}^2\left(-2a_{1}+4f_{2X}+a_{4}\dot{\phi}_{0}^2\right),\\
M^2\alpha_L &= -3 (a_{1}+a_{2})\dot{\phi}_{0}^2,\\
M^2\alpha_H &= 2\left(2f_{2X}-a_{1}\right)\dot{\phi}_{0}^2,\\
HM^2\alpha_M &= 2\dot{\phi}_{0}\left(f_{2\phi}+a_{1\phi}\dot{\phi}_{0}^2+2\left(a_{1}-f_{2X}-a_{1X}\dot{\phi}_{0}^2\right)\ddot{\phi}_{0}\right),\\
M^2\alpha_T &= -2a_{1}\dot{\phi}_{0}^2,\label{eq:param_T}\\
2HM^2\alpha_B &= 2f_{2\phi}\dot{\phi}_{0}-2H\left(2a_{1}+3a_{2}-2f_{2X}\right)\dot{\phi}_{0}^2-2\left(2f_{2\phi X}+Q_{X}\right)\dot{\phi}_{0}^3+H\left(-3a_{3}+4a_{1X}+12a_{2X}\right)\dot{\phi}_{0}^4
\notag \\ &
+\left(2\left(a_{1}-2a_{2}-6f_{2X}\right)\dot{\phi}_{0}+\left(3a_{3}-2a_{4}+4a_{2X}+8f_{2XX}\right)\dot{\phi}_{0}^3+2\left(a_{5}-a_{3X}\right)\dot{\phi}_{0}^5\right)\ddot{\phi}_{0},
\end{align}
%
%%%%%%%%%%%%%%%%%%%%%%%%%%%%%%%%%%%%%%%%%%%%%%%%%%%%%%%%%%%%%%%%%%%%%%%%%%%%%%%
and
%%%%%%%%%%%%%%%%%%%%%%%%%%%%%%%%%%%%%%%%%%%%%%%%%%%%%%%%%%%%%%%%%%%%%%%%%%%%%%%
%
\begin{align}
M^2H^2\alpha_K &= 2\left(3H^2\left(a_{1}-4f_{2X}-3K_{1}\left(a_{2}+2f_{2X}\right)\right)-P_{X}+Q_{\phi}\right)\dot{\phi}_{0}^2+6H\left(-3a_{2\phi}+2Q_{X}\right)\dot{\phi}_{0}^3
\notag \\ &
+\left(3H^2\left(9a_{3}-10a_{1X}-18a_{2X}+16f_{2XX}+K_{1}\left(5a_{3}+4a_{2X}+8f_{2XX}\right)\right)+4P_{XX}-2Q_{\phi X}\right)\dot{\phi}_{0}^4
\notag \\ &
+3H\left(4a_{2\phi X}+5a_{3\phi}-4Q_{XX}\right)\dot{\phi}_{0}^5-6H^2\left(-2a_{1XX}-6a_{2XX}+(3+K_{1})a_{3X}\right)\dot{\phi}_{0}^6-6H a_{3\phi X}\dot{\phi}_{0}^7
\notag \\ &
+4\dot{\phi}_{0}\left(-3(a_{1}+a_{2})H-2\left(a_{1\phi}+a_{2\phi}\right)\dot{\phi}_{0}+3H\left(2(a_{3}+a_{4})+a_{1X}+a_{2X}\right)\dot{\phi}_{0}^2
\right.\notag \\ &\quad\left.
+\left(a_{1\phi X}+a_{2\phi X}+3\left(a_{3\phi}+a_{4\phi}\right)\right)\dot{\phi}_{0}^3-3H\left(3a_{5}+a_{3X}+a_{4X}\right)\dot{\phi}_{0}^4-\left(a_{3\phi X}+a_{4\phi X}+4a_{5\phi}\right)\dot{\phi}_{0}^5
\right.\notag \\ &\quad\left.
+3H a_{5X}\dot{\phi}_{0}^6+a_{5\phi X}\dot{\phi}_{0}^7\right)\ddot{\phi}_{0}
\notag \\ &
+\left(4(a_{1}+a_{2})+2\left(3(a_{3}+a_{4})+5a_{1X}+5a_{2X}\right)\dot{\phi}_{0}^2-2\left(12a_{5}+2a_{1XX}+2a_{2XX}+9a_{3X}+9a_{4X}\right)\dot{\phi}_{0}^4
\right.\notag \\ &\quad\left.
+\left(4a_{3XX}+4a_{4XX}+26a_{5X}\right)\dot{\phi}_{0}^6-4a_{5XX}\dot{\phi}_{0}^8\right)\ddot{\phi}_{0}^2
\notag \\ &
+\left(-8(a_{1}+a_{2})\dot{\phi}_{0}+4\left(3(a_{3}+a_{4})+a_{1X}+a_{2X}\right)\dot{\phi}_{0}^3-4\left(4a_{5}+a_{3X}+a_{4X}\right)\dot{\phi}_{0}^5+4a_{5X}\dot{\phi}_{0}^7\right)\dddot{\phi}_{0}.\label{eq:param_aK}
\end{align}
%
%%%%%%%%%%%%%%%%%%%%%%%%%%%%%%%%%%%%%%%%%%%%%%%%%%%%%%%%%%%%%%%%%%%%%%%%%%%%%%%
Here, all the functions are evaluated at the background values, that is, $\phi =\phi_0(t)$ and $X=-\dot\phi_0^2(t)$, 
and we defined the dimensionless time-derivatives of the Hubble parameter,
%%%%%%%%%%%%%%%%%%%%%%%%%%%%%%%%%%%%%%%%%%%%%%%%%%%%%%%%%%%%%%%%%%%%
%
\begin{align}
 K_n := \frac{1}{H^{n+1}}\frac{d^nH}{dt^n}.
\end{align}
%
%%%%%%%%%%%%%%%%%%%%%%%%%%%%%%%%%%%%%%%%%%%%%%%%%%%%%%%%%%%%%%%%%%%%
Before showing the Euler-Lagrange equations for the perturbed variables in terms of the EFT parameters, 
we should discuss the dependence on the background energy density $\rho_s$ and pressure $p_s$, which are 
related to the background Euler-Lagrange quantities, $\mathcal{E}_N$ and $\mathcal{E}_a$, 
through Eq.~\eqref{eq:background_Na}.
Even when the above relations, Eqs.~\eqref{eq:param_M}--\eqref{eq:param_aK}, are taken into account, 
one finds that there are several residuals in the equations derived in the full DHOST Lagrangian (\ref{eq:DHOST_action}), 
compared with those from Eq.~\eqref{eq:qlag}. 
Following the EFT point of view, these residuals should be rewritten in terms of the background quantities. 
We actually confirm that all the residuals can be identified to be a function of $\rho_s$, $p_s$ and 
their time-derivatives, and the results of the full DHOST can be consistently reproduced.~\footnote{We expect that 
the above equations including these missing terms can be consistently derived from the full EFT action 
taking into account the terms describing the background \cite{Gubitosi:2012hu,Bloomfield:2012ff}.}
The explicit expression of the Euler-Lagrange equations with the background term corrections are summarised in Appendix \ref{appsec:eq_metric}.

Let us consider the type-I degeneracy condition in the context of the effective description of the DHOST theory.
In the EFT language, the fully nonlinear type-I degeneracy condition Eq.~\eqref{eq:degeneracy_DHOST} reduces to 
the simpler conditions for the EFT parameters as
%%%%%%%%%%%%%%%%%%%%%%%%%%%%%%%%%%%%%%%%%%%%%%%%%%%%%%%%%%%%%%%%%%%%%%%%%%%%%%%
%
\begin{align}
\alpha_L=0, \quad \beta_2=-6\beta_1^2, \quad \beta_3=-2\beta_1\left[2(1+\alpha_H)+\beta_1(1+\alpha_T)\right],
\label{eq:degeneracy_EFT}
\end{align}
%
%%%%%%%%%%%%%%%%%%%%%%%%%%%%%%%%%%%%%%%%%%%%%%%%%%%%%%%%%%%%%%%%%%%%%%%%%%%%%%%
reducing the number of the free EFT parameters to six. With this reduced degeneracy conditions,
the Euler-Lagrange equations for $\Psi$, $\Phi$, $\xi$, $\eta$ and $\pi$ 
in the Newton gauge are given as
%%%%%%%%%%%%%%%%%%%%%%%%%%%%%%%%%%%%%%%%%%%%%%%%%%%%%%%%%%%%%%%%%%%%%%%%%%%%%%%
%
\begin{align}
-\frac{1}{M^2}\mathcal{E}_\Psi &=
-6\beta_{1}^2\ddot{\Psi}
-6H\beta_{1}\left((3+\alpha_{M})\beta_{1}+\frac{2}{H}\dot{\beta}_{1}\right)\dot{\Psi}
-\frac{2}{a^2}\beta_{1}(2+2\alpha_{H}+(1+\alpha_{T})\beta_{1})\triangle\Psi
\notag \\ &
+H^2\left(6+12\alpha_{B}-\alpha_{K}-6(3+K_{1}+\alpha_{M})\beta_{1}-\frac{6\dot{\beta}_{1}}{H}+\frac{2\rho_{s}}{H^2M^2}\right)\Psi
+6\beta_{1}\ddot{\Phi}
\notag \\ &
+6H\left(-(1+\alpha_{B}-(3+\alpha_{M})\beta_{1})+\frac{\dot{\beta}_{1}}{H}\right)\dot{\Phi}
+\frac{2}{a^2}(1+\alpha_{H})\triangle\Phi
-6\beta_{1}^2\dddot{\pi}
+6H\beta_{1}\left(1-(3+\alpha_{M})\beta_{1}-\frac{2}{H}\dot{\beta}_{1}\right)\ddot{\pi}
\notag \\ &
+H^2(6\alpha_{B}-\alpha_{K}+6K_{1}\beta_{1})\dot{\pi}
-\frac{2}{a^2}\beta_{1}(1+2\alpha_{H}+(1+\alpha_{T})\beta_{1})\triangle\dot{\pi}
+\frac{2H}{a^2}\left(-\alpha_{B}+\alpha_{H}+\beta_{1}+\alpha_{M}\beta_{1}+\frac{\dot{\beta}_{1}}{H}\right)\triangle\pi
\notag \\ &
+6H^3\left(-K_{1}(1+\alpha_{B})+(K_{2}+K_{1}(3+\alpha_{M}))\beta_{1}+\frac{K_{1}}{H}\dot{\beta}_{1}+\frac{\dot{\rho}_{s}}{6H^3M^2}\right)\pi
\label{eq:DHOST_Eq1}
,\\
-\frac{1}{M^2}\mathcal{E}_\Phi &=
6\beta_{1}\ddot{\Psi}
+6H\left(1+\alpha_{B}+(3+\alpha_{M})\beta_{1}+\frac{\dot{\beta}_{1}}{H}\right)\dot{\Psi}
+\frac{2}{a^2}(1+\alpha_{H})\triangle\Psi
\notag \\ &
+6H^2\left((1+\alpha_{B})(3+K_{1}+\alpha_{M})+\frac{\dot{\alpha}_{B}}{H}-\frac{p_{s}+\rho_{s}}{2H^2M^2}\right)\Psi
-6\ddot{\Phi}
-6H(3+\alpha_{M})\dot{\Phi}
+\frac{2}{a^2}(1+\alpha_{T})\triangle\Phi
\notag \\ &
-\frac{6p_{s}}{M^2}\Phi
+6\beta_{1}\dddot{\pi}
+6H\left(\alpha_{B}+(3+\alpha_{M})\beta_{1}+\frac{\dot{\beta}_{1}}{H}\right)\ddot{\pi}
+\frac{2}{a^2}\alpha_{H}\triangle\dot{\pi}
+\frac{2}{a^2}H(-\alpha_{M}+\alpha_{T})\triangle\pi
\notag \\ &
+6H^2\left(K_{1}(-1+\alpha_{B})+\alpha_{B}(3+\alpha_{M})+\frac{\dot{\alpha}_{B}}{H}-\frac{p_{s}+\rho_{s}}{2H^2M^2}\right)\dot{\pi}
-6H^3\left(K_{2}+K_{1}(3+\alpha_{M})+\frac{\dot{p}_{s}}{2H^3M^2}\right)\pi
\label{eq:DHOST_Eq2}
,\\
-\frac{1}{M^2}\mathcal{E}_\xi &=
2\beta_{1}\triangle\dot{\Psi}
+2H(1+\alpha_{B})\triangle\Psi
-2\triangle\dot{\Phi}
+2\beta_{1}\triangle\ddot{\pi}
+2H\alpha_{B}\triangle\dot{\pi}
-2H^2\left(K_{1}+\frac{p_{s}+\rho_{s}}{2H^2M^2}\right)\triangle\pi
\label{eq:DHOST_Eq3}
,\\
-\frac{1}{M^2}\mathcal{E}_\eta &=
-\frac{1}{3a^2}\left[
  (1+\alpha_{H})\triangle\triangle\Psi
  +(1+\alpha_{T})\triangle\triangle\Phi
  +\alpha_{H}\triangle\triangle\dot{\pi}
  -H(\alpha_{M}-\alpha_{T})\triangle\triangle\pi
\right],
\label{eq:DHOST_Eq4}
\end{align}
%
%%%%%%%%%%%%%%%%%%%%%%%%%%%%%%%%%%%%%%%%%%%%%%%%%%%%%%%%%%%%%%%%%%%%%%%%%%%%%%%
and
%%%%%%%%%%%%%%%%%%%%%%%%%%%%%%%%%%%%%%%%%%%%%%%%%%%%%%%%%%%%%%%%%%%%%%%%%%%%%%%
%
\begin{align}
-\frac{1}{M^2}\mathcal{E}_\pi &=
6\beta_{1}^2\dddot{\Psi}
+6H\beta_{1}\left(1+2(3+\alpha_{M})\beta_{1}+\frac{4\dot{\beta}_{1}}{H}\right)\ddot{\Psi}
\notag \\ &
+6H^2\left(-\alpha_{B}+\frac{1}{6}\alpha_{K}+\beta_{1}(6+K_{1}+2\alpha_{M}+(3+\alpha_{M})(3+K_{1}+\alpha_{M})\beta_{1})
\right. \notag \\ & \left.\quad
+\frac{1}{H}\left(\beta_{1}^2\dot{\alpha}_{M}+2(1+2(3+\alpha_{M})\beta_{1})\dot{\beta}_{1}\right)
+\frac{2}{H^2}\left(\dot{\beta}_{1}^2+\beta_{1}\ddot{\beta}_{1}\right)\right)\dot{\Psi}
+\frac{2}{a^2}\beta_{1}(1+2\alpha_{H}+(1+\alpha_{T})\beta_{1})\triangle\dot{\Psi}
\notag \\ &
+\frac{2H}{a^2}\left(-\alpha_{B}+\alpha_{H}+2(1+\alpha_{H})(1+\alpha_{M})\beta_{1}+(1+\alpha_{M})(1+\alpha_{T})\beta_{1}^2+\frac{\beta_{1}}{H}\left(2\dot{\alpha}_{H}+\beta_{1}\dot{\alpha}_{T}\right)
\right. \notag \\ & \left.\quad
+\frac{2}{H}(1+\alpha_{H}+\beta_{1}+\alpha_{T}\beta_{1})\dot{\beta}_{1}\right)\triangle\Psi
+H^3\left(
\frac{\dot\rho_s}{M^2H^3}+(\alpha_K-6\alpha_B)(3+\alpha_M)
\right. \notag \\ & \left.\quad
+6\left(K_{2}+(3+\alpha_{M})^2\right)\beta_{1}
+2K_{1}(-3-9\alpha_{B}+\alpha_{K}+9(3+\alpha_{M})\beta_{1})
\right. \notag \\ & \left.\quad
-\frac{6}{H}\dot{\alpha}_{B}+\frac{\dot{\alpha}_{K}}{H}+\frac{6}{H}\beta_{1}\dot{\alpha}_{M}+\frac{12}{H}(3+K_{1}+\alpha_{M})\dot{\beta}_{1}+\frac{6}{H^2}\ddot{\beta}_{1}\right)\Psi
-6\beta_{1}\dddot{\Phi}
\notag \\ &
+6H\left(\alpha_{B}-2(3+\alpha_{M})\beta_{1}-\frac{2\dot{\beta}_{1}}{H}\right)\ddot{\Phi}
+6H^2\left(\frac{p_{s}+\rho_{s}}{2H^2M^2}+K_{1}+3\alpha_{B}+K_{1}\alpha_{B}+\alpha_{B}\alpha_{M}
\right. \notag \\ & \left.\quad
-(3+\alpha_{M})(3+K_{1}+\alpha_{M})\beta_{1}
+\frac{\dot{\alpha}_{B}}{H}-\frac{\beta_{1}\dot{\alpha}_{M}}{H}-\frac{2}{H}(3+\alpha_{M})\dot{\beta}_{1}-\frac{\ddot{\beta}_{1}}{H^2}\right)\dot{\Phi}
\notag \\ &
-\frac{2}{a^2}\alpha_{H}\triangle\dot{\Phi}
-\frac{2H}{a^2}\left(\alpha_{M}+\alpha_{H}(1+\alpha_{M})-\alpha_{T}+\frac{\dot{\alpha}_{H}}{H}\right)\triangle\Phi
+6\beta_{1}^2\ddddot{\pi}
+12\beta_{1}H\left((3+\alpha_{M})\beta_{1}+\frac{2\dot{\beta}_{1}}{H}\right)\dddot{\pi}
\notag \\ &
+H^2\left(\alpha_{K}+6\beta_{1}(-2K_{1}+(3+\alpha_{M})(3+K_{1}+\alpha_{M})\beta_{1})+\frac{6\beta_{1}}{H}\left(\beta_{1}\dot{\alpha}_{M}+4(3+\alpha_{M})\dot{\beta}_{1}\right)+\frac{12}{H^2}\left(\dot{\beta}_{1}^2+\beta_{1}\ddot{\beta}_{1}\right)\right)\ddot{\pi}
\notag \\ &
+H^3\left(\alpha_{K}(3+2K_{1}+\alpha_{M})-12(K_{2}+K_{1}(3+\alpha_{M}))\beta_{1}+\frac{\dot{\alpha}_{K}}{H}-\frac{12}{H}K_{1}\dot{\beta}_{1}\right)\dot{\pi}
\notag \\ &
+\frac{2}{a^2}\beta_{1}(2\alpha_{H}+(1+\alpha_{T})\beta_{1})\triangle\ddot{\pi}
+\frac{1}{a^2}\Bigl(2\beta_{1}\left(H(1+\alpha_{M})(2\alpha_{H}+(1+\alpha_{T})\beta_{1})+2\dot{\alpha}_{H}+\beta_{1}\dot{\alpha}_{T}\right)
\notag \\ &
\quad
+4(\alpha_{H}+(1+\alpha_{T})\beta_{1})\dot{\beta}_{1}\Bigr)\triangle\dot{\pi}
+\frac{2H^2}{a^2}\biggl(K_{1}+(1+K_{1}+\alpha_{M})(\alpha_{B}-\alpha_{H})-\alpha_{M}+\alpha_{T}
\notag \\ & 
\quad
-(1+\alpha_{M})(1+K_{1}+\alpha_{M})\beta_{1}
+\frac{1}{H}\left(\dot{\alpha}_{B}-\dot{\alpha}_{H}-\beta_{1}\dot{\alpha}_{M}-2(1+\alpha_{M})\dot{\beta}_{1}\right)-\frac{\ddot{\beta}_{1}}{H^2}+\frac{p_{s}+\rho_{s}}{2H^2M^2}\biggr)\triangle\pi
\notag \\ &
+6H^4\left(K_{2}\alpha_{B}-(K_{3}+2K_{2}(3+\alpha_{M}))\beta_{1}+K_{1}(3+\alpha_{M})(\alpha_{B}-(3+\alpha_{M})\beta_{1})+K_{1}^2(1+\alpha_{B}-(3+\alpha_{M})\beta_{1})
\right. \notag \\ & \left.\quad
+\frac{1}{H}\left(K_{1}\left(\dot{\alpha}_{B}-\beta_{1}\dot{\alpha}_{M}\right)-2(K_{2}+K_{1}(3+\alpha_{M}))\dot{\beta}_{1}\right)-\frac{K_{1}}{H^2}\ddot{\beta}_{1}-\frac{\dot{p}_{s}+\dot{\rho}_{s}}{2H^3M^2}-\frac{\ddot{\rho}_{s}}{6H^4M^2}\right)\pi
,
\label{eq:DHOST_Eq5}
\end{align}
%
%%%%%%%%%%%%%%%%%%%%%%%%%%%%%%%%%%%%%%%%%%%%%%%%%%%%%%%%%%%%%%%%%%%%%%%%%%%%%%%
where $\rho_s$ and $p_s$ are evaluated through the background evolution equations Eq.~\eqref{eq:background_Na}.
Plugging Eqs.~\eqref{eq:param_M}--\eqref{eq:param_aK} and Eq.~\eqref{eq:background_Na} into 
Eqs.~\eqref{eq:DHOST_Eq1}--\eqref{eq:DHOST_Eq5},
one can straightforwardly reproduce the Euler-Lagrange equations derived from the full DHOST theory.

%%==============================================================
%%==============================================================
\subsection{Perturbed matter energy-momentum tensor}
%%==============================================================
%%==============================================================

The perturbations of the energy-momentum tensor given in Eq.~(\ref{eq:def_T}) are calculated as
%%%%%%%%%%%%%%%%%%%%%%%%%%%%%%%%%%%%%%%%%%%%%%%%%%%%%%%%%%%%%%%%%%%%%%%%%%%%%%%
%
\begin{alignat}{3}
\delta T^{00} &= \sum_I\rho_I\left(\delta_I - 2\Psi\right), 
&\quad \delta T^{ij} &= \frac{1}{a^2}\sum_I\left(\delta p_I -2p_I\Phi\right)\delta^{ij},
&\quad T^{0i} &= \frac{1}{a}\sum_I (\rho_I+p_I)v^i_I.
\end{alignat}
%
%%%%%%%%%%%%%%%%%%%%%%%%%%%%%%%%%%%%%%%%%%%%%%%%%%%%%%%%%%%%%%%%%%%%%%%%%%%%%%%
For the baryons and CDM, the pressure and its perturbation satisfy
$p_{b,c}=\delta p_{b,c}=0$, whereas those for the photons and neutrinos 
satisfy $p_{\gamma,\nu}=\rho_{\gamma,\nu}/3$ and $\delta p_{\gamma,\nu}=\delta\rho_{\gamma,\nu}/3$.
Then, performing the Fourier transformation for Eqs.~(\ref{eq:DHOST_Eq1})-(\ref{eq:DHOST_Eq4}), 
we find,
%%%%%%%%%%%%%%%%%%%%%%%%%%%%%%%%%%%%%%%%%%%%%%%%%%%%%%%%%%%%%%%%%%%%%%%%%%%%%%%
%
\begin{align}
-\frac{1}{M^2}\mathcal{E}_\Psi &= \mathcal{S}_\Psi :=
  -3H^2\left(\Omega_{c}\delta_{c}+\Omega_{b}\delta_{b}+\Omega_\gamma\delta_\gamma+\Omega_\nu\delta_\nu\right)+ \frac{2\rho_s}{M^2}\Psi
\label{eq:eq_psi}
,\\
-\frac{1}{M^2}\mathcal{E}_\Phi &=\mathcal{S}_\Phi :=
  3H^2\left(\Omega_\gamma\delta_\gamma+\Omega_\nu\delta_\nu\right) - \frac{6p_s}{M^2}\Phi
\label{eq:eq_phi}
,\\
-\frac{1}{M^2}\mathcal{E}_\xi &=\mathcal{S}_\xi :=
 kaH^2\left(3\Omega_{c}V_{c}+3\Omega_{b}V_{b}+4\Omega_\gamma V_\gamma+4\Omega_\nu V_\nu\right)
,\\
-\frac{1}{M^2}\mathcal{E}_\eta &=\mathcal{S}_\eta :=
 4H^2k^2\left(\Omega_{\gamma}\Theta_{\gamma 2}+\Omega_{\nu}\Theta_{\nu 2}\right),
\end{align}
%
%%%%%%%%%%%%%%%%%%%%%%%%%%%%%%%%%%%%%%%%%%%%%%%%%%%%%%%%%%%%%%%%%%%%%%%%%%%%%%%
where $\Omega_I:=\rho_I/3M^2H^2$ and $\mathcal{E}_A$ are given in Eqs.~\eqref{eq:DHOST_Eq1}--\eqref{eq:DHOST_Eq4}.
In the last equation, $\Theta_{\gamma \ell}$ and $\Theta_{\nu \ell}$ are the multipoles of
the temperature fluctuations of photons and neutrinos, respectively, and in the first three equations,
$\delta_{\gamma} = 4\Theta_{\gamma 0}$ and $V_{\gamma}=-3\Theta_{\gamma 1}$
where we define the velocity potential $V_I:=-ik_iv_I^i/k$.

As for the scalar field, the governing equation is given as
%%%%%%%%%%%%%%%%%%%%%%%%%%%%%%%%%%%%%%%%%%%%%%%%%%%%%%%%%%%%%%%%%%%%%%%%%%%%%%%
%
\begin{align}
-\frac{1}{M^2}\mathcal{E}_\pi = 0,
\label{eq:eq_pi}
\end{align}
%
%%%%%%%%%%%%%%%%%%%%%%%%%%%%%%%%%%%%%%%%%%%%%%%%%%%%%%%%%%%%%%%%%%%%%%%%%%%%%%%
with Eq.~(\ref{eq:DHOST_Eq5}).

%%==============================================================
%%==============================================================
\subsection{Reduction of higher-derivative equations}
%%==============================================================
%%==============================================================

For later convenience, we define $\widetilde{\mathcal{E}}_\xi := k^{-2}\mathcal{E}_\xi/M^2$,
$\widetilde{\mathcal{E}}_\eta := -k^{-4}\mathcal{E}_\eta/M^2$,
and $\widetilde{\mathcal{E}}_i := -\mathcal{E}_i/M^2$ for other equations, and
%%%%%%%%%%%%%%%%%%%%%%%%%%%%%%%%%%%%%%%%%%%%%%%%%%%%%%%%%%%%%%%%%%%%
%
\begin{align}
\widetilde{\mathcal{S}}_\xi &:= -\frac{1}{k^2}\mathcal{S}_\xi
 = -a\frac{H^2}{k}\left(3\Omega_{c}V_{c}+3\Omega_{b}V_{b}+4\Omega_\gamma V_\gamma+4\Omega_\nu V_\nu\right)
\label{eq:DHOST_src_xi}
,\\
\widetilde{\mathcal{S}}_\eta &:= \frac{1}{k^4}\mathcal{S}_\eta
 = 4\frac{H^2}{k^2}\left(\Omega_{\gamma}\Theta_{\gamma 2}+\Omega_{\nu}\Theta_{\nu 2}\right),
\label{eq:DHOST_src_eta}
\end{align}
%
%%%%%%%%%%%%%%%%%%%%%%%%%%%%%%%%%%%%%%%%%%%%%%%%%%%%%%%%%%%%%%%%%%%%
in the Fourier space. Then the evolution equations become simpler form,
$\widetilde{\mathcal{E}}_A=\widetilde{\mathcal{S}}_A$ for $A=\Psi,\Phi,\xi,\eta$,
and $\widetilde{\mathcal{E}}_{\pi}=0$.
%We need another equation $\widetilde{\mathcal{E}}_{\pi}=0$.
This equation, however, contains time-derivatives of $\Psi, \Phi$ and $\pi$ up to 
the fourth-order as given in Eq.~(\ref{eq:DHOST_Eq5}).
As the kinetic matrix of the highest derivatives of $\Psi, \Phi$ and $\pi$ are
degenerated, we can eliminate such higher-order derivative terms. 
To do so, we define
%%%%%%%%%%%%%%%%%%%%%%%%%%%%%%%%%%%%%%%%%%%%%%%%%%%%%%%%%%%%%%%%%%%%%
%%
\begin{align}
\widetilde{\mathcal{E}}_G:=-\widetilde{\mathcal{E}}_\pi+\beta_1\dot{\widetilde{\mathcal{E}}}_\Phi+(H(-\alpha_B+(3+\alpha_M)\beta_1)+2\dot{\beta}_1)\widetilde{\mathcal{E}}_\Phi, 
\end{align}
%%
%%%%%%%%%%%%%%%%%%%%%%%%%%%%%%%%%%%%%%%%%%%%%%%%%%%%%%%%%%%%%%%%%%%%%
which reads
%%%%%%%%%%%%%%%%%%%%%%%%%%%%%%%%%%%%%%%%%%%%%%%%%%%%%%%%%%%%%%%%%%%%
%
\begin{align}
\widetilde{\mathcal{E}}_G &=
\frac{2}{a^2}(\alpha_{H}+(1+\alpha_{T})\beta_{1})\left(\triangle\dot{\Phi}-\beta_{1}\triangle\dot{\Psi}-\beta_{1}\triangle\ddot{\pi}\right)
+\frac{6}{M^2}\left(Hp_{s}(\alpha_{B}-3\beta_{1})-\beta_{1}\dot{p}_{s}-2p_{s}\dot{\beta}_{1}\right)\Phi
\notag \\ &
+\frac{2H}{a^2}\left(-\alpha_{B}(1+\alpha_{T})+\gamma_{3}\right)\triangle\Phi
-\frac{2H}{a^2}\left(\alpha_{B}\alpha_{H}+\beta_{1}\gamma_{3}\right)\triangle\dot{\pi}
-\frac{2H}{a^2}\left(\alpha_{H}(1+\alpha_{B})+\beta_{1}(1+\gamma_{3}+\alpha_{T})\right)\triangle\Psi
\notag \\ &
+6H^3
\left(-\gamma_{1}+\gamma_{2}+\frac{2\dot{\alpha}_{B}\dot{\beta}_{1}}{H^2}+\frac{\beta_{1}\ddot{\alpha}_{B}}{H^2}+\left(\alpha_{B}-3\beta_{1}-\frac{2\dot{\beta}_{1}}{H}\right)\frac{p_{s}+\rho_{s}}{2H^2M^2}-\beta_{1}\frac{2\dot{p}_{s}+\dot{\rho}_{s}}{2H^3M^2}\right)\dot{\pi}
\notag \\ &
+6H^3
\left(K_{1}(1+\alpha_{B})-\gamma_{1}+\gamma_{2}+\frac{\dot{\alpha}_{B}}{H}+\frac{1}{H^2}\left(2\dot{\alpha}_{B}\dot{\beta}_{1}+\beta_{1}\ddot{\alpha}_{B}-\ddot{\beta}_{1}\right)-\beta_{1}\frac{\dot{p}_{s}+\dot{\rho}_{s}}{2H^3M^2}+\left(1+\alpha_{B}-3\beta_{1}-\frac{2\dot{\beta}_{1}}{H}\right)\frac{\rho_{s}+p_{s}}{2H^2M^2}\right)\Psi
\notag \\ &
-6H^2\left(K_{1}(1+\alpha_{B})+\frac{\dot{\alpha}_{B}}{H}-\frac{\ddot{\beta}_{1}}{H^2}+(1+2\beta_{1})\frac{p_{s}}{2H^2M^2}+\frac{\rho_{s}}{2H^2M^2}\right)\dot{\Phi}
\notag \\ &
+\frac{2H^2}{a^2}\left(\gamma_{3}-K_{1}\left(1-\alpha_{H}+\alpha_{B}-(1+\alpha_{T})\beta_{1}\right)-\alpha_{B}(1+\alpha_{T})-\frac{1}{H}\dot{\alpha}_{B}+\frac{\ddot{\beta}_{1}}{H^2}-\frac{p_{s}+\rho_{s}}{2H^2M^2}\right)\triangle\pi
\notag \\ &
-6H^2\left(\alpha_{B}^2+\frac{1}{6}\alpha_{K}-K_{1}\beta_{1}(1+2\alpha_{B})-\alpha_{B}(3+\alpha_{M})\beta_{1}-\frac{1}{H}\left(2\beta_{1}\dot{\alpha}_{B}+\alpha_{B}\dot{\beta}_{1}\right)+\frac{\beta_{1}\ddot{\beta}_{1}}{H^2}+\beta_{1}\frac{p_{s}+\rho_{s}}{2H^2M^2}\right)\left(\ddot{\pi}+\dot{\Psi}\right)
\notag \\ &
-6H^4\left(K_{1}^2(1+\alpha_{B})+\frac{K_{1}}{H}\dot{\alpha}_{B}-\frac{K_{1}}{H^2}\ddot{\beta}_{1}
-\left(1+\alpha_{B}-3\beta_{1}-\frac{2\dot{\beta}_{1}}{H}\right)\frac{\dot{p}_{s}}{2H^3M^2}-\frac{\dot{\rho}_{s}}{2H^3M^2}-\frac{\ddot{\rho}_{s}-3\beta_{1}\ddot{p}_{s}}{6H^4M^2}\right)\pi
\label{eq:DHOST_ELeq_G}
.
\end{align}
%
%%%%%%%%%%%%%%%%%%%%%%%%%%%%%%%%%%%%%%%%%%%%%%%%%%%%%%%%%%%%%%%%%%%%
where
%%%%%%%%%%%%%%%%%%%%%%%%%%%%%%%%%%%%%%%%%%%%%%%%%%%%%%%%%%%%%%%%%%%%
%
\begin{align}
\gamma_{1} &:= \left(\alpha_{B}^2+\frac{1}{6}\alpha_{K}\right)(3+\alpha_{M})-\alpha_{B}\left(K_{2}+(3+\alpha_{M})^2\right)\beta_{1}+\frac{K_{1}}{3}\Bigl(\alpha_{K}+3\alpha_{B}\bigl(-1+\alpha_{B}-3(3+\alpha_{M})\beta_{1}\bigr)\Bigr)
,\\
\gamma_{2} &:= \frac{1}{H}\left[2(3+K_{1}+\alpha_{M})\beta_{1}\dot{\alpha}_{B}-\frac{1}{6}\dot{\alpha}_{K}-\alpha_{B}\left(\dot{\alpha}_{B}-\beta_{1}\dot{\alpha}_{M}-2(3+K_{1}+\alpha_{M})\dot{\beta}_{1}\right)\right]
,\\
\gamma_{3} &:= \alpha_{M}-\alpha_{T}+(1+\alpha_M)\Bigl(\alpha_H+(1+\alpha_T)\beta_1\Bigr)%+\alpha_{H}(1+\alpha_{M})+(1+\alpha_{M})(1+\alpha_{T})\beta_{1}
+\frac{1}{H}\left(\dot{\alpha}_{H}+\beta_{1}\dot{\alpha}_{T}+2(1+\alpha_{T})\dot{\beta}_{1}\right)
.
\end{align}
%
%%%%%%%%%%%%%%%%%%%%%%%%%%%%%%%%%%%%%%%%%%%%%%%%%%%%%%%%%%%%%%%%%%%%
This expression will be used after the Fourier transformation.
The corresponding source term is given as
$\widetilde{\mathcal{S}}_G:=\beta_1\dot{\widetilde{\mathcal{S}}}_\Phi+(H(-\alpha_B+(3+\alpha_M)\beta_1)+2\dot{\beta}_1)\widetilde{\mathcal{S}}_\Phi$.
The resultant field equations for $\Psi, \Phi$ and $\pi$ 
contain the time-derivatives up to the
second-order for $\pi$ and the first-order for $\Psi$ and
$\Phi$. 
We note that the highest order of the derivative
depends on the gravity theory of interest.

%%==============================================================
%%==============================================================
\subsection{Evolution equations to solve}
%%==============================================================
%%==============================================================

In what follows, we explain how to solve the set of equations in the Type-I DHOST theory.
As the unknown variables in the gravity sector are $\Psi, \Phi$ and $\pi$,
we need three independent equations. In this study, we choose 
%%%%%%%%%%%%%%%%%%%%%%%%%%%%%%%%%%%%%%%%%%%%%%%%%%%%%%%%%%%%%%%%%%%%
%
\begin{align}
\widetilde{\mathcal{E}}_\eta(\Phi, \Psi, \dot{\pi}, \pi) &=
\widetilde{\mathcal{S}}_\eta(\Theta_{\rm r2})
\label{eq:DHOST_eq_eta}
,\\
\widetilde{\mathcal{E}}_\xi(\dot{\Phi}, \dot{\Psi}, \ddot{\pi}, \Psi, \dot{\pi}, \pi) &= \widetilde{\mathcal{S}}_\xi( V_I )
\label{eq:DHOST_eq_xi}
 ,\\
\widetilde{\mathcal{E}}_G(\dot{\Phi},\dot{\Psi},\ddot{\pi},\Phi, \Psi, \dot{\pi}, \pi) &= \widetilde{\mathcal{S}}_G(\Phi,\delta_I,V_I)
\label{eq:DHOST_eq_G}
,
\end{align}
%
%%%%%%%%%%%%%%%%%%%%%%%%%%%%%%%%%%%%%%%%%%%%%%%%%%%%%%%%%%%%%%%%%%%%
where the right-hand sides of the first two equations are defined in 
Eqs.~(\ref{eq:DHOST_src_xi})(\ref{eq:DHOST_src_eta}), and we shortly write 
$\delta_I=\{\delta_{b},\delta_{c},\delta_\gamma,\delta_\nu\}$,
$V_I=\{V_{b},V_{c},V_\gamma,V_\nu\}$ and 
$\Theta_{{\rm r}i}=\{\Theta_{\gamma i},\Theta_{\nu i}\}$.
The first time-derivative of Eq.~(\ref{eq:DHOST_eq_eta}) becomes
%%%%%%%%%%%%%%%%%%%%%%%%%%%%%%%%%%%%%%%%%%%%%%%%%%%%%%%%%%%%%%%%%%%%
%
\begin{align}
\dot{\widetilde{\mathcal{E}}}_\eta(\dot{\Phi}, \dot{\Psi}, \ddot{\pi},
 \Phi, \Psi, \dot{\pi}, \pi) &= \dot{\widetilde{\mathcal{S}}}_\eta(V_I, \Theta_{\rm r2}, \Theta_{\rm r3} ),
\label{eq:DHOST_eq_deta}
\end{align}
%
%%%%%%%%%%%%%%%%%%%%%%%%%%%%%%%%%%%%%%%%%%%%%%%%%%%%%%%%%%%%%%%%%%%%
where we have used the Boltzmann equation for $\Theta_{\gamma 2}$ and $\Theta_{\nu 2}$.
Since the coefficient matrix of $\dot{\Phi}, \dot{\Psi}$ and $\ddot{\pi}$ in
Eqs.~(\ref{eq:DHOST_eq_xi}), (\ref{eq:DHOST_eq_G}) and (\ref{eq:DHOST_eq_deta})
is invertible, we can solve these equations with respect to $\dot{\Phi}$, $\dot{\Psi}$ and $\ddot{\pi}$ and obtain
%%%%%%%%%%%%%%%%%%%%%%%%%%%%%%%%%%%%%%%%%%%%%%%%%%%%%%%%%%%%%%%%%%%%
%
\begin{align}
 \dot{\Phi} &= \mathcal{F}_{\Phi}(\Phi, \Psi, \dot{\pi},
 \pi; \delta_I, V_I, \Theta_{\rm r2}, \Theta_{\rm r3})
\label{eq:DHOST_eq_dPhi}
 ,\\
 \dot{\Psi} &= \mathcal{F}_{\Psi}(\Phi, \Psi, \dot{\pi}, \pi; \delta_I, V_I, \Theta_{\rm r2}, \Theta_{\rm r3})
\label{eq:DHOST_eq_dPsi}
,\\
 \ddot{\pi} &= \mathcal{F}_{\pi}(\Phi, \Psi, \dot{\pi}, \pi; \delta_I, V_I, \Theta_{\rm r2}, \Theta_{\rm r3}).
\label{eq:DHOST_eq_ddpi}
\end{align}
%
%%%%%%%%%%%%%%%%%%%%%%%%%%%%%%%%%%%%%%%%%%%%%%%%%%%%%%%%%%%%%%%%%%%%
These equations can be straightforwardly derived, though
the right-hand sides of these equations are too long to show here. 
Once one solves this set of equations numerically, one can obtain the time evolution of $\Phi,\Psi,\pi$.
Unfortunately, however, it is failed
since the equation for $\dot{\Psi}$ seems to be unstable at late time.
%We use a constraint equation for $\Psi$ and $\Phi$.
The easiest way to avoid the numerical instability
is to replace Eq.~(\ref{eq:DHOST_eq_dPsi}) by a constraint equation Eq.~(\ref{eq:DHOST_eq_eta}),
and we compute $\Psi$ from Eq.~(\ref{eq:DHOST_eq_eta}) after updating $\Phi$ and $\pi$ by
solving Eqs.~(\ref{eq:DHOST_eq_dPhi}) and (\ref{eq:DHOST_eq_ddpi}).

%%%%%%%%%%%%%%%%%%%%%%%%%%%%%%%%%%%%%%%%%%%%%%%%%%%%%%%%%%%%%%%%%%%%%%%%%%%%%%%
%%%%%%%%%%%%%%%%%%%%%%%%%%%%%%%%%%%%%%%%%%%%%%%%%%%%%%%%%%%%%%%%%%%%%%%%%%%%%%%
\section{Numerical setup}
\label{sec:setup}
%%%%%%%%%%%%%%%%%%%%%%%%%%%%%%%%%%%%%%%%%%%%%%%%%%%%%%%%%%%%%%%%%%%%%%%%%%%%%%%
%%%%%%%%%%%%%%%%%%%%%%%%%%%%%%%%%%%%%%%%%%%%%%%%%%%%%%%%%%%%%%%%%%%%%%%%%%%%%%%

We developed a Boltzmann solver implementing the framework of the DHOST theory. Our numerical
code {\tt cmb2nd}\footnote{This Boltzmann code is not public yet, but we have confirmed that the numerical results with it precisely agree with those from CAMB ({\tt https://camb.info/}). See also Refs.~\cite{Hiramatsu:2018nfa,Hiramatsu:2018vfw} in which one of the authors of this paper used the same code.} solves the Boltzmann equations for photons, $\Theta_{\gamma\ell}$,
and massless neutrinos, $\Theta_{\nu\ell}$, the continuity equations and 
the Euler equations for baryons and CDM, $\delta_{b},V_{b},\delta_{c},V_{c}$,
the modified Einstein equations for $\Psi,\Phi$ and the field equation for $\pi$
given in Eqs.~(\ref{eq:DHOST_eq_dPhi}) and (\ref{eq:DHOST_eq_ddpi}) with Eq.~(\ref{eq:DHOST_eq_eta}).
One can find the basic equations in the matter sector in a standard textbook, e.g., Ref.~\cite{Dodelson:2003ft}.

Respecting the scaling of the EFT parameters in Eq.~(\ref{eq:scaling1}), we can totally neglect
the scalar field and its influence on the metric perturbations in the early time.
Hence we can impose the same initial conditions for the perturbative quantities as those given in GR,
%%%%%%%%%%%%%%%%%%%%%%%%%%%%%%%%%%%%%%%%%%%%%%%%%%%%%%%%%%%%%%%%%%%%
%
\begin{align}
\Psi = -\frac{10}{4f_{\nu}+15}\zeta, \quad
\Phi = -\frac{4f_{\nu}+10}{4f_{\nu}+15}\zeta, \quad
\Theta_{\nu 2} = -\frac{1}{12f_{\nu}+45}\frac{k^2}{\HH^2}\zeta, \quad
\pi = 0,
\end{align}
%
%%%%%%%%%%%%%%%%%%%%%%%%%%%%%%%%%%%%%%%%%%%%%%%%%%%%%%%%%%%%%%%%%%%%
where $\zeta$ is the curvature perturbation generated during inflation and $f_\nu=\rho_\nu/(\rho_\gamma+\rho_\nu)$.

To follow the same setup as in Ref.~\cite{DAmico:2016ntq}, 
we assume the $\Lambda$CDM background as a demonstration
and $\alpha_{B,0},\alpha_{T,0}<0$, $\beta_{1,0},\alpha_{H,0},\alpha_{M,0}>0$ and $\alpha_{K,0}=1$. 
The choices of signature of the EFT parameters and their values are restricted to a certain range
arising from the avoidance of the superluminality, ghost instability and gradient instability of 
the scalar perturbation \cite{Langlois:2017mxy} (see also \cite{DAmico:2016ntq} in the GLPV theory). 
To put constrains on these parameters from the real observations,
we have to take care of the appropriate range. Our aim in the present study, however, is to
demonstrate the impact of these parameters on the angular power spectra.
Hence we adopt the above weak assumptions on the EFT parameters.

At this stage, we can freely choose the present values of the EFT parameters, $\alpha_{i,0}$ 
for $i=K,B,T,M,H$ and $\beta_{1,0}$.
However the time-dependent functions $\alpha_i(t)$ and $\beta_i(t)$ are
primarily described by the arbitrary functions $\{P(\phi,X),Q(\phi,X),f_2(\phi,X),a_{i}(\phi,X)\}$ 
introduced in the original DHOST Lagrangian (see Eq.~(\ref{eq:DHOST_action})), 
and thus the EFT parameters 
should be related with each other.
As we shall explain later, we also demonstrate this situation by adopting 
a model proposed by Crisostomi and Koyama \cite{Crisostomi:2017pjs} (CK) 
in which there is a cosmological solution exhibiting the late-time 
self-acceleration regime. In this model,
$\alpha_i$ for $i=K,B,M,H$ and $\beta_1$ are described by four constants
$c_2,c_3,c_4,\beta$, while $\alpha_T$ is fixed to be 0.

%%%%%%%%%%%%%%%%%%%%%%%%%%%%%%%%%%%%%%%%%%%%%%%%%%%%%%%%%%%%%%%%%%%%%%%%%%%%%%%
%%%%%%%%%%%%%%%%%%%%%%%%%%%%%%%%%%%%%%%%%%%%%%%%%%%%%%%%%%%%%%%%%%%%%%%%%%%%%%%
\section{Results in EFT framework}
\label{sec:results_eft}
%%%%%%%%%%%%%%%%%%%%%%%%%%%%%%%%%%%%%%%%%%%%%%%%%%%%%%%%%%%%%%%%%%%%%%%%%%%%%%%
%%%%%%%%%%%%%%%%%%%%%%%%%%%%%%%%%%%%%%%%%%%%%%%%%%%%%%%%%%%%%%%%%%%%%%%%%%%%%%%

\begin{figure}[t]
\centering{
 \includegraphics[width=5cm]{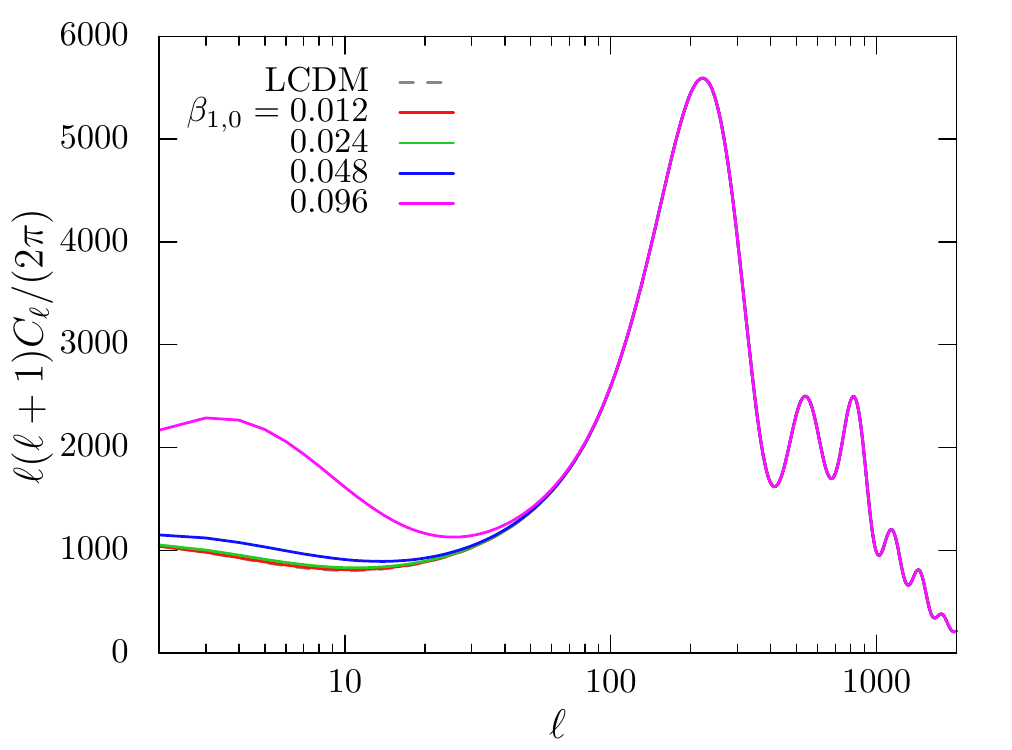}
 \includegraphics[width=5cm]{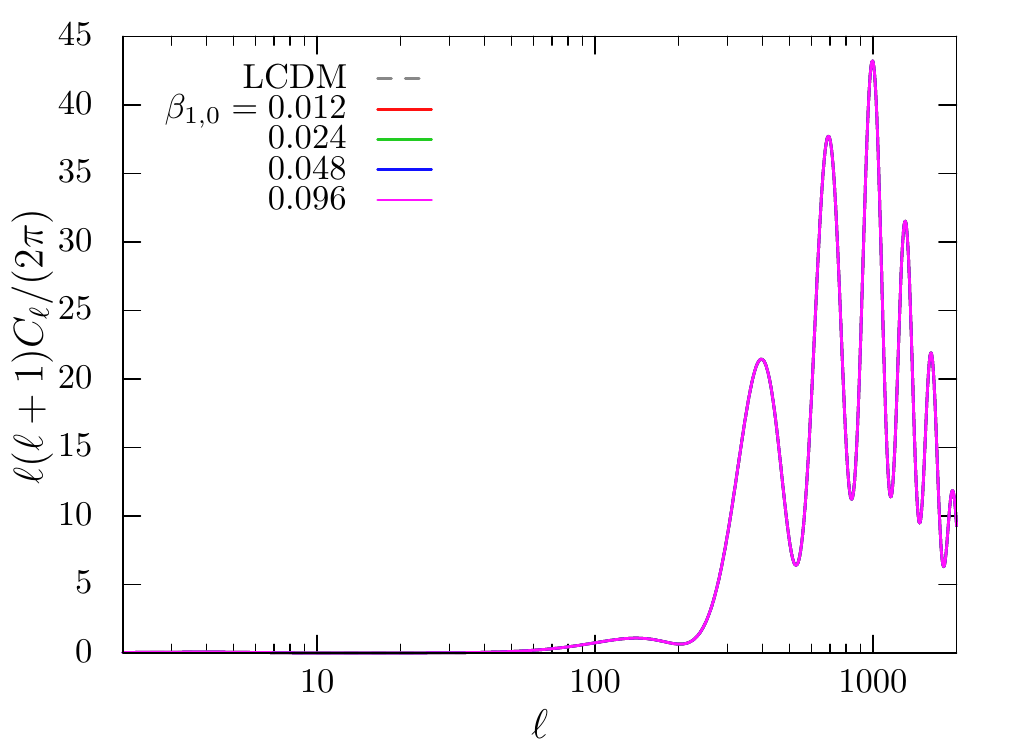}
 \includegraphics[width=5cm]{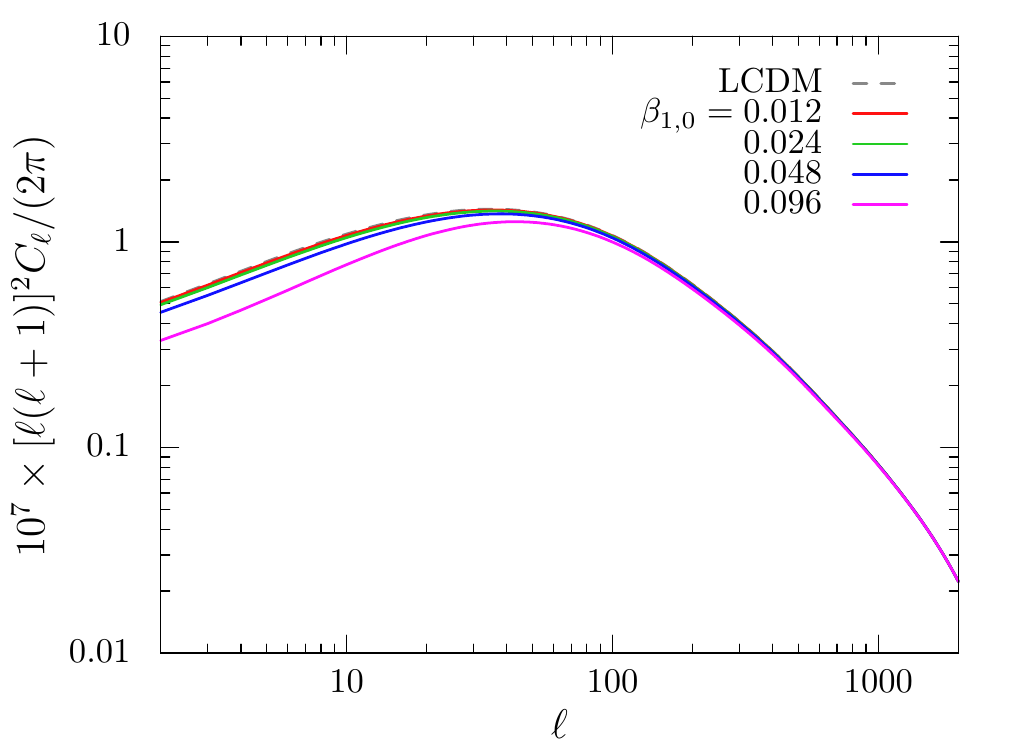}
}
\centering{
 \includegraphics[width=5cm]{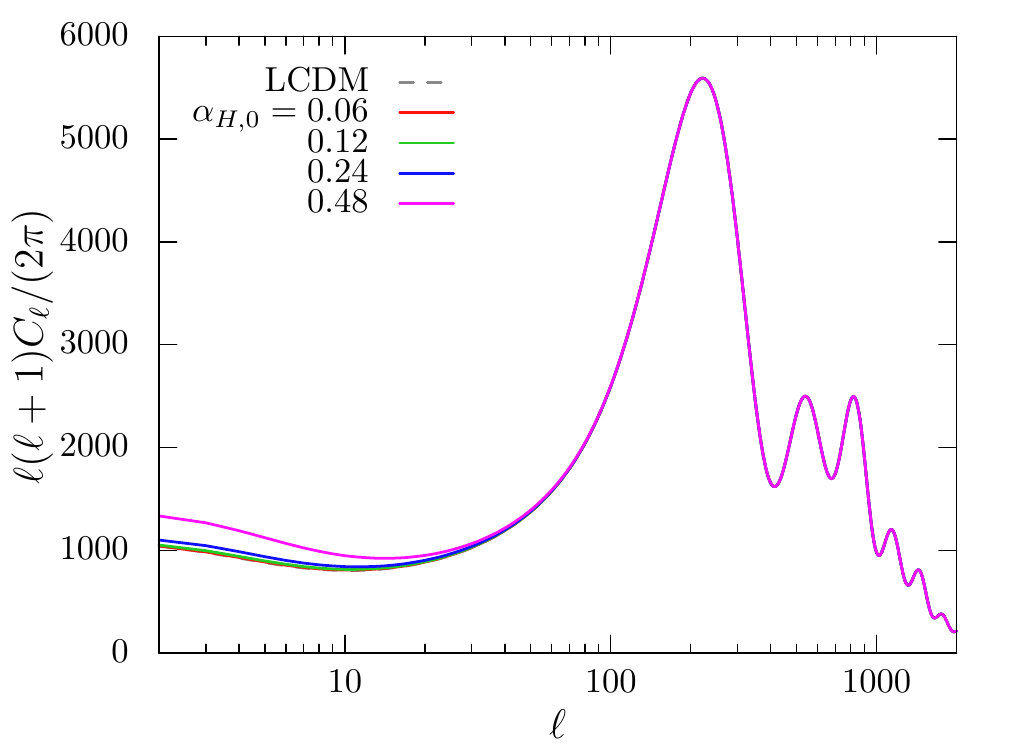}
 \includegraphics[width=5cm]{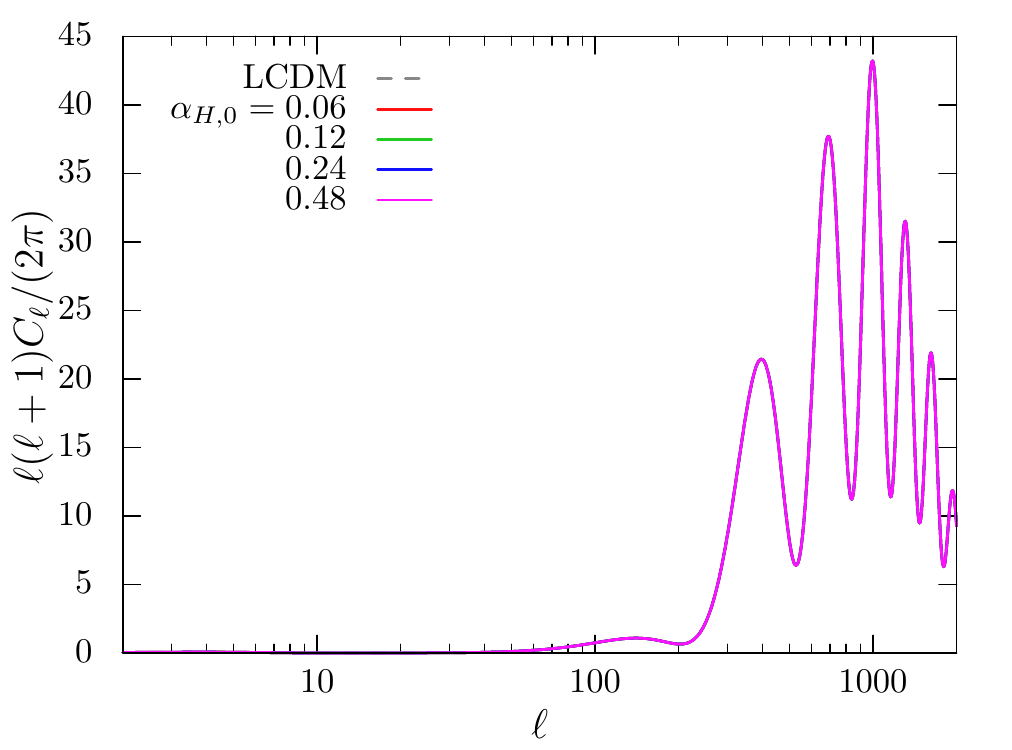}
 \includegraphics[width=5cm]{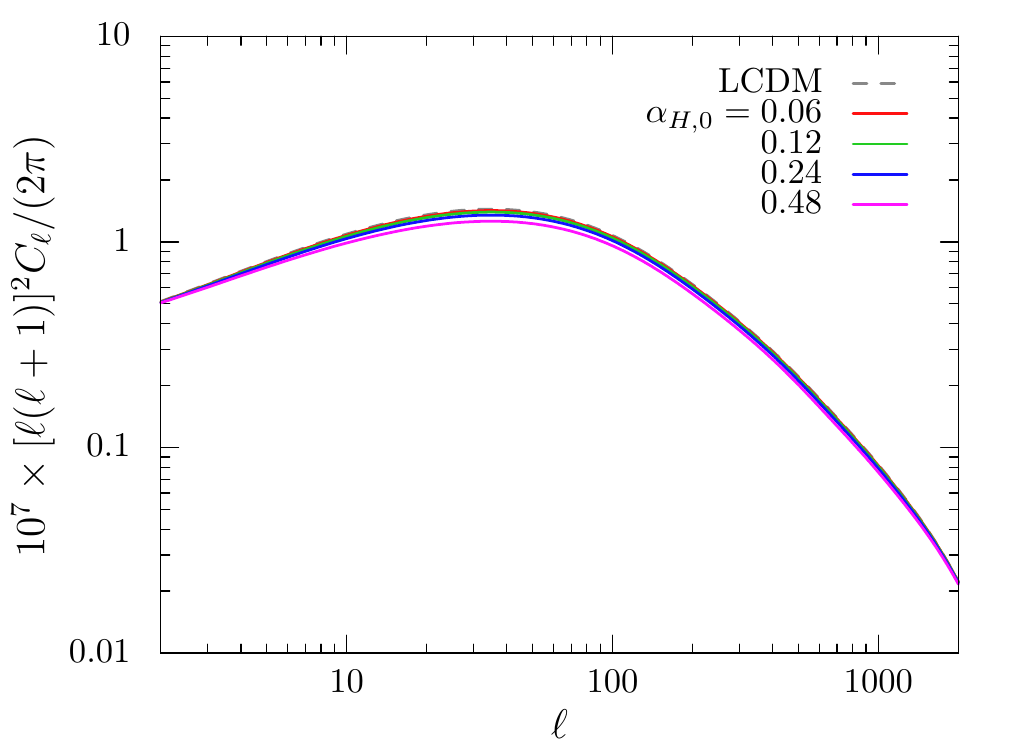}
}
\centering{
 \includegraphics[width=5cm]{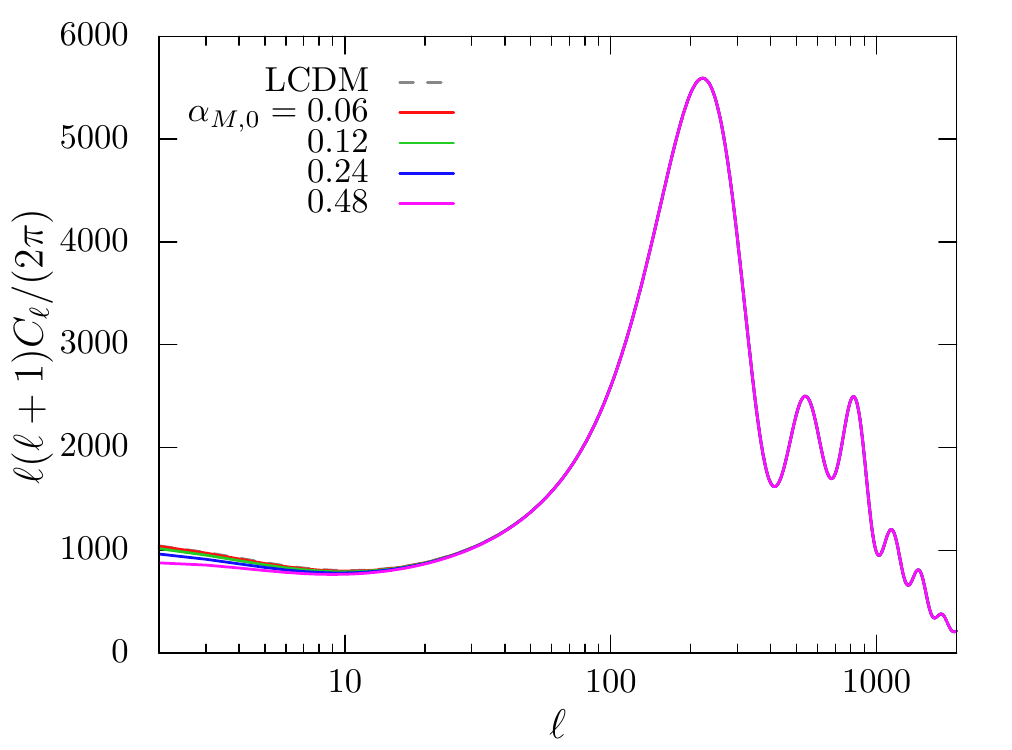}
 \includegraphics[width=5cm]{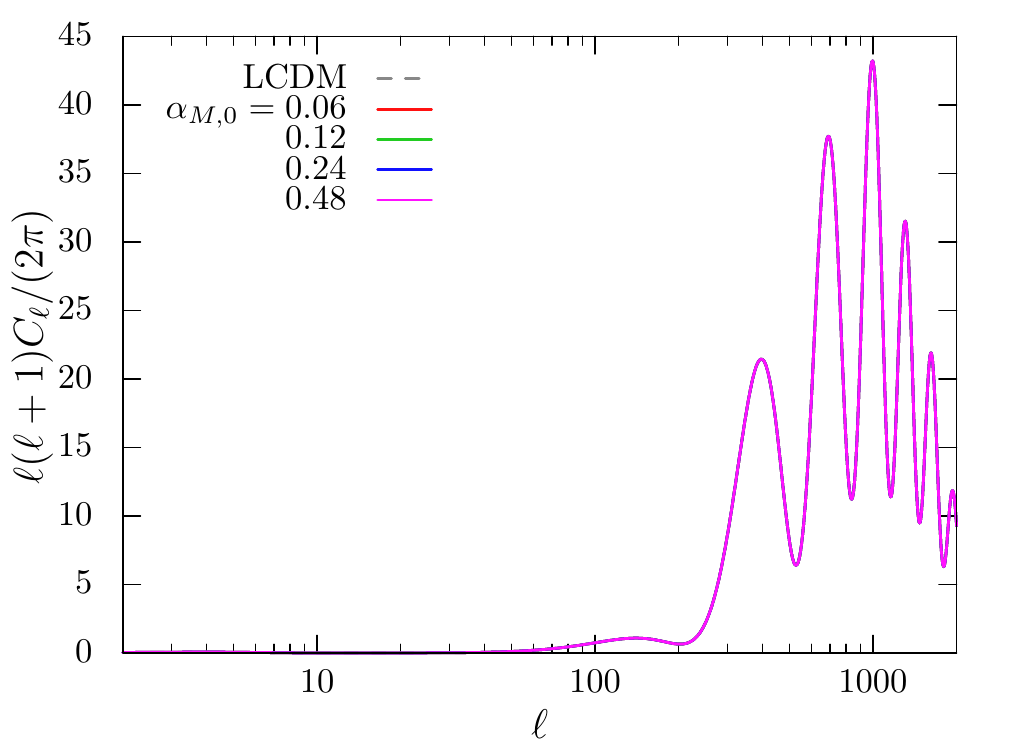}
 \includegraphics[width=5cm]{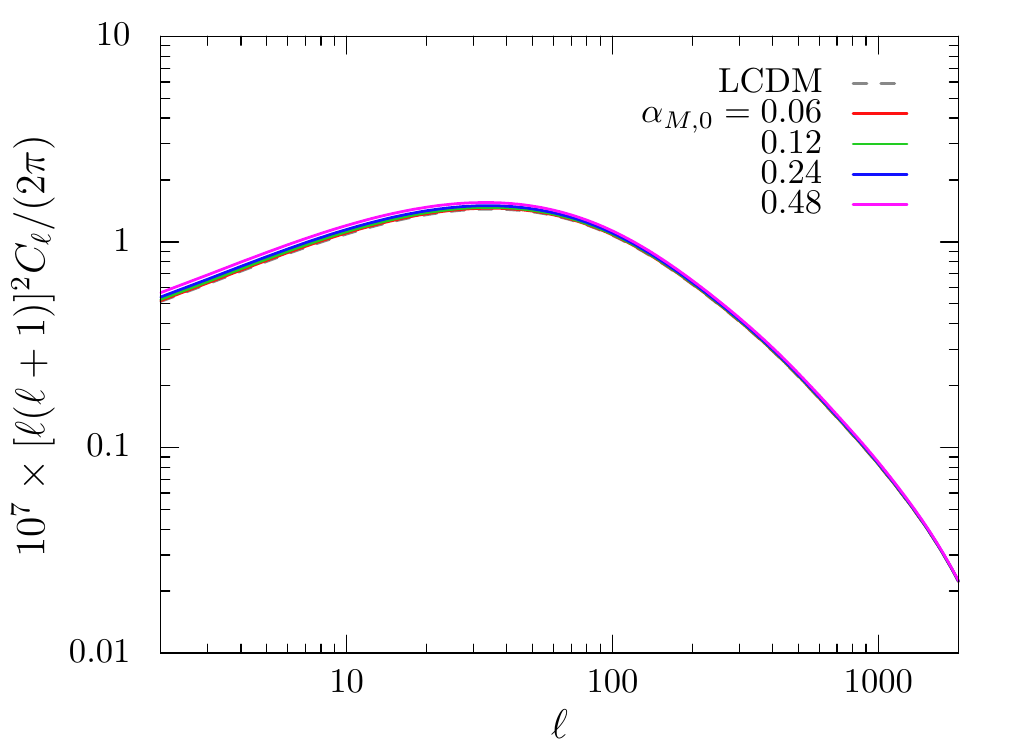}
}
\centering{
 \includegraphics[width=5cm]{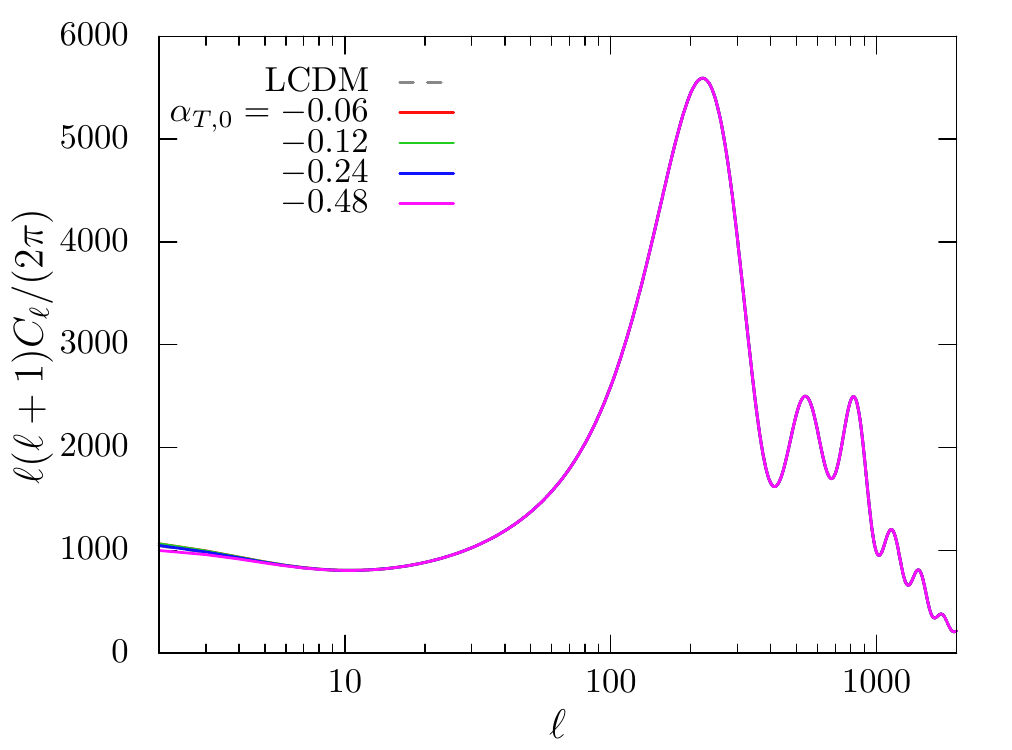}
 \includegraphics[width=5cm]{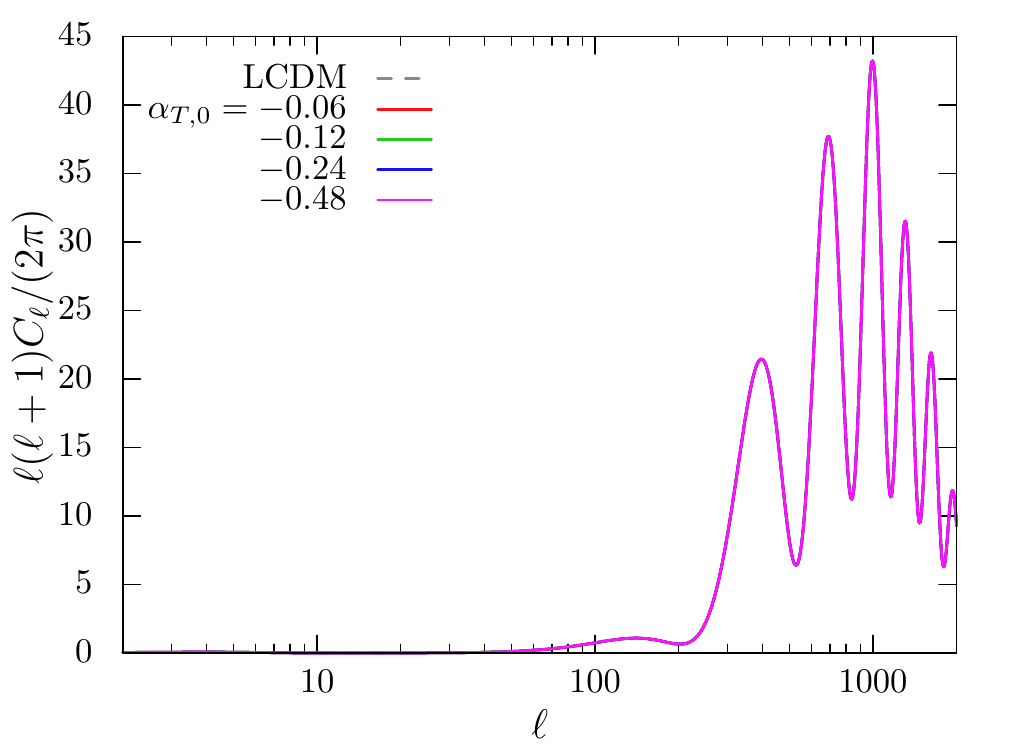}
 \includegraphics[width=5cm]{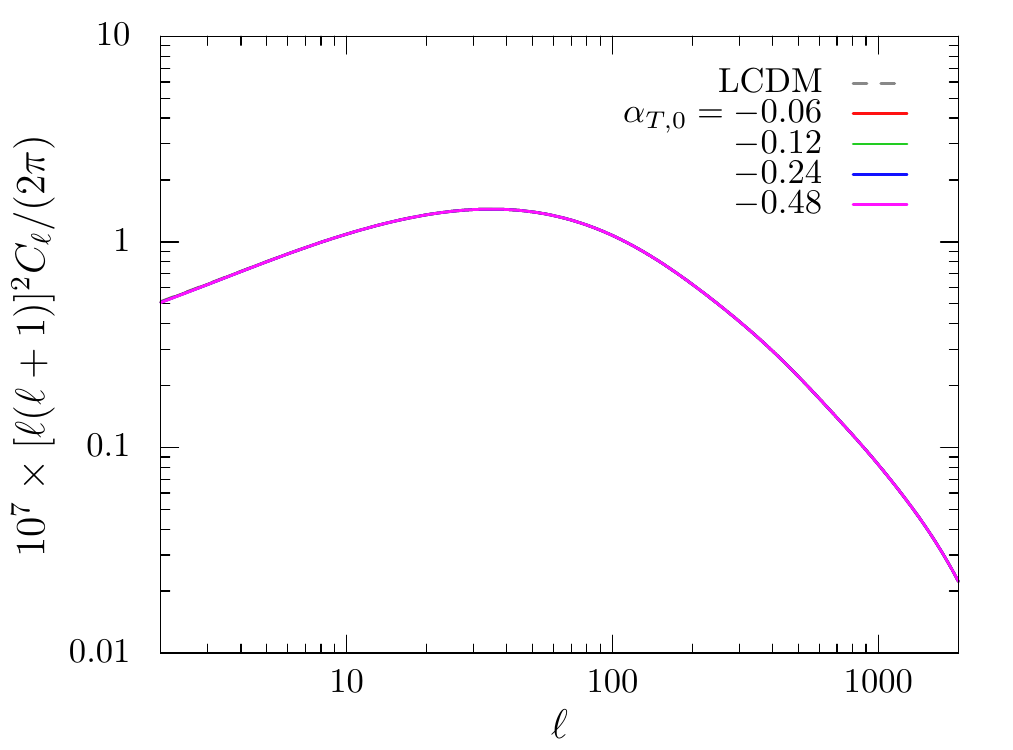}
}
\centering{
 \includegraphics[width=5cm]{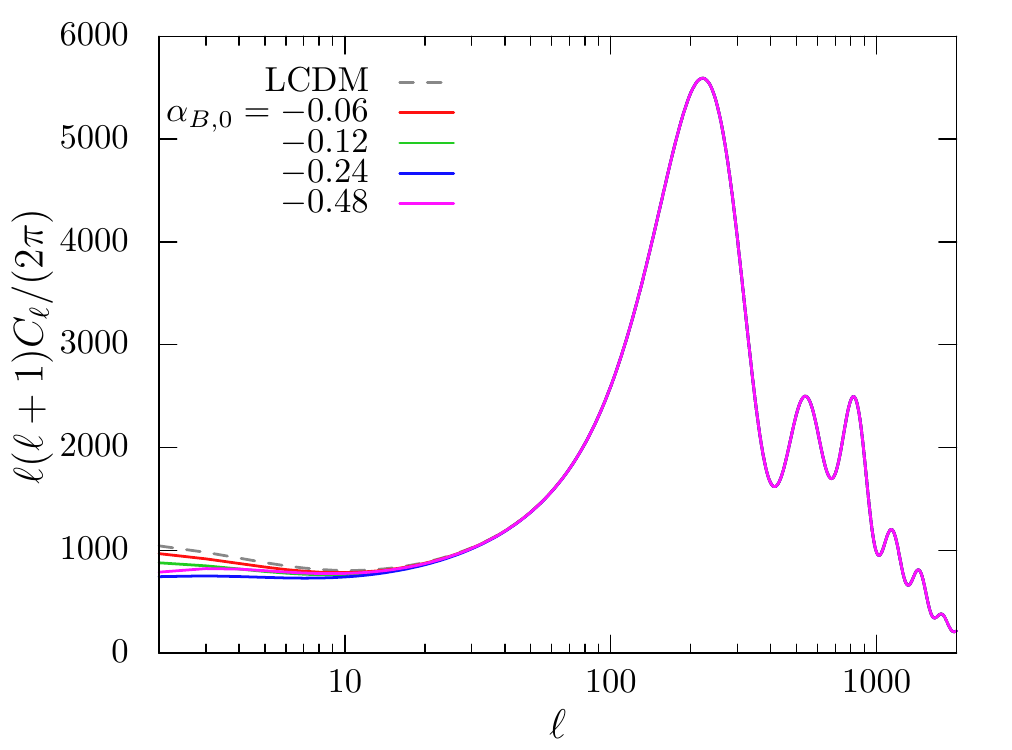}
 \includegraphics[width=5cm]{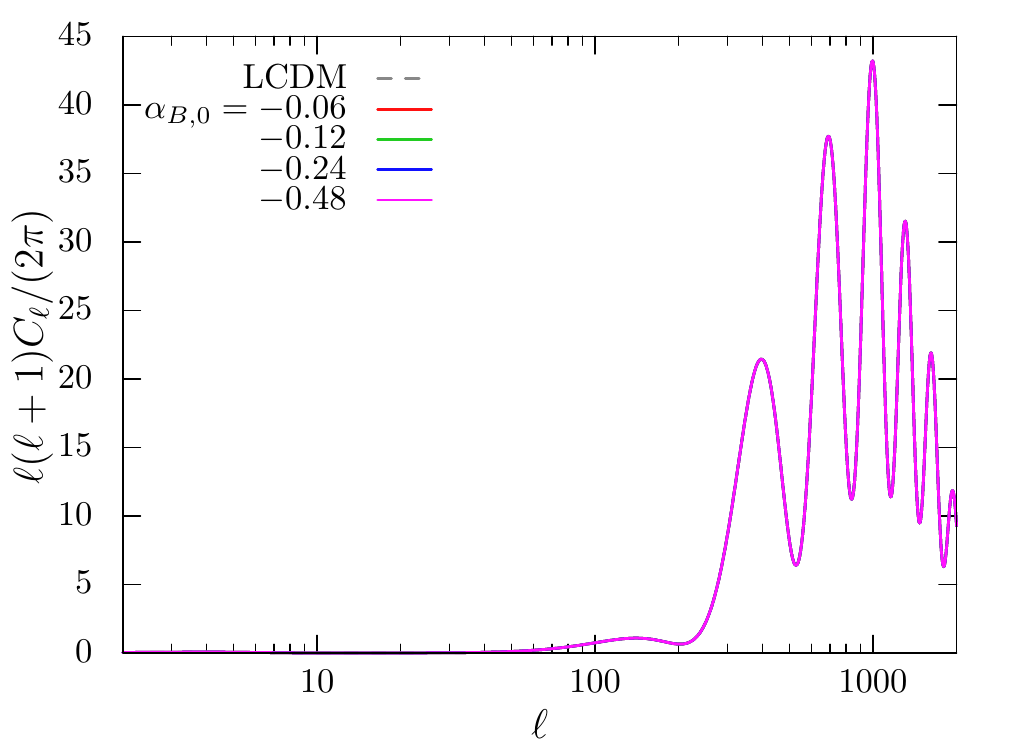}
 \includegraphics[width=5cm]{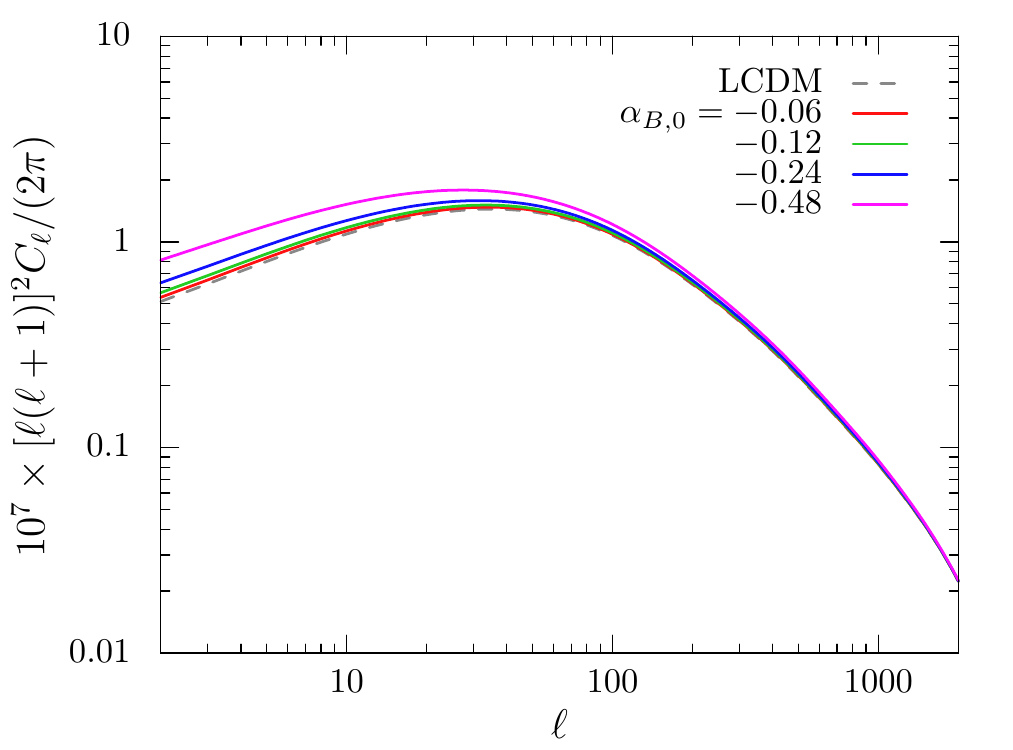}
}
\caption{Angular power spectra with varying $\beta_{1,0}$, $\alpha_{H,0}$, $\alpha_{M,0}$,
$\alpha_{T,0}$ and $\alpha_{B,0}$ from top to bottom. From left to right, we show the angular power spectra
of temperature ($C^{TT}_\ell$), E-mode ($C^{EE}_\ell$) and lensing potential ($C^{\phi\phi}_{\ell}$).}
\label{fig:TT}
\end{figure}

\begin{figure}[t]
\centering{
 \includegraphics[width=5cm]{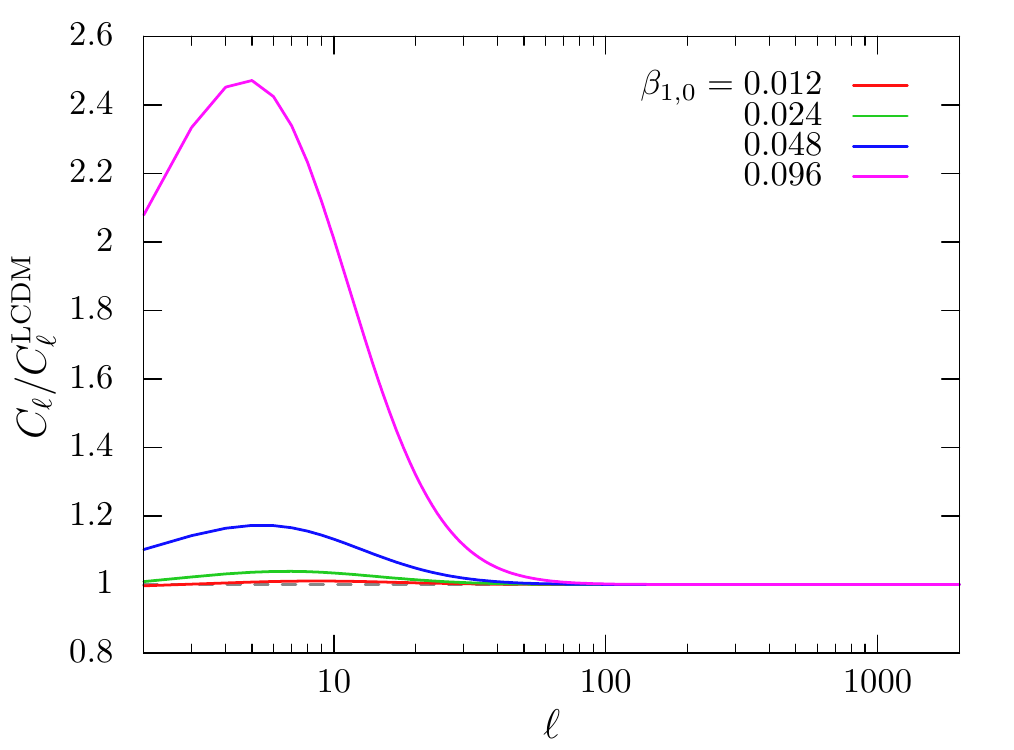}
 \includegraphics[width=5cm]{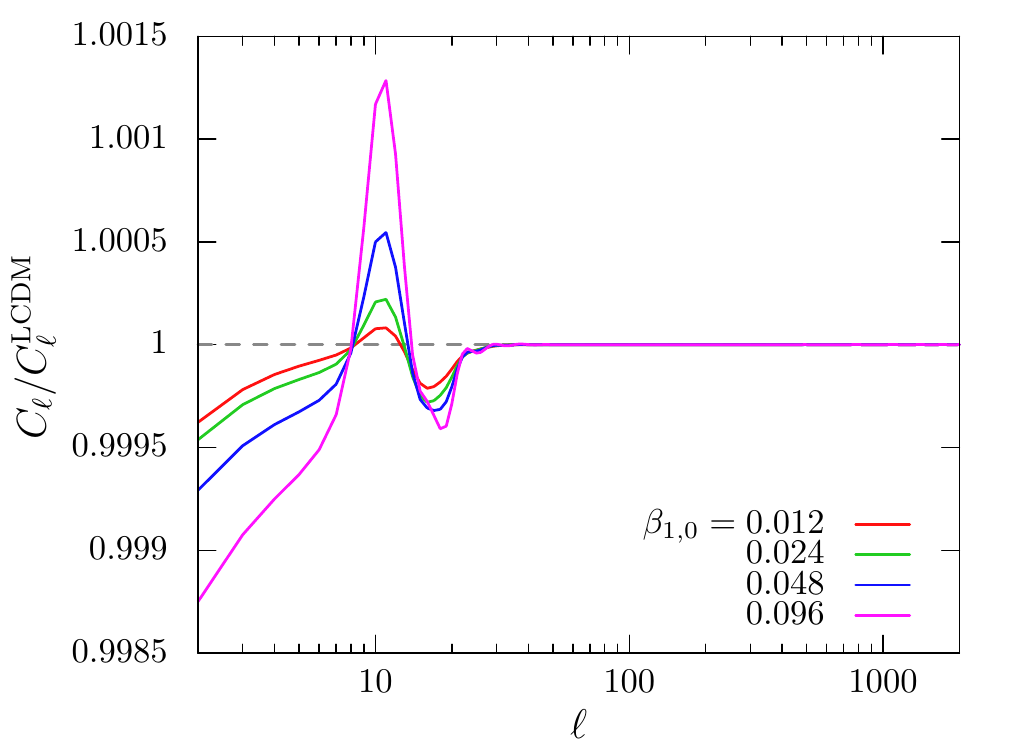}
 \includegraphics[width=5cm]{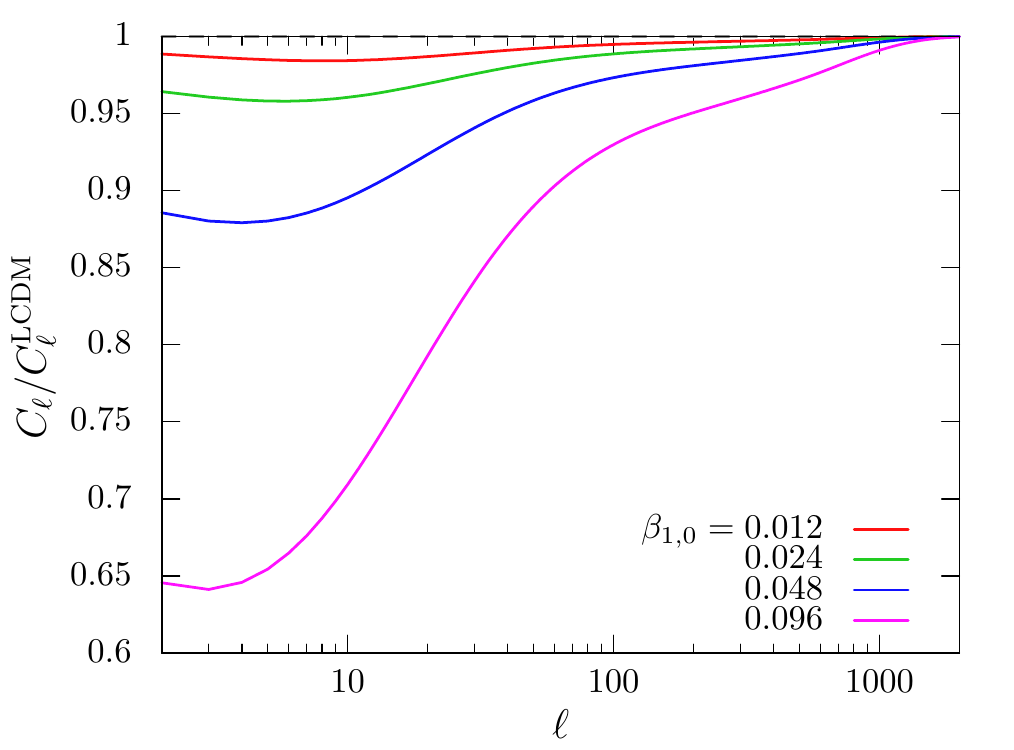}
}
\centering{
 \includegraphics[width=5cm]{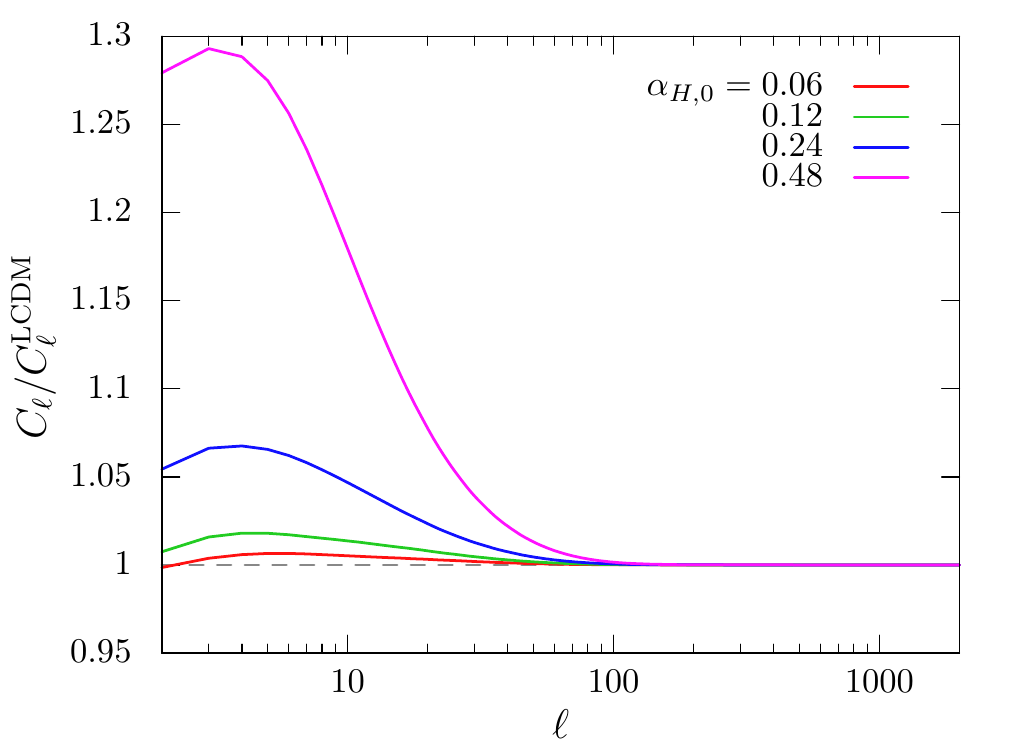}
 \includegraphics[width=5cm]{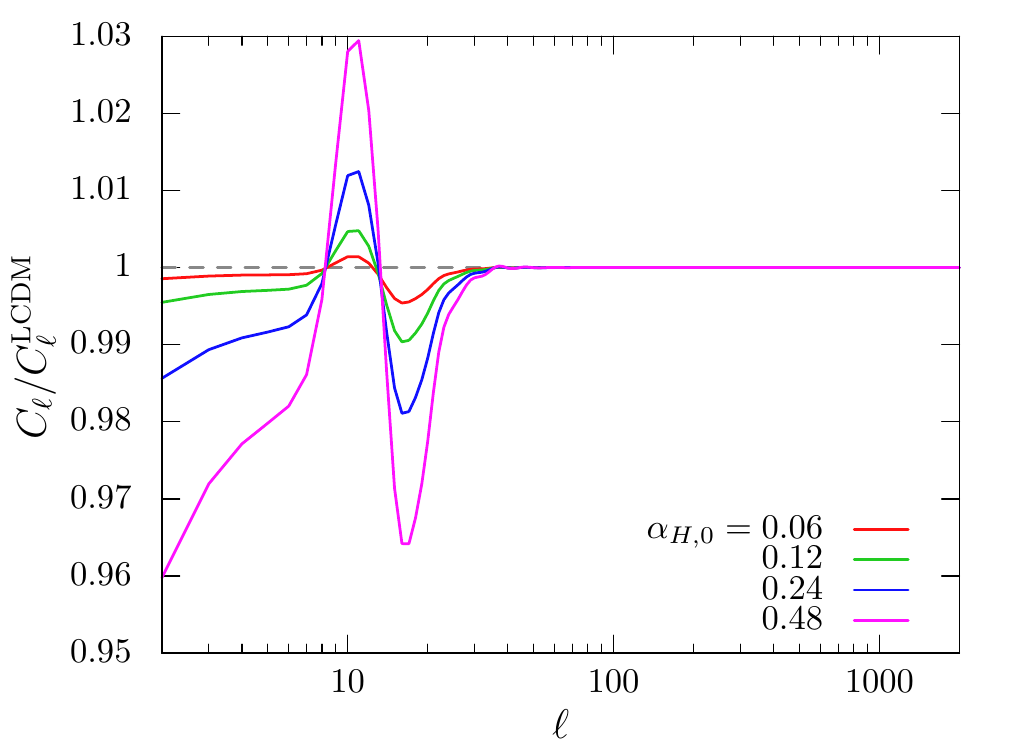}
 \includegraphics[width=5cm]{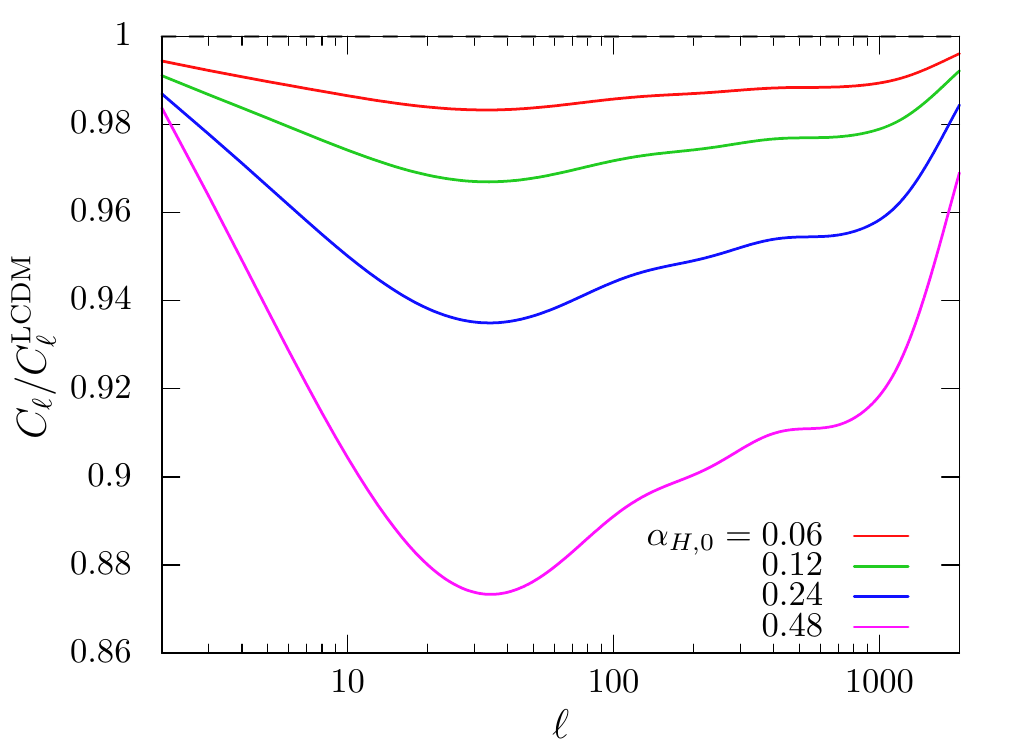}
}
\centering{
 \includegraphics[width=5cm]{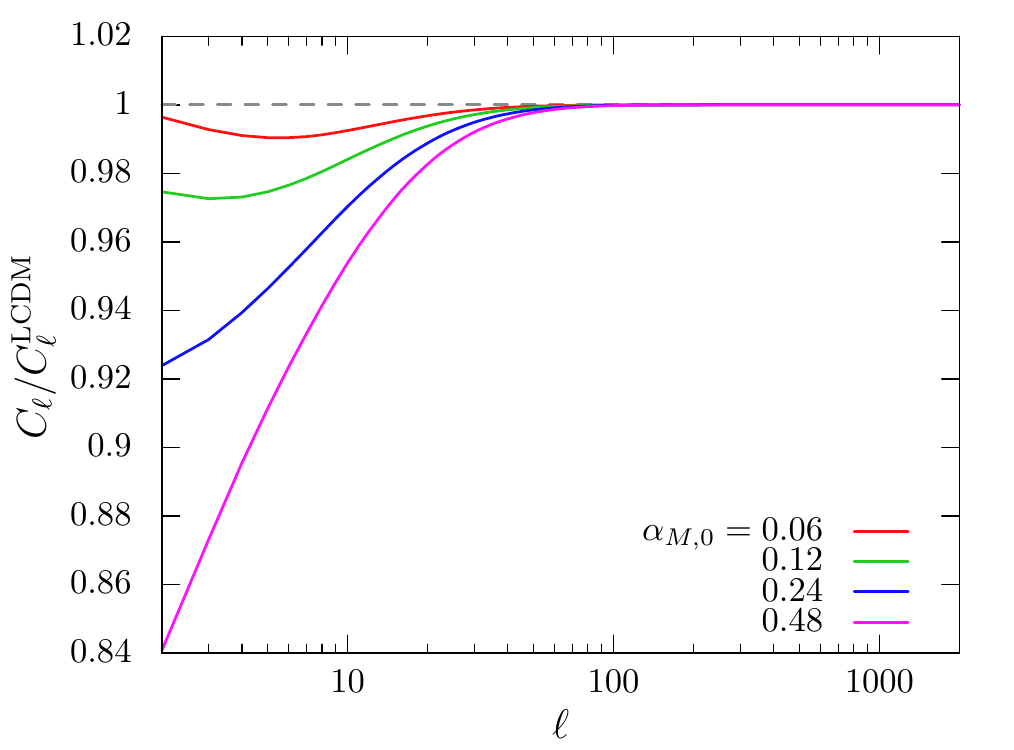}
 \includegraphics[width=5cm]{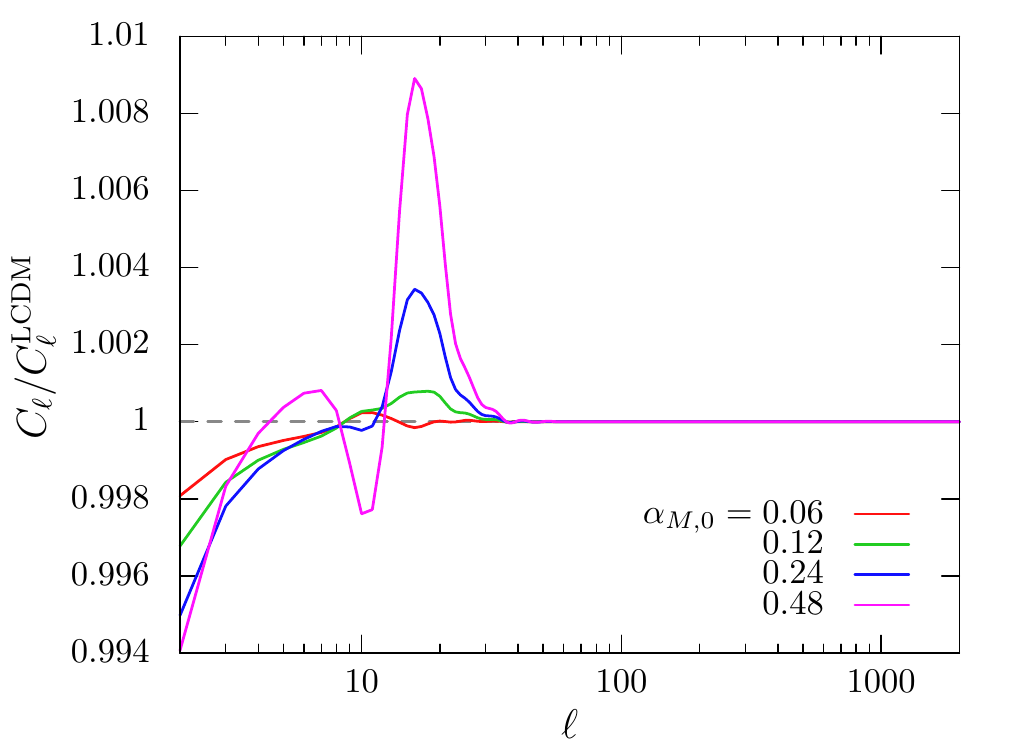}
 \includegraphics[width=5cm]{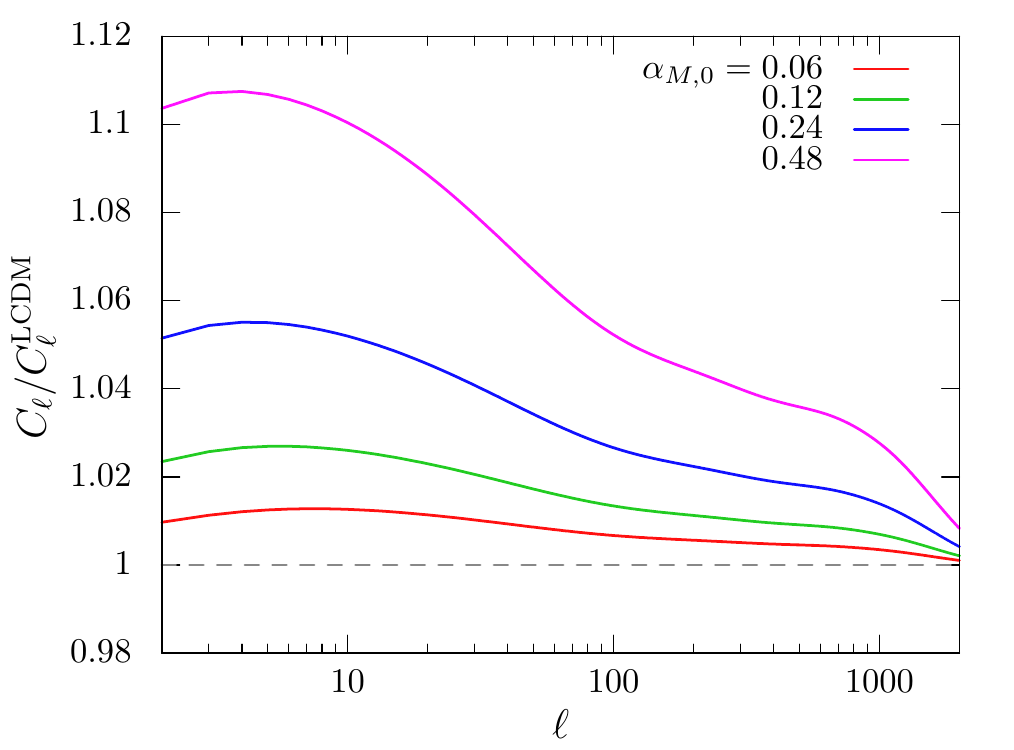}
}
\centering{
 \includegraphics[width=5cm]{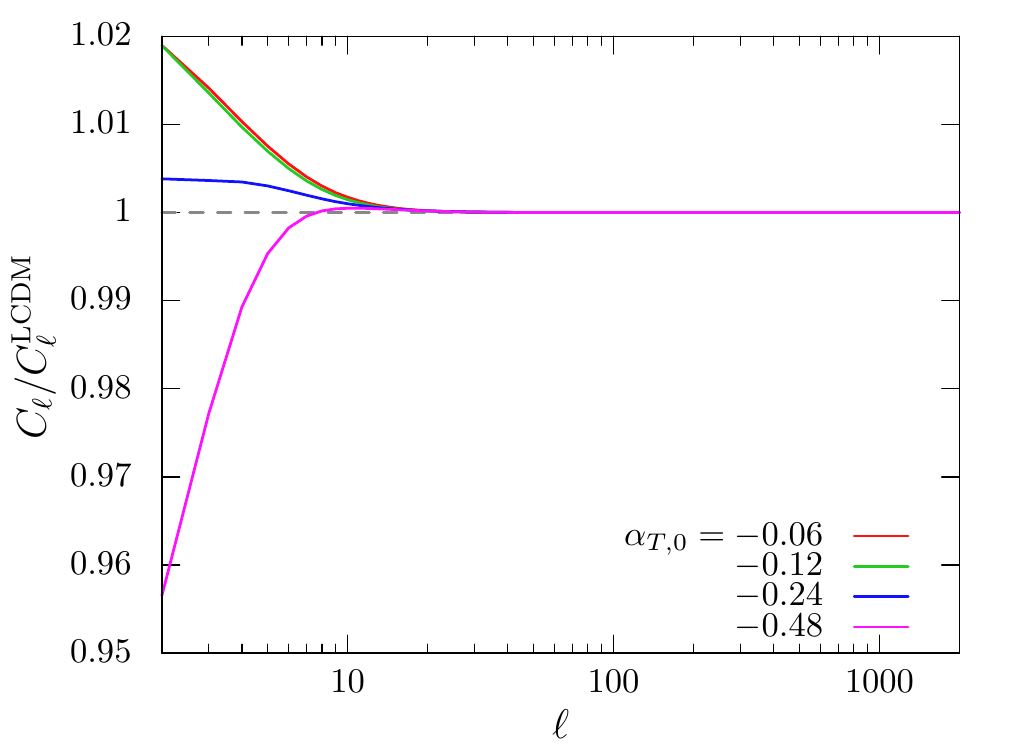}
 \includegraphics[width=5cm]{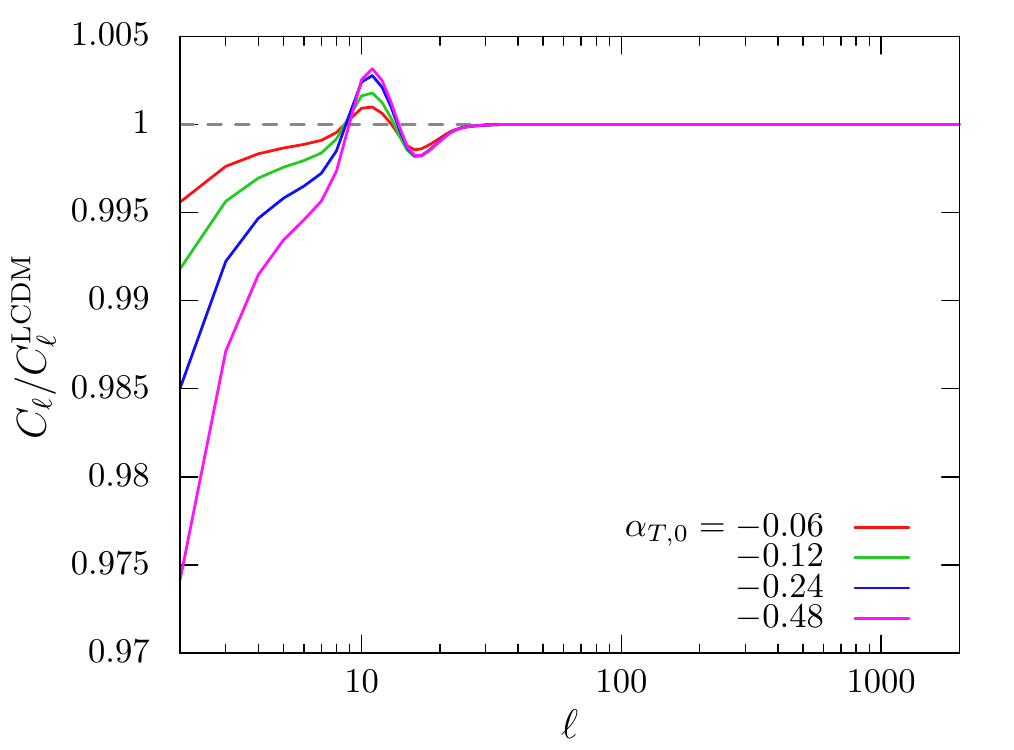}
 \includegraphics[width=5cm]{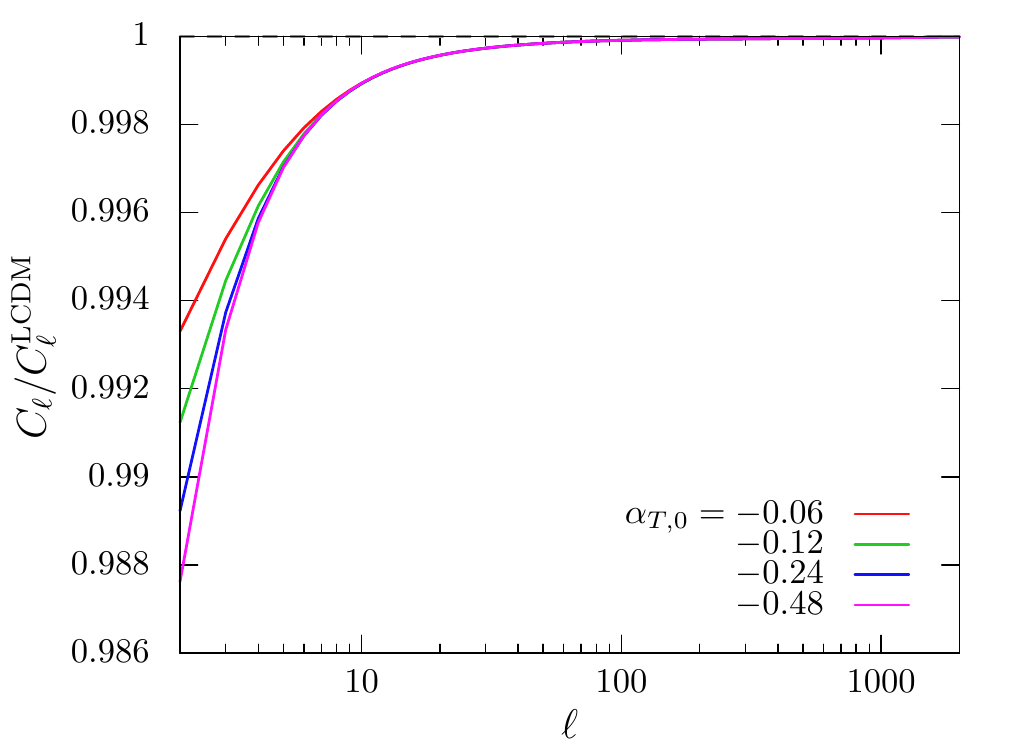}
}
\centering{
 \includegraphics[width=5cm]{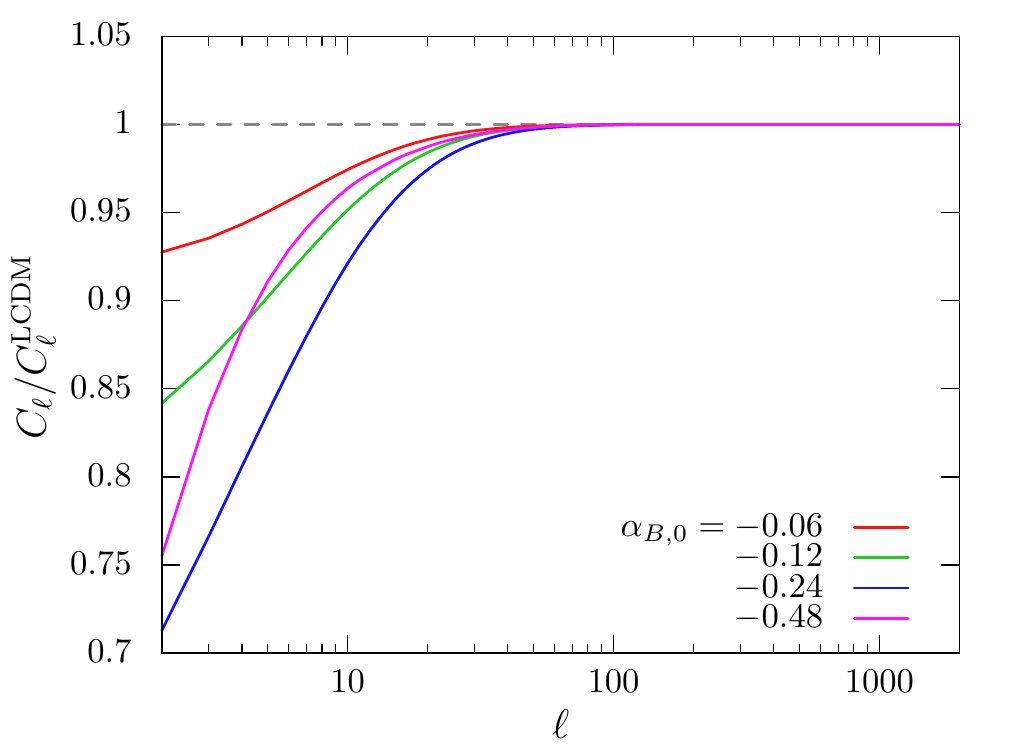}
 \includegraphics[width=5cm]{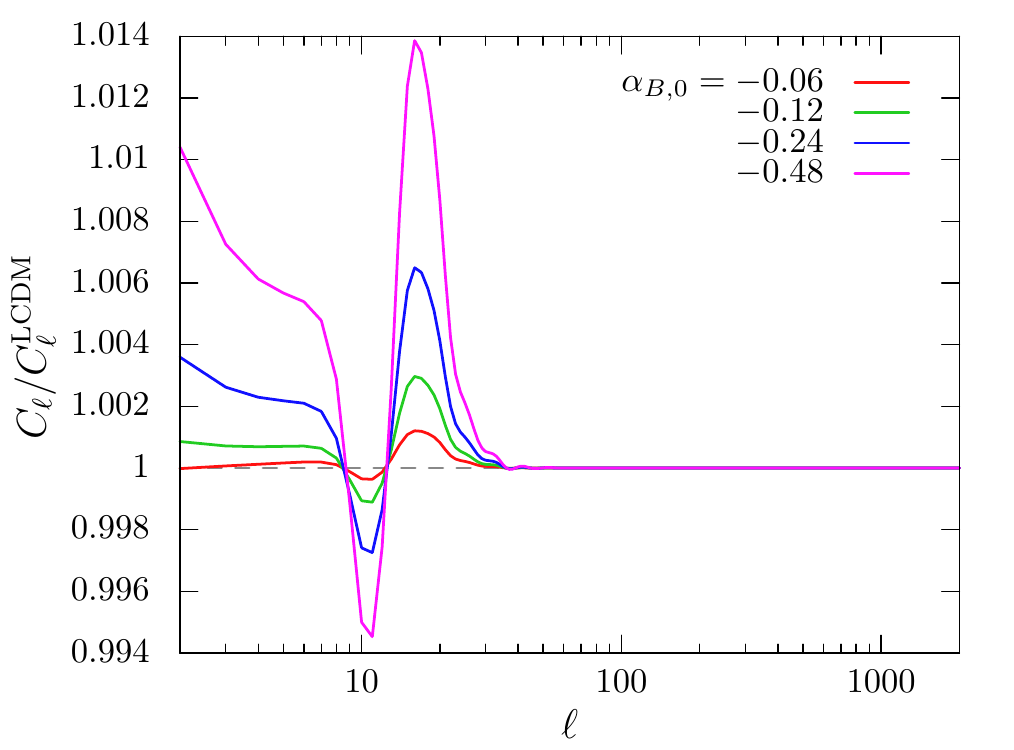}
 \includegraphics[width=5cm]{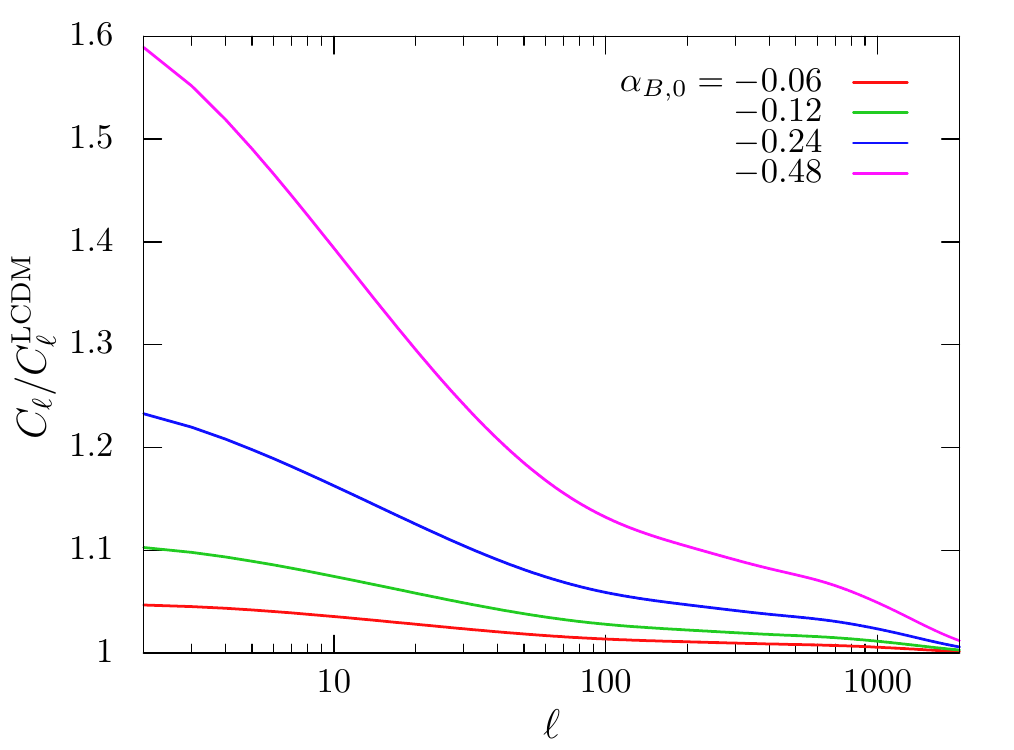}
}
\caption{The same power spectra in Fig.~\ref{fig:TT} divided by those in $\Lambda$CDM.}
\label{fig:change}
\end{figure}

In Fig.~\ref{fig:TT}, we show the angular power spectra of the CMB temperature
anisotropies, $C^{TT}_\ell$ (left), 
E-mode $C^{EE}_\ell$ (middle) and the lensing potential $C^{\phi\phi}_\ell$ (right).
To magnify the changes from the $\Lambda$CDM case, we also show the power spectra
divided by those in $\Lambda$CDM model in Fig.~\ref{fig:change}.
From the top to bottom, we show the parameter dependence on
$\beta_{1,0}$, $\alpha_{H,0}$, $\alpha_{M,0}$, $\alpha_{T,0}$ and $\alpha_{B,0}$.
We vary $\alpha_{i,0}$ with the order of $\mathcal{O}(0.1)$, and $\beta_{1,0}$
with that of $\mathcal{O}(0.01)$.
In the present parametrisation, a small change of $\beta_1$ yields a significant effect on the 
power spectra. 

We find that these parameters affect the angular power spectrum of the
temperature anisotropies only on the large scales through the integrated 
Sachs-Wolfe effect as expected.
In contrast, as the scalar metric perturbations does not directly couple to the 
photon's E-mode polarisation with $\ell=2$, 
the changes of $C^{EE}_{\ell}$ are highly suppressed.
In this sense, the information from the E-mode does not
improve the constrains on the EFT parameters as far as we focus on the scalar metric perturbations.
Taking a look at the power spectrum of the lensing potential
provided in the top-right panel in Fig.~\ref{fig:change} and the panel below this,
the large influences from the two beyond-Horndeski parameters, $\alpha_H$ and $\beta_1$, 
appear on the different scales; non-zero $\alpha_H$ yields the significant change 
from $\Lambda$CDM at $\ell\sim 30$, while $\beta_1$ does at $\ell\sim 3$.
This fact indicates that, in principle, we can distinguish the effects from
each other using the lensing potential, whereas it is difficult to do it
only from the temperature anisotropies since the two effects are quite similar
as shown in the left panels in Fig.~\ref{fig:change}.
Note that the angular power spectra, $C^{TT}_\ell$ and $C^{\phi\phi}_\ell$,
depending on $\alpha_{H,0}$ (the second panels from the top)
and $\alpha_{B,0}$ (the bottom panels) in Fig.~\ref{fig:TT} and Fig.~\ref{fig:change} 
reproduce the results in the pioneering work by D'Amico et al. \cite{DAmico:2016ntq}
(and see also \cite{Traykova:2019oyx}).

To understand the suppression of $C^{\phi\phi}_\ell$ with $\beta_1>0$ on large scales,
we define the following quantity
~\cite{Gleyzes:2015pma,DAmico:2016ntq},
%%%%%%%%%%%%%%%%%%%%%%%%%%%%%%%%%%%%%%%%%%%%%%%%%%%%%%%%%%%%%%%%%%%%
%
\begin{align}
 \mu_{\rm WL} := \frac{2\nabla^2(\Psi-\Phi)}{3a^2H^2\Omega_{m}\delta_{m}}.
\end{align}
%
%%%%%%%%%%%%%%%%%%%%%%%%%%%%%%%%%%%%%%%%%%%%%%%%%%%%%%%%%%%%%%%%%%%%
Here we focus on the case that matter is non-relativistic: 
$\Omega_{m}\delta_{m}\approx\Omega_{b}\delta_{b}+\Omega_{c}\delta_{c}$ with $\Omega_{m}\approx\Omega_{b}+\Omega_{c}$
Since $\mu_{\rm WL}=2$ for the case of the $\Lambda$CDM, $\mu_{\rm WL}-2$ characterises the deviation from the $\Lambda$CDM
in weak lensing observations.
To evaluate this quantity, we study the quasi-static evolution of the perturbations inside the sound horizon scale. 
Under such approximation, it is enough to consider the highest spatial derivative contributions
in Eqs.~(\ref{eq:eq_psi}), (\ref{eq:eq_phi}) and (\ref{eq:eq_pi}).
Combining the governing equation of the density fluctuations,
$\ddot{\delta}_M + 2H\dot{\delta}_M - \frac{1}{a^2}\nabla^2\Psi = 0,$
we obtain $\mu_{\rm WL}$ as a function of $\alpha_i$ and $\beta_1$ in addition to $\Omega_{m}$.

If $\alpha_H$ is non-zero and the others set to be zero, 
we recover the result in Ref.~\cite{DAmico:2016ntq},
%%%%%%%%%%%%%%%%%%%%%%%%%%%%%%%%%%%%%%%%%%%%%%%%%%%%%%%%%%%%%%%%%%%%
%
\begin{align}
 \mu_{\rm WL}-2 = \frac{\alpha_H(8-9\Omega_{m}(1+\Omega_{m}))}{2+3\Omega_{m}(1-\alpha_H)},
\end{align}
%
%%%%%%%%%%%%%%%%%%%%%%%%%%%%%%%%%%%%%%%%%%%%%%%%%%%%%%%%%%%%%%%%%%%%
where we have assumed $\dot\delta_m\approx H\delta_{m}$ for simplicity.
When the denominator of the right-hand side is close to zero, 
the deviation from the $\Lambda$CDM, $\mu_{\rm WL}-2$, 
can be very large. However it is the case only if $\alpha_H\sim \mathcal{O}(1)$. 

The situation drastically changes in the case with
$\beta_1 \ne 0$. If $\beta_1$ is the only non-zero parameter, we obtain
%%%%%%%%%%%%%%%%%%%%%%%%%%%%%%%%%%%%%%%%%%%%%%%%%%%%%%%%%%%%%%%%%%%%
%
\begin{align}
 \mu_{\rm WL}-2 = \frac{6\beta_1[2(1-\Omega_{m})(2+3\Omega_{m}(11-45\Omega_{m}))-9\beta_1\Omega_{m}^2(22+\Omega_{m}(19+3\Omega_{m}))]}{[-2+3\Omega_{m}(3+3\Omega_{m}(-3+\beta_1)+8\beta_1)][-2+9\Omega_{m}(1-3\Omega_{m}+(3+\Omega_{m})\beta_1)]}.
\label{eq:mu_DHOST}
\end{align}
%
%%%%%%%%%%%%%%%%%%%%%%%%%%%%%%%%%%%%%%%%%%%%%%%%%%%%%%%%%%%%%%%%%%%%
In this case, $\mu_{\rm WL}-2$ can be very large even if $\beta_1\sim \mathcal{O}(0.1)$.
That is why small $\beta_1$ has a large impact on $C^{\phi\phi}_\ell$ comparing with the other cases
as shown in Fig.~\ref{fig:change}.

\newcommand{\hyphen}{\mathchar`-}

Next, we estimate the 1-sigma error in estimating the EFT parameters 
in the Fisher analysis
to quantify the significance of the changes from $\Lambda$CDM case. 
We consider CMB-S4~\cite{Abazajian:2016yjj} as a representative future CMB observations,
which is assumed to be the sky coverage, $f_{\rm sky}=0.4$, 
the noise level in measuring temperature anisotropies, $w_T^{-1/2}= 1\mu {\rm K}\cdot$arcmin 
and that for E-mode polarisation, $w_P^{-1/2} = \sqrt{2}\mu {\rm K}\cdot$arcmin with 
a beam width of $1$~arcmin.
%, and measure an EFT parameter assuming 
%that all other parameters are fixed to be zero.
The Fisher matrix is defined as
%%%%%%%%%%%%%%%%%%%%%%%%%%%%%%%%%%%%%%%%%%%%%%%%%%%%%%%%%%%%%%%%%%%%
%
\begin{align}
 F_{ij} &= \sum_{XY}\sum_{\ell}
    \frac{\partial C_{\ell}^X}{\partial\theta_i}
    (\mathscr{C}_\ell^{-1})^{XY}
    \frac{\partial C_{\ell}^Y}{\partial\theta_j},
\end{align}
%
%%%%%%%%%%%%%%%%%%%%%%%%%%%%%%%%%%%%%%%%%%%%%%%%%%%%%%%%%%%%%%%%%%%%
where $X,Y=TT,TE,EE,T\phi,E\phi,\phi\phi$ and
%%%%%%%%%%%%%%%%%%%%%%%%%%%%%%%%%%%%%%%%%%%%%%%%%%%%%%%%%%%%%%%%%%%%
%
\begin{align}
\mathscr{C}_\ell
= \frac{2}{2\ell+1}
\begin{pmatrix}
 \left(C^{\rm TT}_{\ell}\right)^2 & C^{\rm TE}_{\ell} C^{\rm TT}_{\ell} & \left(C^{\rm TE}_{\ell}\right)^2 & C^{T\phi}_{\ell} C^{\rm TT}_{\ell} & C^{\rm TE}_{\ell} C^{T\phi}_{\ell} & \left(C^{T\phi}_{\ell}\right)^2 \\
 C^{\rm TE}_{\ell} C^{\rm TT}_{\ell} & \mathcal{C}^{TE,TE}_\ell & C^{EE}_{\ell} C^{\rm TE}_{\ell} & \mathcal{C}^{TE,T\phi}_\ell & \mathcal{C}^{TE,\phi E}_\ell & C^{E\phi}_{\ell} C^{T\phi}_{\ell} \\
 \left(C^{\rm TE}_{\ell}\right)^2 & C^{EE}_{\ell} C^{\rm TE}_{\ell} & \left(C^{EE}_{\ell}\right)^2 & C^{E\phi}_{\ell} C^{\rm TE}_{\ell} & C^{EE}_{\ell} C^{E\phi}_{\ell} & \left(C^{E\phi}_{\ell}\right)^2 \\
 C^{T\phi}_{\ell} C^{\rm TT}_{\ell} & \mathcal{C}^{TE,T\phi}_\ell & C^{E\phi}_{\ell} C^{\rm TE}_{\ell} & \mathcal{C}^{TT,\phi\phi}_\ell & \mathcal{C}^{TE,\phi\phi}_\ell & C^{\phi\phi}_{\ell} C^{T\phi}_{\ell} \\
 C^{\rm TE}_{\ell} C^{T\phi}_{\ell} & \mathcal{C}^{TE,\phi E} & C^{EE}_{\ell} C^{E\phi}_{\ell} & \mathcal{C}^{TE,\phi\phi}_\ell & \mathcal{C}^{EE,\phi\phi}_\ell & C^{E\phi}_{\ell} C^{\phi\phi}_{\ell} \\
 \left(C^{T\phi}_{\ell}\right)^2 & C^{E\phi}_{\ell} C^{T\phi}_{\ell} & \left(C^{E\phi}_{\ell}\right)^2 & C^{\phi\phi}_{\ell} C^{T\phi}_{\ell} & C^{E\phi}_{\ell} C^{\phi\phi}_{\ell} & \left(C^{\phi\phi}_{\ell}\right)^2
\end{pmatrix}
\end{align}
%
%%%%%%%%%%%%%%%%%%%%%%%%%%%%%%%%%%%%%%%%%%%%%%%%%%%%%%%%%%%%%%%%%%%%
with
%%%%%%%%%%%%%%%%%%%%%%%%%%%%%%%%%%%%%%%%%%%%%%%%%%%%%%%%%%%%%%%%%%%%
%
\begin{align}
\mathcal{C}^{ab,cd}_\ell := \frac{1}{2}\left(C^{ab}_\ell C^{cd}_\ell+C^{ac}_\ell C^{bd}_\ell\right).
\end{align}
%
%%%%%%%%%%%%%%%%%%%%%%%%%%%%%%%%%%%%%%%%%%%%%%%%%%%%%%%%%%%%%%%%%%%%
The model parameters, $\theta_i$, are given as
$\theta_i:=\{\alpha_{B,0},\alpha_{T,0},\alpha_{M,0},\alpha_{H,0},\beta_{1,0}\}$,
and the 1-sigma error of each parameter is given by
$\sigma_{\theta_i} = 1/\sqrt{(F^{-1})_{ii}}$.
The derivative $\partial C_{\ell}^X/\partial\theta_i$ is computed as
%%%%%%%%%%%%%%%%%%%%%%%%%%%%%%%%%%%%%%%%%%%%%%%%%%%%%%%%%%%%%%%%%%%%
%
\begin{align}
 \left.\frac{\partial C^X_\ell}{\partial\theta_i}\right|_{\theta_i=\overline{\theta}_i}
 \approx
 \frac{-C^X_\ell(\overline{\theta}_i+2\delta\theta_i)+4C^X_\ell(\overline{\theta}_i+\delta\theta_i)-3C^X_\ell(\overline{\theta}_i)}{2\delta\theta_i},
\end{align}
%
%%%%%%%%%%%%%%%%%%%%%%%%%%%%%%%%%%%%%%%%%%%%%%%%%%%%%%%%%%%%%%%%%%%%
where the fiducial value in our study is $\overline{\theta}_i=0$.
In Table~\ref{tab:single}, we show the 1-sigma errors
in estimating $\alpha_{i,0}$ for $i=B,T,M,H$ and $\beta_{1,0}$ with keeping $\alpha_{K,0}=1$.
From this rough estimation, we find that the all of the EFT parameters including $\beta_1$ 
could be constrained to the order of $\mathcal{O}(0.1)$ with the CMB-S4 observations.
Moreover the inclusion of the CMB lensing data would improve the constraints by one order of magnitude except for $\alpha_T$.

\begin{table}[t]
 \begin{tabular}[t]{c|ccc}
\hline
 parameter   & TT & TT$+$pol & TT$+$pol+lens \\ \hline\hline
$\beta_{1}$  & 0.27 & 0.21 & 0.013 \\
$\alpha_{H}$ & 1.5  & 0.90 & 0.012 \\
$\alpha_{M}$ & 1.0  & 0.74 & 0.030 \\
$\alpha_{T}$ & 1.2  & 0.47 & 0.37  \\
$\alpha_{B}$ & 0.24 & 0.18 & 0.017 \\ \hline
 \end{tabular}
\caption{1-sigma error for the estimation of the EFT parameters, $\alpha_{i,0}$ with the CMB-S4 observations, The fiducial model is $\Lambda$CDM. }
\label{tab:single}
\end{table}

%%%%%%%%%%%%%%%%%%%%%%%%%%%%%%%%%%%%%%%%%%%%%%%%%%%%%%%%%%%%%%%%%%%%%%%%%%%%%%%
%%%%%%%%%%%%%%%%%%%%%%%%%%%%%%%%%%%%%%%%%%%%%%%%%%%%%%%%%%%%%%%%%%%%%%%%%%%%%%%
\section{Demonstration in a specific model in DHOST theory}
\label{sec:results_ck}
%%%%%%%%%%%%%%%%%%%%%%%%%%%%%%%%%%%%%%%%%%%%%%%%%%%%%%%%%%%%%%%%%%%%%%%%%%%%%%%
%%%%%%%%%%%%%%%%%%%%%%%%%%%%%%%%%%%%%%%%%%%%%%%%%%%%%%%%%%%%%%%%%%%%%%%%%%%%%%%

%%==============================================================
%%==============================================================
\subsection{Background evolution and EFT parameters}
%%==============================================================
%%==============================================================

Up to here, we treat the EFT parameters as free functions of time.
In the DHOST theory, however, these parameters are described by the arbitrary functions
$P(\phi,X)\,,Q(\phi,X)\,,f_2(\phi,X)$ and $a_i(\phi,X)$ for $i=1,\ldots,5$.
They are thus related with each other, and the cosmic expansion history also depends on
these functions.
We demonstrate the case in which the arbitrary functions are parameterised
so that the resultant cosmic expansion is self-accelerated at late time.
To do it, we adopt a parametrisation proposed by Crisostomi and Koyama
where the propagation speed of gravitational waves strictly coincides with the speed of light, 
$\alpha_T=0$. \cite{Crisostomi:2017pjs}.

To solve Eqs.~(\ref{eq:background_phi0}) and (\ref{eq:background_Na}), 
we fix the arbitrary functions, $P$, $Q$, $f_2$ and $a_i$.
The condition, $\alpha_T=0$, reads $a_1=a_2=0$ from the first condition in 
Eq.~(\ref{eq:degeneracy_DHOST}) and Eq.~(\ref{eq:param_T})~\cite{Langlois:2017dyl}.
Respecting this additional condition and the degeneracy condition given in Eq.~(\ref{eq:degeneracy_DHOST}), 
one finds that the remaining free functions are $P,Q,f_2$ and $a_3$.
In Ref.~\cite{Crisostomi:2017pjs}, the authors propose the following parametrisation,
%%%%%%%%%%%%%%%%%%%%%%%%%%%%%%%%%%%%%%%%%%%%%%%%%%%%%%%%%%%%%%%%%%%%%%%%%%%%%%%
%
\begin{align}
 P &= c_2X, \quad
 Q = \frac{c_3}{\Lambda^3}X, \quad
 f_2 = \frac{\Mpl^2}{2}+ c_4\frac{X^2}{\Lambda^6}, \quad
 B_1 := \frac{X}{4f_2}\left(4f_{2X} + a_3X\right) = -\frac{X^2}{\frac{\Mpl^2}{2}+c_4\frac{X^2}{\Lambda^6}}\frac{\beta}{4\Lambda^6}.
\end{align}
%
%%%%%%%%%%%%%%%%%%%%%%%%%%%%%%%%%%%%%%%%%%%%%%%%%%%%%%%%%%%%%%%%%%%%%%%%%%%%%%%
This model has the shift symmetry, $\phi \to \phi+{\rm const.}$, and 
is parameterised by four constants, $c_2, c_3, c_4$
and $\beta$. The new energy scale $\Lambda$ is given as
$\Lambda = (\Mpl H^2)^{1/3}$.

Rescaling the time coordinate and the scalar field 
as $t\rightarrow \Mpl^{1/2}\Lambda^{-3/2}t$ and $\phi_0\rightarrow \Mpl\phi_{0}$,
we can reduce the equations (\ref{eq:background_phi0}) and (\ref{eq:background_Na})
to those without any scales.
The acceleration equation in Eq.~(\ref{eq:background_Na})
can be solved with respect to $\dot{H}$.
Using this, we can eliminate $\dot{H}$ and $\ddot{H}$ 
in Eqs.~(\ref{eq:background_phi0}) and (\ref{eq:background_Na}).
Eventually, these equations can be expressed
in a simpler form as
$U_1(\chi,a)\dot{\chi} +U_2(\chi,a) = 0\,,\  \dot{a}/a=U_3(\chi,a)$ 
where $\chi := \dot{\phi}_0$. We do not explicitly show $U_i$, but they are given as
functions of the cosmological parameters as well as the model parameters $c_i,\beta$.
The initial value of $\chi$ is not sensitive to the final results since $\chi$
follows its attractor solution in the later time.
After solving these equations, we can then rewrite the EFT parameters, 
$\alpha_i(t)$ and $\beta_1(t)$, in terms of $\chi(t)$, $c_i$ and $\beta$.

\begin{figure}[!ht]
\centering{
 \includegraphics[width=8cm]{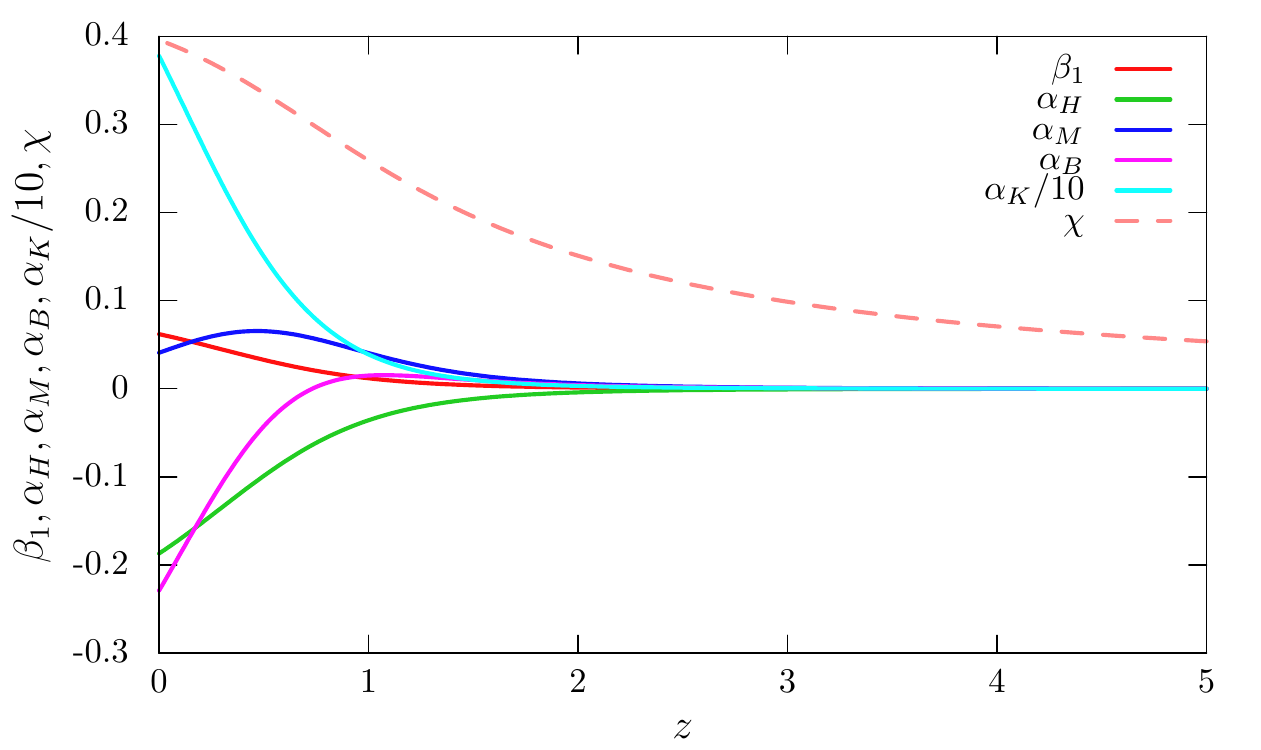}
}
\caption{Time evolution of $\beta_1$, $\alpha_i$ with $i=H,M,B,K$ (solid line) and $\chi(=\dot{\phi}_0)$ (dashed line) in the CK model with $(c_2,c_3,c_4,\beta)=(3.0,5.0,1.0,-5.3)$. As $\alpha_K$ is an order of magnitude larger than the others, we multiply it by $1/10$.}
\label{fig:back_CK}
\end{figure}

The time-evolution of $\alpha_i$, $\beta_1$ and $\chi$ are depicted in Fig.~\ref{fig:back_CK}. As is shown in this figure, the EFT parameters become significant only at small $z$. In particular, in this model, $\beta_{1}=(\beta/16c_4)\alpha_{H}$ is always satisfied.
In the present case, $\beta_{1}, \alpha_{H}$ and $\alpha_{K}$ are monotonically growing in time, while $\alpha_{M}$ and $\alpha_{B}$ are not. The non-monotonic behaviour of the EFT parameters has been pointed out in Ref.~\cite{Arai:2019zul}, where $\beta_{1}$ is 'oscillated' at low $z$.

%%==============================================================
%%==============================================================
\subsection{Angular power spectra}
%%==============================================================
%%==============================================================

\begin{figure}[!ht]
\centering{
 \includegraphics[width=5.5cm]{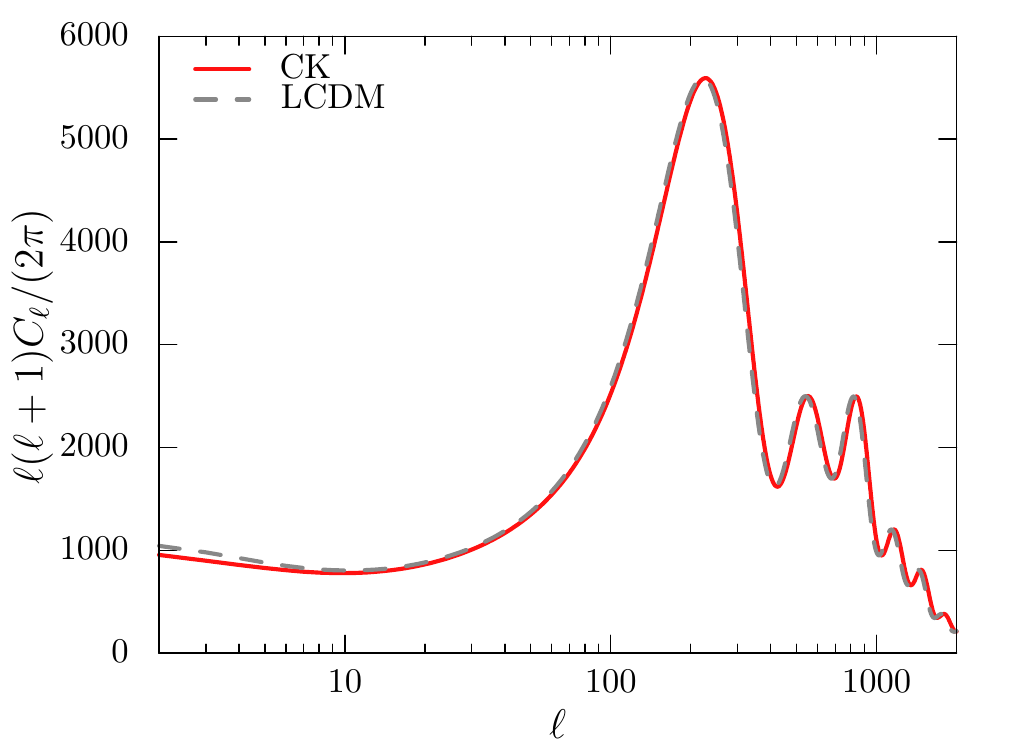}
 \includegraphics[width=5.5cm]{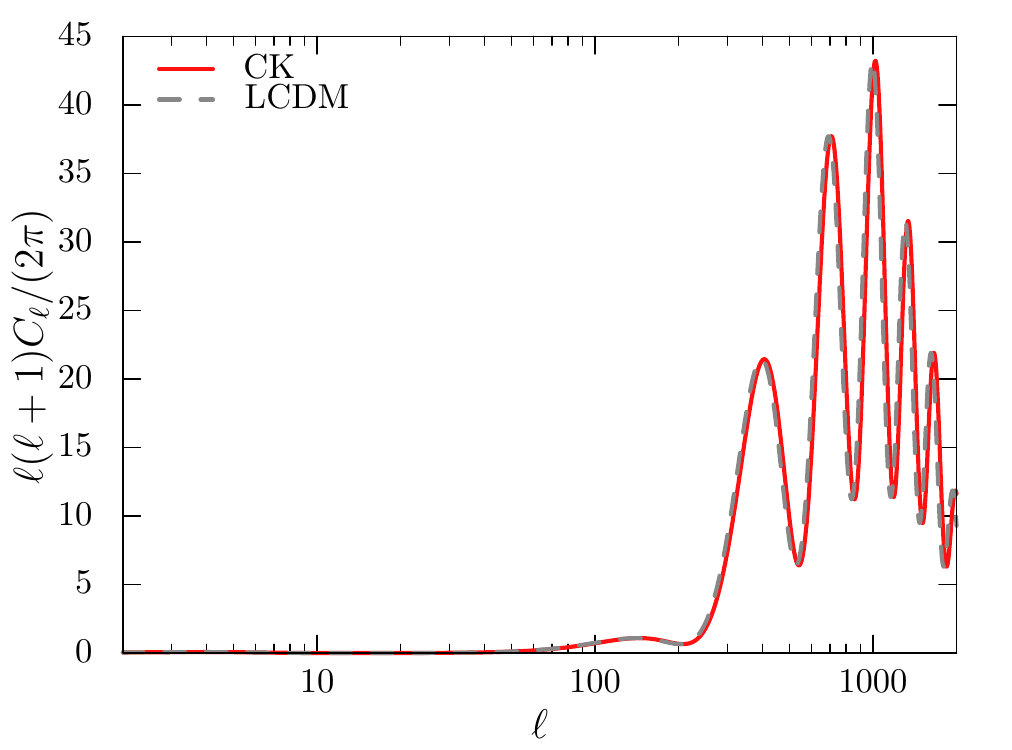}
 \includegraphics[width=5.5cm]{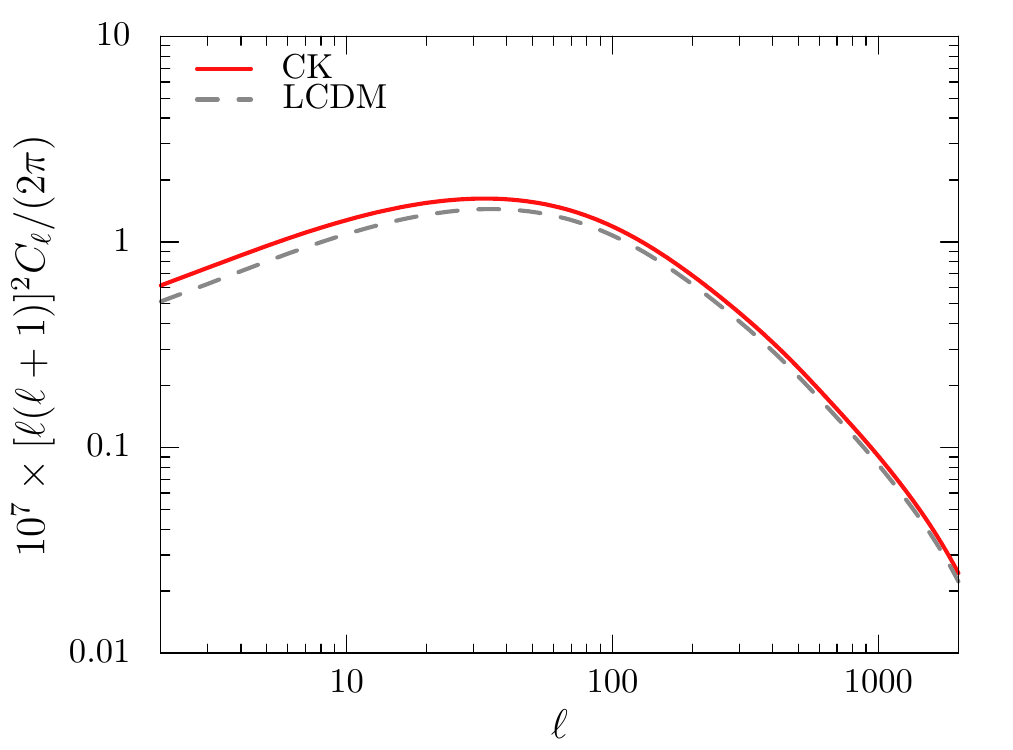}
}
\caption{Angular power spectra in CK model. From left to right, we show the angular power spectra
of temperature ($C^{TT}_\ell$), E-mode ($C^{EE}_\ell$) and lensing potential ($C^{\phi\phi}_{\ell}$).}
\label{fig:TT_CK}
\end{figure}

\begin{figure}[!ht]
\centering{
 \includegraphics[width=5.5cm]{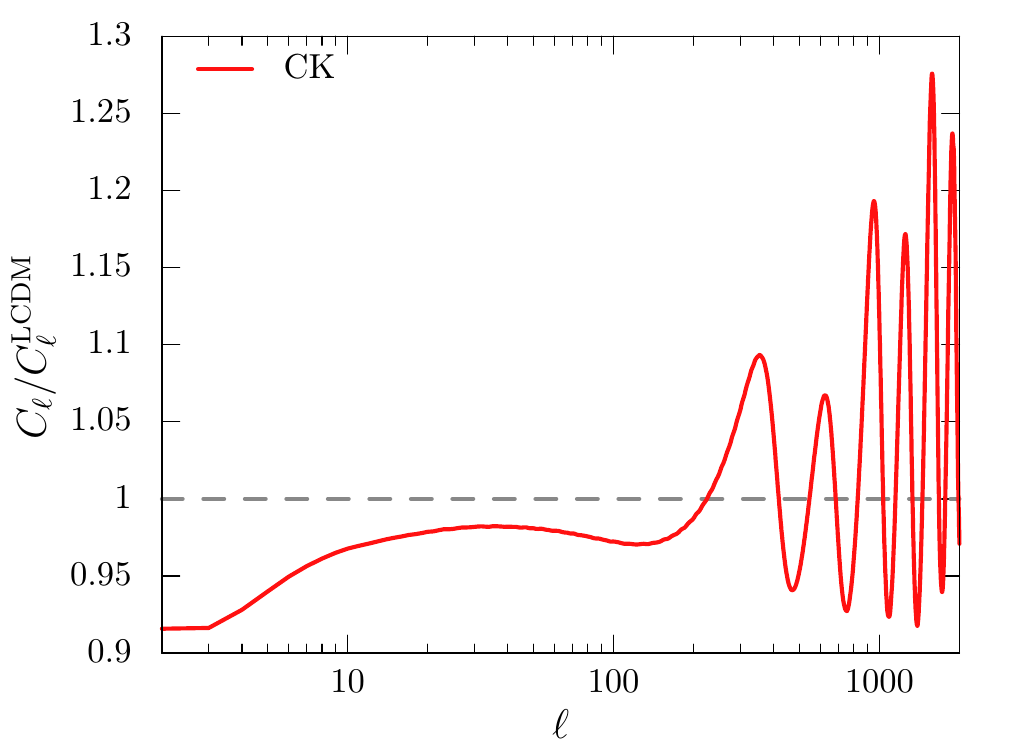}
 \includegraphics[width=5.5cm]{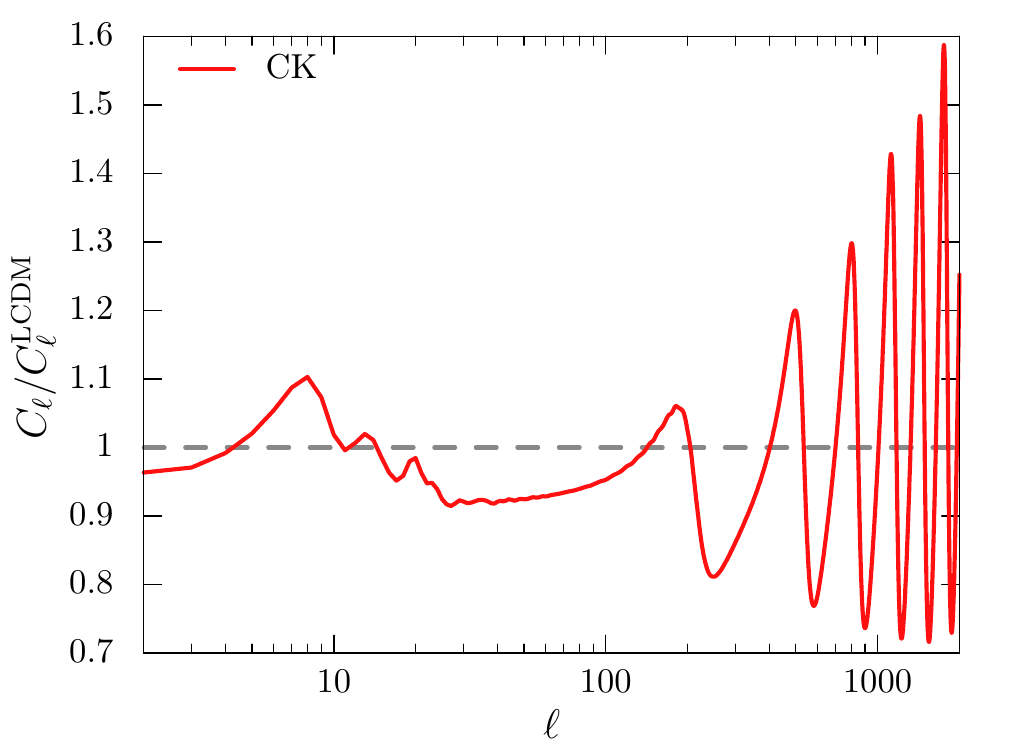}
 \includegraphics[width=5.5cm]{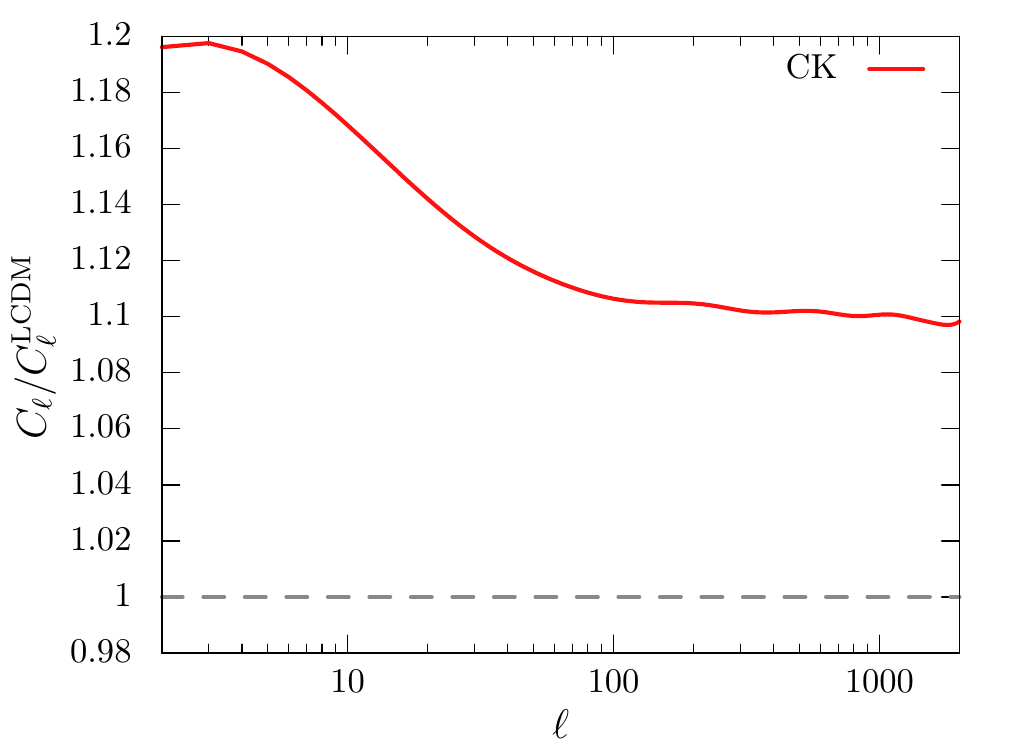}
}
\caption{The same power spectra in Fig.~\ref{fig:TT_CK} divided by those in $\Lambda$CDM.}
\label{fig:change_CK}
\end{figure}

The angular power spectra, $C^{TT}_\ell, C^{EE}_\ell$ and $C^{\phi\phi}_\ell$, in the CK model with $(c_2,c_3,c_4,\beta)=(3.0,5.0,1.0,-5.3)$ are shown in Figs.~\ref{fig:TT_CK} and \ref{fig:change_CK}.
Although $\beta_1$ and $\alpha_H$ deviate from zero more than the range that we show in Fig.~\ref{fig:TT}, $C^{TT}_\ell$ and $C^{EE}_\ell$ are not significantly deviated from those in $\Lambda$CDM on large scales. That is because these parameters are correlated so that the large negative $\alpha_H$ cancel the large positive $\beta_1$.

In contrast, one can observe the large deviation from $\Lambda$CDM on small scales. The choice of parameters, $c_2,c_3,c_4$ and $\beta$, in this demonstration recovers the cosmic expansion history in $\Lambda$CDM as reported in Ref.~\cite{Crisostomi:2017pjs}. There is, however, a small change of expansion history around the beginning of the dark energy epoch at $z \lesssim 1$. This fact induces a small change of the angular diameter distance of the horizon scale at the last scattering surface measured from us, and thus the peak location of the acoustic oscillations on small scales is a little bit shifted.

The small change of the angular diameter distance significantly affects $C^{\phi\phi}_\ell$ over the whole range of $\ell$ that can be observed in the present time. However, it does not immediately lead to the observability of these signals, since we cannot directly observe $C^{\phi\phi}_\ell$, but it is reconstructed from the combination of other observations such as the large-scale structure. We thus envisage that a large error induced from the reconstruction process makes it difficult to constrain the CK model only from $C^{\phi\phi}_\ell$.
Our present study does not intend to mention how well we can constrain the CK model from these angular power spectra. 
Hence we leave the detail analysis for the observability for future study.

%%%%%%%%%%%%%%%%%%%%%%%%%%%%%%%%%%%%%%%%%%%%%%%%%%%%%%%%%%%%%%%%%%%%%%%%%%%%%%%
%%%%%%%%%%%%%%%%%%%%%%%%%%%%%%%%%%%%%%%%%%%%%%%%%%%%%%%%%%%%%%%%%%%%%%%%%%%%%%%
\section{Conclusion}
\label{sec:conclusion}
%%%%%%%%%%%%%%%%%%%%%%%%%%%%%%%%%%%%%%%%%%%%%%%%%%%%%%%%%%%%%%%%%%%%%%%%%%%%%%%
%%%%%%%%%%%%%%%%%%%%%%%%%%%%%%%%%%%%%%%%%%%%%%%%%%%%%%%%%%%%%%%%%%%%%%%%%%%%%%%

In the present study, we have investigated the impact of 
the deviation from GR on the angular power spectra of CMB anisotropies using the 
type-I degenerate higher-order scalar-tensor (DHOST) theory.
We first formulated the linear perturbations in the DHOST theory
and their effective description parameterised by time-varying EFT parameters,
resulting in the governing equation of the metric perturbation $\Phi$ in 
Eq.~(\ref{eq:DHOST_eq_dPhi}) and that of the scalar perturbation 
$\pi:=-\delta\phi/\dot{\phi}_0$ in Eq.~(\ref{eq:DHOST_eq_ddpi}).
Based on the {effective description}, we developed a Boltzmann solver 
implementing the DHOST theory.

We then obtain the angular power spectra of the CMB temperature anisotropies ($C^{TT}_\ell$), 
E-mode polarisation ($C^{EE}_\ell$) and the lensing potential ($C^{\phi\phi}_\ell$) 
using the parametrisation given in Eq.~(\ref{eq:scaling1}).
In Figs.~\ref{fig:TT} and \ref{fig:change}, we show 
these angular power spectra
and those normalised by the spectra in $\Lambda$CDM model
as our main results.
In Eq.~(\ref{eq:mu_DHOST}), we derived the deviation in weak lensing observations 
from the $\Lambda$CDM model, $\mu_{\rm WL}-2$, when $\beta_1\ne 0$ and the other 
EFT parameters are set to be zero.
From this, we found that the deviation becomes significant even if $\beta_1$ is small.
The top-right panel in Fig.~\ref{fig:change} clarifies this fact from our numerical computation, 
and the large change of growth history of the metric perturbations gives a significant impact 
on the CMB temperature anisotropies as shown in the top-left panel in Fig.~\ref{fig:change}.
On the other hand, the E-mode polarisation is not so sensitive to this, since the 
polarisation mode does not directly couple to gravity but is affected only through the 
quadrupole moment of the temperature anisotropy.

We estimate the 1-sigma error in estimating the EFT parameters by computing the Fisher
matrix assuming the CMB-S4 observations. The results are summarised in 
Table \ref{tab:single}.

Finally, we demonstrate a specific model proposed by Ref.~\cite{Crisostomi:2017pjs}
which is a subclass of the DHOST theory with $\alpha_T=0$.
In our EFT approach, the background is fixed to be $\Lambda$CDM, while
in this specific model all of the EFT parameters as well as the cosmic expansion history are 
consistently determined from the time-evolution of the background scalar field $\phi_0(t)$.
The resultant angular power spectra with $(c_2,c_3,c_4,\beta)=(3.0,5.0,1.0,-5.3)$,
a parameter set proposed in Ref.~\cite{Crisostomi:2017pjs} realising the self-accelerating Universe, 
are shown in Figs.~\ref{fig:TT_CK} and \ref{fig:change_CK}.
As there are degeneracies among the EFT parameters, the parameters can vary in a larger range 
keeping the cosmic expansion history similar to that in $\Lambda$CDM model as shown in Fig.~\ref{fig:back_CK}.
This is not the case when only one of the parameters can be varied with $\alpha_{K,0}=1$.
In this specific model, 
we find the $8\%$ suppression from $\Lambda$CDM in the temperature anisotropies
on large scales, and $\mathcal{O}(10)\%$ deviation on small scales 
caused by the small change of the 
angular diameter distance to the last-scattering surface due to the tiny change of the cosmic expansion
history around the transition to the dark energy domination epoch.
As for the lensing potential, there are huge deviation from the $\Lambda$CDM model over the whole range of 
angular scales. However this fact does not immediately conclude that it is easy to put a strong constraint 
on the deviation from $\Lambda$CDM, since the lensing potential should be reconstructed through a statistical process.
In addition, it depends on how to parameterise the arbitrary functions in the DHOST theory to put constrains
on the deviation from $\Lambda$CDM.
Hence we leave the quantitative study for the future.

\acknowledgments{
 We would like to thank Tsutomu Kobayashi, Shinji Mukohyama, Kazuya Koyama, Atsushi Naruko, 
 Kazufumi Takahashi, Shin'ichi Hirano for giving fruitful comments and helpful discussion.
 This work was supported in part by MEXT/JSPS KAKENHI Grants No. JP17H06359, JP19H01891 (D. Y.).
}

\appendix

%%%%%%%%%%%%%%%%%%%%%%%%%%%%%%%%%%%%%%%%%%%%%%%%%%%%%%%%%%%%%%%%%%%%%%%%%%%%%%%
%%%%%%%%%%%%%%%%%%%%%%%%%%%%%%%%%%%%%%%%%%%%%%%%%%%%%%%%%%%%%%%%%%%%%%%%%%%%%%%
\section{Euler-Lagrange equations for the background variables}
\label{appsec:background}
%%%%%%%%%%%%%%%%%%%%%%%%%%%%%%%%%%%%%%%%%%%%%%%%%%%%%%%%%%%%%%%%%%%%%%%%%%%%%%%
%%%%%%%%%%%%%%%%%%%%%%%%%%%%%%%%%%%%%%%%%%%%%%%%%%%%%%%%%%%%%%%%%%%%%%%%%%%%%%%

The variations of the Lagrangian in the gravity sector defined in Eq.~(\ref{eq:def_EA}) 
with respect to $N,a$ and $\phi$ are computed as
%%%%%%%%%%%%%%%%%%%%%%%%%%%%%%%%%%%%%%%%%%%%%%%%%%%%%%%%%%%%%%%%%%%%
%
\begin{align}
\mathcal{E}_N &:=
P+6f_{2} H^2+6H f_{2\phi}\dot{\phi}_0+\left(-9a_{1} H^2+2P_{X}-Q_{\phi}+12f_{2X}\left(2H^2+\dot{H}\right)
+a_{2}\left(-9H^2+6\dot{H}\right)\right)\dot{\phi}_0^2
\notag \\ &
+6H\left(a_{2\phi}-Q_{X}\right)\dot{\phi}_0^3+\left(6H^2\left(a_{1X}+3a_{2X}\right)-3a_{3}\left(3H^2+\dot{H}\right)\right)\dot{\phi}_0^4-3H a_{3\phi}\dot{\phi}_0^5
\notag \\ &
+\left(6H\left(a_{1}-2f_{2X}\right)\dot{\phi}_0+2\left(a_{1\phi}+a_{2\phi}\right)\dot{\phi}_0^2-3 (a_{3}+2a_{4}) H\dot{\phi}_0^3-2\left(a_{3\phi}+a_{4\phi}\right)\dot{\phi}_0^4+6a_{5} H\dot{\phi}_0^5+2a_{5\phi}\dot{\phi}_0^6\right)\ddot{\phi}_0
\notag \\ &
+\left(-a_{1}-a_{2}+\left(-a_{3}-a_{4}-2\left(a_{1X}+a_{2X}\right)\right)\dot{\phi}_0^2+\left(3a_{5}+2\left(a_{3X}+a_{4X}\right)\right)\dot{\phi}_0^4-2a_{5X}\dot{\phi}_0^6\right)\ddot{\phi}_0^2
\notag \\ &
+\left(2 (a_{1}+a_{2})\dot{\phi}_0-2 (a_{3}+a_{4})\dot{\phi}_0^3+2a_{5}\dot{\phi}_0^5\right)\dddot{\phi}_0
\label{eq:var_N}
,\\
-\frac{a}{3}\mathcal{E}_a &:= 
-P-2f_{2}\left(3H^2+2\dot{H}\right)-4H f_{2\phi}\dot{\phi}_0+\left(-2f_{2\phi\phi}-Q_{\phi}+(a_{1}+3a_{2})\left(3H^2+2\dot{H}\right)\right)\dot{\phi}_0^2+2H\left(a_{1\phi}+3a_{2\phi}\right)\dot{\phi}_0^3
\notag \\ &
+\left(-2f_{2\phi}+4H\left(a_{1}+3a_{2}+2f_{2X}\right)\dot{\phi}_0+2\left(a_{2\phi}+4f_{2\phi X}+Q_{X}\right)\dot{\phi}_0^2-4H\left(a_{1X}+3a_{2X}\right)\dot{\phi}_0^3-a_{3\phi}\dot{\phi}_0^4\right)\ddot{\phi}_0
\notag \\ &
+\left(-a_{1}+a_{2}+4f_{2X}+\left(-2a_{3}+a_{4}-4\left(a_{2X}+2f_{2XX}\right)\right)\dot{\phi}_0^2+\left(-a_{5}+2a_{3X}\right)\dot{\phi}_0^4\right)\ddot{\phi}_0^2
\notag \\ &
+\left(2\left(a_{2}+2f_{2X}\right)\dot{\phi}_0-a_{3}\dot{\phi}_0^3\right)\dddot{\phi}_0
\label{eq:var_a}
,
\end{align}
%
%%%%%%%%%%%%%%%%%%%%%%%%%%%%%%%%%%%%%%%%%%%%%%%%%%%%%%%%%%%%%%%%%%%%
and
%%%%%%%%%%%%%%%%%%%%%%%%%%%%%%%%%%%%%%%%%%%%%%%%%%%%%%%%%%%%%%%%%%%%
%
\begin{align}
\mathcal{E}_{\phi} &:= 
P_{\phi}+6f_{2\phi}\left(2H^2+\dot{H}\right)
\notag \\ &
+6\left(12H^3f_{2X}+H P_{X}-H Q_{\phi}+14H f_{2X} \dot{H}-a_{1}\left(3H^3+2H \dot{H}\right)+2f_{2X} \ddot{H}+a_{2}\left(3H \dot{H}+\ddot{H}\right)\right)\dot{\phi}_0
\notag \\ &
+\left(2P_{\phi X}-3H^2\left(a_{1\phi}-9a_{2\phi}-8f_{2\phi X}+6Q_{X}\right)-Q_{\phi\phi}-6\left(-2\left(a_{2\phi}+f_{2\phi X}\right)+Q_{X}\right) \dot{H}\right)\dot{\phi}_0^2
\notag \\ &
+\left(6H\left(a_{2\phi\phi}-Q_{\phi X}+\left(a_{1X}+3a_{2X}\right)\left(3H^2+2\dot{H}\right)\right)-3a_{3}\left(9H\left(H^2+\dot{H}\right)+\ddot{H}\right)\right)\dot{\phi}_0^3
\notag \\ &
+\left(6H^2\left(a_{1\phi X}+3a_{2\phi X}-3a_{3\phi}\right)-6a_{3\phi} \dot{H}\right)\dot{\phi}_0^4-3H a_{3\phi\phi}\dot{\phi}_0^5
\notag \\ &
+\left(2\left(P_{X}-Q_{\phi}+3a_{1}\left(2H^2+\dot{H}\right)+6f_{2X}\left(2H^2+\dot{H}\right)+a_{2}\left(9H^2+6\dot{H}\right)\right)+6H\left(2a_{1\phi}+5a_{2\phi}-2Q_{X}\right)\dot{\phi}_0
\right. \notag \\ & \left.\quad
+\Bigl(30H^2a_{1X}+2\left(a_{1\phi\phi}+a_{2\phi\phi}-2P_{XX}+Q_{\phi X}\right)+6a_{2X}\left(9H^2-2\dot{H}\right)-24f_{2XX}\left(2H^2+\dot{H}\right)
\right.
\notag \\ &
\left.\quad\quad
-3 (5a_{3}+2a_{4})\left(3H^2+\dot{H}\right)\Bigr)\dot{\phi}_0^2
%\right. \notag \\ & \left.\quad
+3H\left(-9a_{3\phi}-4\left(a_{2\phi X}+a_{4\phi}\right)+4Q_{XX}\right)\dot{\phi}_0^3
\right. \notag \\ & \left.\quad
+\left(6H^2\left(-2a_{1XX}-6a_{2XX}+3a_{3X}\right)-2\left(a_{3\phi\phi}+a_{4\phi\phi}\right)+6a_{3X} \dot{H}+6a_{5}\left(3H^2+\dot{H}\right)\right)\dot{\phi}_0^4
\right. \notag \\ & \left.\quad
+6H\left(a_{3\phi X}+2a_{5\phi}\right)\dot{\phi}_0^5+2a_{5\phi\phi}\dot{\phi}_0^6\right)\ddot{\phi}_0
\notag \\ &
+\left(3\left(a_{1\phi}+a_{2\phi}\right)-18H\left(a_{3}+a_{4}+a_{1X}+a_{2X}\right)\dot{\phi}_0-3\left(2\left(a_{1\phi X}+a_{2\phi X}\right)+3a_{3\phi}+3a_{4\phi}\right)\dot{\phi}_0^2
\right. \notag \\ & \left.\quad
+18H\left(2a_{5}+a_{3X}+a_{4X}\right)\dot{\phi}_0^3+\left(6\left(a_{3\phi X}+a_{4\phi X}\right)+15a_{5\phi}\right)\dot{\phi}_0^4-18H a_{5X}\dot{\phi}_0^5-6a_{5\phi X}\dot{\phi}^6\right)\ddot{\phi}_0^2
\notag \\ &
+\left(-2\left(a_{3}+a_{4}+a_{1X}+a_{2X}\right)+2\left(6a_{5}+2\left(a_{1XX}+a_{2XX}\right)+5a_{3X}+5a_{4X}\right)\dot{\phi}_0^2
\right. \notag \\ & \left.\quad
-2\left(2\left(a_{3XX}+a_{4XX}\right)+9a_{5X}\right)\dot{\phi}_0^4+4a_{5XX}\dot{\phi}_0^6\right)\ddot{\phi}_0^3
\notag \\ &
+\left(12 (a_{1}+a_{2}) H+4\left(a_{1\phi}+a_{2\phi}\right)\dot{\phi}_0-12 (a_{3}+a_{4}) H\dot{\phi}_0^2-4\left(a_{3\phi}+a_{4\phi}\right)\dot{\phi}_0^3+12a_{5} H\dot{\phi}_0^4+4a_{5\phi}\dot{\phi}_0^5
\right. \notag \\ & \left.\quad
+\left(-8\left(a_{3}+a_{4}+a_{1X}+a_{2X}\right)\dot{\phi}_0+8\left(2a_{5}+a_{3X}+a_{4X}\right)\dot{\phi}_0^3-8a_{5X}\dot{\phi}_0^5\right)\ddot{\phi}\right)\dddot{\phi}_0
\notag \\ &
+\left(2 (a_{1}+a_{2})-2 (a_{3}+a_{4})\dot{\phi}_0^2+2a_{5}\dot{\phi}_0^4\right)\ddddot{\phi}_0,
\end{align}
%
%%%%%%%%%%%%%%%%%%%%%%%%%%%%%%%%%%%%%%%%%%%%%%%%%%%%%%%%%%%%%%%%%%%%
where we set $N=1$, and the subscripts $\phi$ and $X$ stand for the derivative with respect to them.
Here all the functions are evaluated at the background values, $\phi=\phi_0(t)$ and $X=-\dot\phi_0^2(t)$.

%%%%%%%%%%%%%%%%%%%%%%%%%%%%%%%%%%%%%%%%%%%%%%%%%%%%%%%%%%%%%%%%%%%%%%%%%%%%%%%
%%%%%%%%%%%%%%%%%%%%%%%%%%%%%%%%%%%%%%%%%%%%%%%%%%%%%%%%%%%%%%%%%%%%%%%%%%%%%%%
\section{Quadratic Lagrangian in Newtonian gauge}
\label{appsec:qlag_term}
%%%%%%%%%%%%%%%%%%%%%%%%%%%%%%%%%%%%%%%%%%%%%%%%%%%%%%%%%%%%%%%%%%%%%%%%%%%%%%%
%%%%%%%%%%%%%%%%%%%%%%%%%%%%%%%%%%%%%%%%%%%%%%%%%%%%%%%%%%%%%%%%%%%%%%%%%%%%%%%

The quadratic Lagrangian defined in Eq.~(\ref{eq:qlag}) 
after $\pi$ is recovered by the coordinate transformation $t\to t+\pi(t,\xx)$ is
given as
%%%%%%%%%%%%%%%%%%%%%%%%%%%%%%%%%%%%%%%%%%%%%%%%%%%%%%%%%%%%%%%%%%%%%%%%%%%%%%%
%
\begin{align}
\frac{2}{M^2}\mathcal{L}^{(2)}_0&=
-H^2(12\alpha_{B}-\alpha_{K}+6(1+\alpha_{L}))\Psi^2
-12H\beta_{1}\dot{\Psi}\Psi
+12H^2K_{1}\beta_{1}\dot{\Psi}\pi
+2H^2(-6\alpha_{B}+\alpha_{K})\dot{\pi}\Psi 
\notag \\ &
+12H^3K_{1}(1+\alpha_{B}+\alpha_{L})\pi\Psi 
+H^2\alpha_{K}\dot{\pi}^2
+12H(1+\alpha_{B}+\alpha_{L})\dot{\Phi}\Psi 
+12\beta_{1}\dot{\Phi}\dot{\Psi}
+\beta_{2}\dot{\Psi}^2
-12H\beta_{1}\ddot{\pi}\Psi
\notag \\ &
+2\beta_{2}\ddot{\pi}\dot{\Psi}
-6(1+\alpha_{L})\dot{\Phi}^2
+12H\alpha_{B}\dot{\Phi}\dot{\pi}
-12H^2K_{1}(1+\alpha_{L})\dot{\Phi}\pi
+12\beta_{1}\dot{\Phi}\ddot{\pi}
\notag \\ &
+4\left(
-H(1+\alpha_{B}+\alpha_{L})\Psi
-\beta_{1}\dot{\Psi}
+(1+\alpha_{L})\dot{\Phi}
-\frac{\alpha_{L}}{6}\triangle\xi
-H\alpha_{B}\dot{\pi}
+H^2K_{1}(1+\alpha_{L})\pi
-\beta_{1}\ddot{\pi}
\right)\triangle\xi
\notag \\ &
+\frac{1}{6}(\triangle\dot{\eta})^2
-\frac{2}{3}\triangle\xi\triangle\dot{\eta}
+12H^3K_{1}\alpha_{B}\dot{\pi}\pi
+\beta_{2}\ddot{\pi}^2
+12H^2K_{1}\beta_{}1\ddot{\pi}\pi
-6H^4K_{1}^2(1+\alpha_{L})\pi^2
,\\
\frac{2}{M^2}\mathcal{L}^{(2)}_2&=
-\beta_{3}\left[\Psi\triangle\Psi+2\Psi\triangle\dot{\pi}+\dot{\pi}\triangle\dot{\pi}\right]
\notag \\ &
-2\left[2(1+\alpha_{H})\Psi+2(1+\alpha_{H})\dot{\pi}+(1+\alpha_{T})\Phi+2(1+\alpha_{T})H\pi\right]\triangle\Phi
\notag \\ &
+\frac{2}{3}\left((1+\alpha_{T})\triangle\Phi+ H(1+\alpha_{T})\triangle\pi-\frac{1}{12}(1+\alpha_{T})\triangle\triangle\eta
+(1+\alpha_{H})\triangle\Psi+(1+\alpha_{H})\triangle\dot{\pi}\right)\triangle\eta
\notag \\ &
+\left(4(-1+\alpha_{B}-\alpha_{H})H\dot{\pi}-4(1+\alpha_{L})\dot{\Phi}+4(\alpha_{B}-\alpha_{H}+\alpha_{L})H\Psi
\right. \notag \\ & \left.\quad
-2(1+2K_{1}(1+\alpha_{L})+\alpha_{T})H^2\pi+4\beta_{1}\ddot{\pi}+4\beta_{1}\dot{\Psi}+\frac{4\alpha_{L}}{3}\triangle\xi
+\frac{2}{3}\triangle\dot{\eta}\right)\triangle\pi
,\\
\frac{2}{M^2}\mathcal{L}^{(2)}_4
&=
  -\frac{2}{3}\alpha_{L}(\triangle\pi)^2
,
\end{align}
%
%%%%%%%%%%%%%%%%%%%%%%%%%%%%%%%%%%%%%%%%%%%%%%%%%%%%%%%%%%%%%%%%%%%%%%%%%%%%%%%
where we integrate by part with respect to the spatial coordinates.

%%%%%%%%%%%%%%%%%%%%%%%%%%%%%%%%%%%%%%%%%%%%%%%%%%%%%%%%%%%%%%%%%%%%%%%%%%%%%%%
%%%%%%%%%%%%%%%%%%%%%%%%%%%%%%%%%%%%%%%%%%%%%%%%%%%%%%%%%%%%%%%%%%%%%%%%%%%%%%%
\section{Euler-Lagrange equations for metric and scalar perturbations}
\label{appsec:eq_metric}
%%%%%%%%%%%%%%%%%%%%%%%%%%%%%%%%%%%%%%%%%%%%%%%%%%%%%%%%%%%%%%%%%%%%%%%%%%%%%%%
%%%%%%%%%%%%%%%%%%%%%%%%%%%%%%%%%%%%%%%%%%%%%%%%%%%%%%%%%%%%%%%%%%%%%%%%%%%%%%%

Varying the effective quadratic Lagrangian Eq.~(\ref{eq:quad_action}) with 
respect to $\Psi, \Phi, \xi, \eta$ and $\pi$, and taking into account the 
terms describing the background,
we obtain the Euler-Lagrange equations,
%%%%%%%%%%%%%%%%%%%%%%%%%%%%%%%%%%%%%%%%%%%%%%%%%%%%%%%%%%%%%%%%%%%%%%%%%%%%%%%
%
\begin{align}
-\frac{1}{M^2}\mathcal{E}_\Psi &=
\beta_{2}\dddot{\pi}
+\beta_{2}\ddot{\Psi}
+6\beta_{1}\ddot{\Phi}
+\left(H(6\beta_{1}+(3+\alpha_{M})\beta_{2})+\dot{\beta}_{2}\right)\ddot{\pi}
+\left(H(3+\alpha_{M})\beta_{2}+\dot{\beta}_{2}\right)\dot{\Psi}
\notag \\ &
+6\left(-H(1+\alpha_{B}+\alpha_{L})+H(3+\alpha_{M})\beta_{1}+\dot{\beta}_{1}\right)\dot{\Phi}
+H^2(6\alpha_{B}-\alpha_{K}+6K_{1}\beta_{1})\dot{\pi}
\notag \\ &
+\left(H^2(6+12\alpha_{B}-\alpha_{K}+6\alpha_{L}-6(3+K_{1}+\alpha_{M})\beta_{1})+\frac{2\rho_{s}}{M^2}-6H\dot{\beta}_{1}\right)\Psi
+\frac{1}{a^2}(2\beta_{1}+\beta_{3})\triangle\dot{\pi}
\notag \\ &
+\frac{1}{a^2}\beta_{3}\triangle\Psi
+\frac{2}{a^2}(1+\alpha_{H})\triangle\Phi
+\frac{2}{a^2}\left(H(-\alpha_{B}+\alpha_{H}-\alpha_{L}+\beta_{1}+\alpha_{M}\beta_{1})+\dot{\beta}_{1}\right)\triangle\pi
\notag \\ &
+\left(6H^2\left(H(-K_{1}(1+\alpha_{B}+\alpha_{L})+(K_{2}+K_{1}(3+\alpha_{M}))\beta_{1})+K_{1}\dot{\beta}_{1}\right)+\frac{\dot{\rho}_{s}}{M^2}\right)\pi
\label{eq:DHOSTEFT1},\\
-\frac{1}{M^2}\mathcal{E}_\Phi &=
6\beta_{1}\dddot{\pi}
+6\beta_{1}\ddot{\Psi}
-6(1+\alpha_{L})\ddot{\Phi}
+6\left(H(\alpha_{B}+(3+\alpha_{M})\beta_{1})+\dot{\beta}_{1}\right)\ddot{\pi}
\notag \\ &
+6\left(H(1+\alpha_{B}+\alpha_{L}+(3+\alpha_{M})\beta_{1})+\dot{\beta}_{1}\right)\dot{\Psi}
+\left(-6H(1+\alpha_{L})(3+\alpha_{M})-6\dot{\alpha}_{L}\right)\dot{\Phi}
\notag \\ &
+\left(6H\left(H(K_{1}(-1+\alpha_{B}-\alpha_{L})+\alpha_{B}(3+\alpha_{M}))+\dot{\alpha}_{B}\right)-\frac{3(p_{s}+\rho_{s})}{M^2}\right)\dot{\pi}
\notag \\ &
+\frac{2}{a^2}(\alpha_{H}-\alpha_{L})\triangle\dot{\pi}
+\frac{2}{a^2}(1+\alpha_{H})\triangle\Psi
+\frac{2}{a^2}(1+\alpha_{T})\triangle\Phi
-\frac{2}{a^2}\left(H(\alpha_{M}+\alpha_{L}(1+\alpha_{M})-\alpha_{T})+\dot{\alpha}_{L}\right)\triangle\pi
\notag \\ &
+\left(6H\left(H(1+\alpha_{B}+\alpha_{L})(3+K_{1}+\alpha_{M})+\dot{\alpha}_{B}+\dot{\alpha}_{L}\right)-\frac{3(p_{s}+\rho_{s})}{M^2}\right)\Psi
-\frac{6p_{s}}{M^2}\Phi 
\notag \\ &
-6H^3\left((1+\alpha_{L})(K_{2}+K_{1}(3+\alpha_{M}))+\frac{1}{H}K_{1}\dot{\alpha}_{L}+\frac{\dot{p}_{s}}{2H^3M^2}\right)\pi
\label{eq:DHOSTEFT2},\\
-\frac{1}{M^2}\mathcal{E}_\xi &=
2\beta_{1}\triangle\ddot{\pi}
+2\beta_{1}\triangle\dot{\Psi}
-2(1+\alpha_{L})\triangle\dot{\Phi}
+2H\alpha_{B}\triangle\dot{\pi}
+2H(1+\alpha_{B}+\alpha_{L})\triangle\Psi 
-\frac{2}{3a^2}\alpha_{L}\triangle\triangle\pi
\notag \\ &
+\left(-2H^2K_{1}(1+\alpha_{L})-\frac{p_{s}+\rho_{s}}{M^2}\right)\triangle\pi
\label{eq:DHOSTEFT3},\\
-\frac{1}{M^2}\mathcal{E}_\eta &=
-\frac{1}{3a^2}\left[
 \alpha_{H}\triangle\triangle\dot{\pi}
+(1+\alpha_{H})\triangle\triangle\Psi
+(1+\alpha_{T})\triangle\triangle\Phi
-H(\alpha_{M}-\alpha_{T})\triangle\triangle\pi
\right],
\label{eq:DHOSTEFT4}
\end{align}
%
%%%%%%%%%%%%%%%%%%%%%%%%%%%%%%%%%%%%%%%%%%%%%%%%%%%%%%%%%%%%%%%%%%%%%%%%%%%%%%%
and the equation for $\pi$ becomes
%%%%%%%%%%%%%%%%%%%%%%%%%%%%%%%%%%%%%%%%%%%%%%%%%%%%%%%%%%%%%%%%%%%%%%%%%%%%%%%
%
\begin{align}
-\frac{1}{M^2}\mathcal{E}_\pi &=
-\beta_{2}\ddddot{\pi}
-\beta_{2}\dddot{\Psi}
-6\beta_{1}\dddot{\Phi}
-2\left(H(3+\alpha_{M})\beta_{2}+\dot{\beta}_{2}\right)\dddot{\pi}
+\left(6H\beta_{1}-2H(3+\alpha_{M})\beta_{2}-2\dot{\beta}_{2}\right)\ddot{\Psi}
\notag \\ &
+\left(6H(\alpha_{B}-2(3+\alpha_{M})\beta_{1})-12\dot{\beta}_{1}\right)\ddot{\Phi}
\notag \\ &
+H^2\left(\alpha_{K}-12K_{1}\beta_{1}-(3+\alpha_{M})(3+K_{1}+\alpha_{M})\beta_{2}-\frac{1}{H}\beta_{2}\dot{\alpha}_{M}-\frac{2}{H}(3+\alpha_{M})\dot{\beta}_{2}-\frac{\ddot{\beta}_{2}}{H^2}\right)\ddot{\pi}
\notag \\ &
-H^2\left(6\alpha_{B}-\alpha_{K}-6(6+K_{1}+2\alpha_{M})\beta_{1}+(3+\alpha_{M})(3+K_{1}+\alpha_{M})\beta_{2}
\right. \notag \\ &\left.\quad
+\frac{1}{H}\left(\beta_{2}\dot{\alpha}_{M}-12\dot{\beta}_{1}+2(3+\alpha_{M})\dot{\beta}_{2}\right)+\frac{\ddot{\beta}_{2}}{H^2}\right)\dot{\Psi}
\notag \\ &
+6H^2\left((3+\alpha_{M})(\alpha_{B}-(3+\alpha_{M})\beta_{1})+K_{1}(1+\alpha_{B}+\alpha_{L}-(3+\alpha_{M})\beta_{1})
\right. \notag \\ &\left.\quad
+\frac{1}{H}\left(\dot{\alpha}_{B}-\beta_{1}\dot{\alpha}_{M}-2(3+\alpha_{M})\dot{\beta}_{1}\right)-\frac{\ddot{\beta}_{1}}{H^2}+\frac{p_{s}+\rho_{s}}{2M^2H^2}\right)\dot{\Phi}
\notag \\ &
+H^2\left(H(\alpha_{K}(3+2K_{1}+\alpha_{M})-12(K_{2}+K_{1}(3+\alpha_{M}))\beta_{1})+\dot{\alpha}_{K}-12K_{1}\dot{\beta}_{1}\right)\dot{\pi}
+\frac{2\alpha_{L}}{3a^4}\triangle\triangle\pi
\notag \\ &
-\frac{1}{a^2}(4\beta_{1}+\beta_{3})\triangle\ddot{\pi}
-\frac{1}{a^2}(2\beta_{1}+\beta_{3})\triangle\dot{\Psi}
-\frac{2}{a^2}(\alpha_{H}-\alpha_{L})\triangle\dot{\Phi}
-\frac{1}{a^2}\left(H(1+\alpha_{M})(4\beta_{1}+\beta_{3})+4\dot{\beta}_{1}+\dot{\beta}_{3}\right)\triangle\dot{\pi}
\notag \\ &
-\frac{1}{a^2}\left(H(2(\alpha_{B}-\alpha_{H}+\alpha_{L})+(1+\alpha_{M})\beta_{3})+\dot{\beta}_{3}\right)\triangle\Psi
-\frac{2}{a^2}\left(H(\alpha_{M}+\alpha_{H}(1+\alpha_{M})-\alpha_{T})+\dot{\alpha}_{H}\right)\triangle\Phi
\notag \\ &
+\frac{2H^2}{a^2}\left(K_{1}+(1+K_{1}+\alpha_{M})(\alpha_{B}-\alpha_{H})+2K_{1}\alpha_{L}-\alpha_{M}+\alpha_{T}-(1+\alpha_{M})(1+K_{1}+\alpha_{M})\beta_{1}
\right. \notag \\ &\left.\quad
+\frac{1}{H}\left(\dot{\alpha}_{B}-\dot{\alpha}_{H}-\beta_{1}\dot{\alpha}_{M}-2(1+\alpha_{M})\dot{\beta}_{1}\right)-\frac{\ddot{\beta}_{1}}{H^2}+\frac{p_{s}+\rho_{s}}{2M^2H^2}\right)\triangle\pi
\notag \\ &
+H^3\left((\alpha_{K}-6\alpha_{B})(3+\alpha_{M})+6\left(K_{2}+(3+\alpha_{M})^2\right)\beta_{1}+2K_{1}(-3-9\alpha_{B}+\alpha_{K}-3\alpha_{L}+9(3+\alpha_{M})\beta_{1})
\right. \notag \\ & \left.\quad
%+\frac{1}{H}\left(-6\dot{\alpha}_{B}+\dot{\alpha}_{K}+6\beta_{1}\dot{\alpha}_{M}+12(3+K_{1}+\alpha_{M})\dot{\beta}_{1}\right)+\frac{6}{H^2}\ddot{\beta}_{1}-\frac{3(p_{s}+\rho_{s})}{M^2H^2}\right)\Psi
+\frac{1}{H}\left(-6\dot{\alpha}_{B}+\dot{\alpha}_{K}+6\beta_{1}\dot{\alpha}_{M}+12(3+K_{1}+\alpha_{M})\dot{\beta}_{1}\right)+\frac{6}{H^2}\ddot{\beta}_{1}+\frac{\dot{\rho}_s}{M^2H^3}\right)\Psi
\notag \\ &
+6H^4\left(K_{2}\alpha_{B}-(K_{3}+2K_{2}(3+\alpha_{M}))\beta_{1}+K_{1}(3+\alpha_{M})(\alpha_{B}-(3+\alpha_{M})\beta_{1})+K_{1}^2(1+\alpha_{B}+\alpha_{L}-(3+\alpha_{M})\beta_{1})
\right. \notag \\ &\left.\quad
+\frac{1}{H}\left(K_{1}\left(\dot{\alpha}_{B}-\beta_{1}\dot{\alpha}_{M}\right)-2(K_{2}+K_{1}(3+\alpha_{M}))\dot{\beta}_{1}\right)-\frac{K_{1}}{H^2}\ddot{\beta}_{1}-\frac{\dot{p}_{s}+\dot{\rho}_{s}}{2M^2H^3}-\frac{\ddot{\rho}_{s}}{6M^2H^4}\right)\pi
.
\label{eq:DHOSTEFT5}
\end{align}
%
%%%%%%%%%%%%%%%%%%%%%%%%%%%%%%%%%%%%%%%%%%%%%%%%%%%%%%%%%%%%%%%%%%%%%%%%%%%%%%%

\end{document}